\documentclass[twocolumn]{aastex631}

\usepackage{amsmath}
\usepackage{hyperref}

\usepackage{booktabs}
\usepackage{graphicx}

\received{23-Sep, 2022}
\accepted{06-Jan, 2023}
\submitjournal{ApJ}
\shorttitle{A Spectroscopic Analysis of a Sample of K2 Planet-Host Stars}
\shortauthors{Loaiza-Tacuri et al.}

\graphicspath{{./}{figures/}}

\begin{document}

\title{A Spectroscopic Analysis of a Sample of K2 Planet-Host Stars: Stellar Parameters, Metallicities and Planetary Radii}

\correspondingauthor{Ver\'onica Loaiza Tacuri}
\email{vtacuri@on.br}

\author[0000-0003-0506-8269]{V. Loaiza-Tacuri}
\affiliation{Observat\'orio Nacional, Rua General Jos\'e Cristino, 77, 20921-400 S\~ao Crist\'ov\~ao, Rio de Janeiro, RJ, Brazil}

\author[0000-0001-6476-0576]{Katia Cunha}
\affiliation{Steward Observatory, University of Arizona, 933 North Cherry Avenue, Tucson, AZ 85721, USA}
\affiliation{Observat\'orio Nacional, Rua General Jos\'e Cristino, 77, 20921-400 S\~ao Crist\'ov\~ao, Rio de Janeiro, RJ, Brazil}

\author[0000-0002-0134-2024]{Verne V. Smith}
\affiliation{NSF's NOIRLab, 950 North Cherry Avenue, Tucson, AZ 85719, USA}

\author{Cintia F. Martinez}
\affiliation{Instituto de Astronomía y Física del Espacio, Universidad de Buenos Aires, Ciudad Universitaria S/N, Buenos Aires, Argentina}

\author{Luan Ghezzi}
\affil{Universidade Federal do Rio de Janeiro, Observatório do Valongo, Ladeira do Pedro Antônio, 43, Rio de Janeiro, RJ 20080-090, Brazil}

\author{Simon C. Schuler}
\affiliation{University of Tampa, 401 W Kennedy Boulevard, Tampa, FL 33606, USA}

\author{Johanna Teske}
\affil{Carnegie Earth \& Planets Lab, 5241 Broad Branch Road, NW, Washington, DC 20015, USA}

\author{Steve B. Howell}
\affiliation{NASA Ames Research Center, Moffett Field, CA 94035, USA}

\begin{abstract}

The physical properties of transiting exoplanets are connected with the physical properties of their host stars. We present a homogeneous spectroscopic analysis based on spectra of FGK-type stars observed with the Hydra spectrograph on the WIYN telescope. We derived effective temperatures, surface gravities, and metallicities, for 81 stars observed by K2 and 33 from Kepler 1. We constructed an Fe I and II line list that is adequate for the analysis of R$\sim$18,000 spectra covering 6050--6350 \AA\ and adopted the spectroscopic technique based on equivalent width measurements. The calculations were done in LTE using Kurucz model atmospheres and the qoyllur-quipu ($q^2$) package. We validated our methodology via analysis of a benchmark solar twin and solar proxies, which are used as the solar reference. We estimated the effects that including Zeeman sensitive Fe I lines have on the derived stellar parameters for young and possibly active stars in our sample and found it not to be significant. Stellar masses and radii were derived by combining the stellar parameters with Gaia EDR3 and V magnitudes and isochrones. The measured stellar radii have 4.2\% median internal precision, leading to a median internal uncertainty of 4.4\% in the derived planetary radii. With our sample of 83 confirmed planets orbiting K2 host stars, the radius gap near R$_{planet}\sim$1.9R$_\oplus$ is detected, in agreement with previous findings. Relations between the planetary radius, orbital period and metallicity are explored and these also confirm previous findings for Kepler 1 systems.

\end{abstract}


%---------------------------------------------
\section{Introduction} \label{sec:intro}
%---------------------------------------------
To date, according to the NASA Exoplanet Archive\footnote{\url{https://exoplanetarchive.ipac.caltech.edu/}} and Exoplanet Catalog\footnote{\url{https://exoplanets.nasa.gov/discovery/exoplanet-catalog/}}, there are  more than $\sim$ 5000 detected exoplanets orbiting $\sim$ 3800 parent stars, with most of these planets ($\sim$3900) discovered via the transit method. 
This impressive number of planets detected via transits so far, is thanks to the Corot misson \citep[37 planets;][]{deleuil2000,deleuil2018}, the more recent TESS mission \citep[249 planets;][]{ricker2015} and mostly thanks to the Kepler Mission \citep[2708 planets;][]{borucki2010, koch2010, borucki2016}, along with the extended K2 mission \citep[537 planets;][]{howell2014}. The original Kepler mission (Kepler 1; operating between 2009 -- 2013) pointed at a single field-of-view in the constellations of Cygnus and Lyra. For the K2 mission (operating between 2014 -- 2018), the Kepler telescope was re-oriented to point at different fields along the ecliptic plane for about 80 days each, with a latency period between them as the spacecraft orbited the Sun. One advantage of the change in targeting strategy during the Kepler mission, due to the loss of two of the guidance gyros in the Kepler telescope, is that K2 observed Galactic targets in regions of the disk that Kepler 1 had not reached.
It is also relevant to note that K2 targeted a larger and more diverse sample compared to Kepler 1.

Determining stellar atmospheric parameters (effective temperatures, metallicities and surface gravities) of exoplanet host stars is crucial to exoplanet studies because host-star parameters must be known with precision in order to derive precise fundamental planetary properties.
One key stellar parameter that needs to be determined as precisely as possible is the stellar radius, as planetary transits essentially measure the ratio of the planet radius to stellar radius. Using a quantitative high-resolution stellar spectroscopic analysis to derive more precise stellar radii, the California Kepler Survey \citep[CKS;][]{petigura2017} made the important discovery of a bimodal distribution in the radii of small planets, where the separation between the peaks falls at R$_{\rm pl}\sim$1.8 R$_{\oplus}$ \citep[see Figure 7 of][]{fulton2017}. The gap in planet radius (now known as the Fulton gap, or also referred to as the radius valley) represents the transition from super-Earths to mini-Neptunes and had been predicted by models \citep{lopez2013,owen2013,ginzburg2018,gupta2019}. This bimodal distribution has been confirmed independently by other studies that reached high-enough precision in their derived radii to uncover and confirm the radius gap \citep{fulton2018,berger2018,vaneylen2018,martinez2019}.
\cite{vaneylen2018} used asteroseismology for a small sample of host stars (75 host stars and 117 planets) and derived a negative slope for the value of the radius gap versus orbital period, while \cite{martinez2019}, using a precise classical high-resolution spectroscopic analysis, measured a similar slope for the radius gap using the larger CKS sample (1232 host stars and 1633 planets).

According to the two theoretical models that predict the radius gap, core-powered mass loss and mass loss by photoevaporation, there is a variation with stellar mass \citep{fulton2018}. 
Photoevaporation models predict that the loss of the outer layers of gaseous planets is associated with the radiation (X-rays and EUV) from their host stars \citep{lopez2013,owen2013}, while core-powered mass loss models predict that the loss of the atmospheric mass of the planet is caused by the energy of the young and hot planetary cores \citep{ginzburg2016,ginzburg2018,gupta2019}. In that sense, several studies have shown that there is a dependence of the radius gap on stellar mass \citep[i.e.,][]{fulton2018,cloutier2020,berger2020b,vaneylen2021}. Importantly, \cite{cloutier2020} showed that the radius gap persists in low-mass stars (M$_{star} = 0.08-0.93$ M$_\odot$). However, \cite{petigura2022} investigated the radius gap and found no evidence that it is a function of the stellar mass of host stars (for M$_{star} = 0.5-1.4$ M$_\odot$). 

Besides stellar radii, the stellar metallicity of host stars is another parameter that is important in studying possible star-planet connections.
Several studies have investigated and found the well-known correlation between the occurrence of giant planets and the host-star metallicity; the formation of giant planets is favored around stars with larger metal content \citep[e.g.,][]{gonzalez1997, santos2004, fischer2005, sousa2011,Ghezzi2010, ghezzi2018, adibekyan2019}. This correlation has played an important role in the exoplanet field, especially in planet formation theory \citep[e.g.,][]{ida2004a,ida2004b,ida2005,nayakshin2010,mordasini2012,mordasini2015,owen2018,venturini2020}.

Unlike the well-established planet–metallicity correlation for giant planets, it is still unclear whether smaller planets (planet radius $R_{\rm pl} < 4 R_\oplus$, or planet mass $M_{\rm pl} < M_{Neptune}$), especially terrestrial planets \citep[R$_{\rm pl} $$<1.7 R_\oplus$,][]{buchhave2014}, also follow a planet–metallicity correlation. Early studies showed that planets with a radius R$_{\rm pl}<4R_\oplus$ display a wide range of metallicities indistinguishable from the distribution of stars without planets \citep[e.g.,][]{buchhave2012,everett2013}. 
Several additional studies have analyzed the small planet-metallicity correlation based on a large sample of transiting exoplanets from the Kepler catalog. These studies used spectroscopic stellar metallicities and concluded that small-rocky planets (R$_{\rm pl}<1.7 R_\oplus$) do not show a preference for metal-rich stars \citep[e.g.,][]{sousa2008,batalha2013,buchhave2014,buchhave2015,Schuler2015,mulders2016,petigura2018,narang2018,adibekyan2019}, while the occurrence rate of larger transiting planets (R$_{\rm pl}\sim1.7 - 3.9 R_\oplus$) show a correlation with metallicity \citep{buchhave2015,mulders2016,narang2018}.  
But \cite{wang2015} suggested a universality around the planet-metallicity correlation, indicating that not only giant planets (R$_{\rm pl}>4R_\oplus$), but also gas dwarf planets (R$_{\rm pl}\sim 1.7 - 3.9 R_\oplus$) and terrestrial planets ($<1.7 R_\oplus$) occur most often in metal-rich stars. \cite{zhu2016} tried to explain this discrepancy, i.e., if the planet-metallicity correlation is universal or not, by suggesting that it is due to a high rate of planet occurrence and low detection efficiency. Aside from the rate of occurrence, various studies analyzed the relationship between orbital period and metallicity, concluding that small and hot planets (orbital period $P\lesssim 10$ days) appear preferentially around metal-rich stars \citep[e.g.,][]{adibekyan2013,beauge2013,dawson2015,Adibekyan2015,mulders2016,petigura2018,wilson2018,dong2018,owen2018}.
Another correlation that has been investigated is the influence of the stellar metallicity on the planetary system architecture \citep{weiss2018, Ghezzi2021}. 

The studies highlighted above demonstrate that it is important to characterize well exoplanet stellar hosts in terms of their stellar parameters and metallicities. For example, via community effort \citep[ExoPAG;][]{gaudi2013}\footnote{\url{https://exoplanets.nasa.gov/exep/exopag/overview/}}, significant progress has been made towards this goal in recent years. We point out, however, that there have been far fewer detailed spectroscopic studies of stars observed in the extended K2 mission when compared to the Kepler 1 mission, with most of K2 star compilations of results so far being based on photometric and asteroseismic analyses, or trigonometric methods \citep[e.g.][]{huber2017,vanderburg2016,berger2020a,boyajian2013}. Together with precise EDR3 Gaia \citep[][]{gaia2020} parallaxes, deriving precise stellar parameters for K2 stars via high-resolution spectra can be considered a necessity for the community.
This study focuses on the homogeneous spectroscopic analysis of a sample of K2 stars using optical spectra obtained with the Hydra Spectrograph on the WIYN 3.5-meter telescope and measurements of selected Fe I and Fe II lines to derive fundamental stellar properties, such as T$_{\rm eff}$, log $g$, [Fe/H], mass, and radius. 
Most of the target K2 stars in this study have confirmed planets and their stellar radii are used to compute precise radii for K2 planets. 

This paper is organized as follows. In Section \ref{sec:obser}, we describe the observations and data reduction. In Section \ref{sec:method}, we discuss the methodology employed in the derivation of the stellar parameters, effective temperatures, surface gravities, metallicities, stellar masses, and radii, along with planetary radii. In Section \ref{sec:result}, our results are presented. Finally, discussions and conclusions are presented in Sections \ref{sec:discus} and \ref{sec:conclusion}, respectively.

%-------------------------------------------
\section{Observations} \label{sec:obser}
%-------------------------------------------
The spectra analyzed in this study were obtained in several observing runs targeting stars from the Ecliptic Plane Input Catalog (EPIC) for the K2 mission and, as a lower priority, Kepler objects of interest (KOI) from the Kepler 1 mission.
The observing runs took place between 2015 and 2019 using the Hydra multi-fiber spectrograph ($R\sim 18,500$) mounted on the WIYN 3.5-meter telescope at Kitt Peak\footnote{The WIYN Observatory is a joint facility of the University of Wisconsin–Madison, Indiana University, NSF’s NOIRLab, the Pennsylvania State University, Purdue University, University of California, Irvine, and the University of Missouri.}. Most of the observed stars are from the K2 mission.

\begin{figure*}
\epsscale{1.1}
\plotone{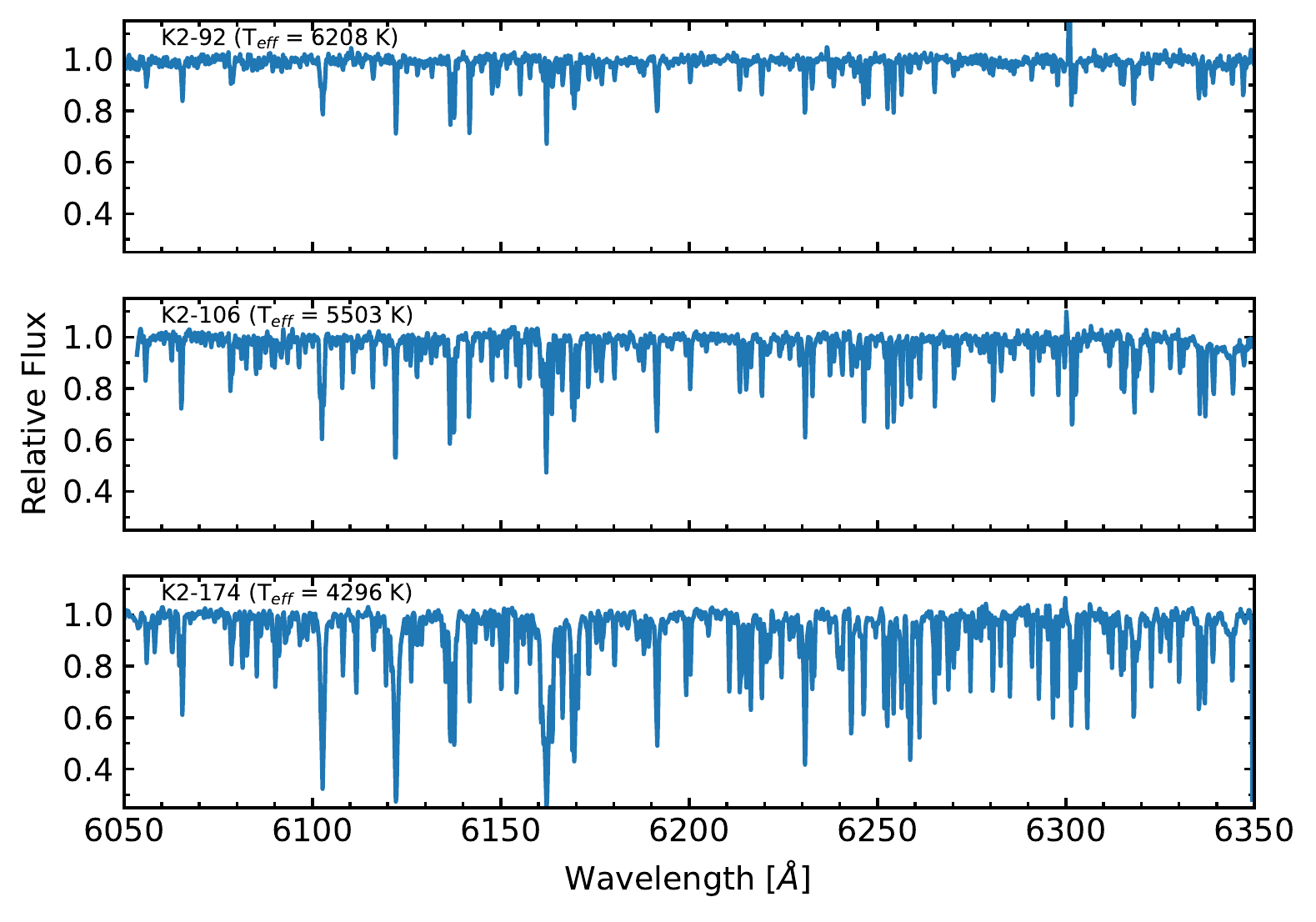}
\caption{Examples of reduced Hydra spectra; the top, middle and lower panels correspond to the target stars K2-92 (T$_{\rm eff}$ = 6208K; F dwarf), K2-106 (T$_{\rm eff}$ = 5503K; G dwarf) and K2-174 (T$_{\rm eff}$ = 4296K; K dwarf), respectively. The increasing spectral-line absorption from spectral type F to G to K is clear, which arise from both a larger number of lines plus increasing line absorption.}
\label{fig:spec_red}
\end{figure*}

The Hydra spectra were reduced using routines in IRAF\footnote{IRAF is distributed by the National Optical Astronomy Observatory, which is operated by the Association of Universities for Research in Astronomy, Inc., under a cooperative agreement with the National Science Foundation.}.
Briefly, we trimmed and overscan-subtracted the individual frames. The bias frames were combined into a master bias and then this was subtracted from the rest of the images. Cosmic rays from object frames were removed using L. A. Cosmic, an IRAF script developed by \cite{vandokkum2001PASP..113.1420V}. After aperture extraction, the one-dimensional spectra were flat fielded and wavelength calibrated.
Finally, to remove the telluric lines between 6270--6300 {\AA} we used the spectrum of a rapidly rotating B-type star. The continuum normalization of all spectra was done with the IRAF task continuum and a spline fit. 
Figure \ref{fig:spec_red} displays continuum normalized spectra showing the entire wavelength range ($\lambda$6050 to 6350\AA) covered by the analyzed Hydra spectra. 
The three stars presented in the different panels of the figure were selected to showcase the change in spectral features due to different effective temperatures. The top panel shows a hotter star with T$_{\rm eff}$= 6208 K (K2-92), the middle panel a star with T$_{\rm eff}$= 5503 K (K2-106), and the bottom panel a cooler star with T$_{\rm eff}$= 4296 K (K2-174). The effective temperatures in the panels are those from this study.

%-----------------------------------
%-----------------------------------

The sample studied here contains 115 stars and the targets were selected based on observability, and with an emphasis on G and K spectral types (Section \ref{sec:discus}). Our prime program was to target K2 stars but due to observing constraints Kepler 1 targets were also observed. The total number of K2 stars analyzed is 81 (69 stars with confirmed planets and 12 stars with candidate planets) and it includes stars from campaigns C0, C1, C3$-$C6, C8 and C10 (some stars from C5 were also observed in C16 and C18). Our sample includes 33 stars identified as Kepler, KOIs, or KIC, plus two asteroids observed as solar proxies (Astraea and Phartenopen). We also observed a solar twin (HIP 81512), which was analyzed previously by  \cite{Ramirez2009,ramirez2013}, as a comparison star. 
The quality of the spectra in this analysis are good and suitable for a precise spectroscopic analysis with most of the spectra having $S/N\sim100$, while about 10$\%$ of the spectra have $50 < S/N < 100$. The S/N were estimated in spectral regions between $\sim 6068 - 6075$ \AA.

In our sample there are seven stars that have been flagged as eclipsing binaries: KOI-6 in \cite{slawson2011AJ....142..160S}, EPIC 202126847 in \cite{lacourse2015MNRAS.452.3561L}, EPIC 210754505 in \cite{barros2016AA...594A.100B}, EPIC 201569483, EPIC 202071289, EPIC 202086968, and K2-10 in \citep{armstrong2015A&A...579A..19A}. 
In order to evaluate if there is evidence of a companion star in the observed spectra, all seven stars have been inspected carefully, finding no trace of contamination by a second set of spectral lines from a companion, although some small level ($\sim$1-2\%) effect may still be present. In any case, these stars do not enter into the calculation of any planetary radii, except for K2-10. We note, that the exoplanet archive classifies K2-10 as a confirmed planet host based on the following studies: \cite{kruse2019}, \cite{vaneylen2016AJ....152..143V}, \cite{vanderburg2016}, \cite{montet2015ApJ...809...25M}, \cite{barros2016AA...594A.100B}, and \cite{crossfield2016}. 
We also note that \cite{lester2021AJ....162...75L} and \cite{howell2021AJ....161..164H}, have shown that in binary systems, smaller planets (R $<2$R$_\oplus$) are not detected as the companion ``third-light'' would fill in their shallow transit. 

In addition to systems that have been identified as eclipsing binaries, eight of the Kepler-1 targets here have been found to have nearby companion stars lying less than 1" away from the primary star.  These stars were taken from \cite{furlan2017AJ....153...71F} and are: Kepler-132, KOI-1119, Kepler-1040, Kepler-396, Kepler-1339, Kepler-1505, Kepler-1525, and Kepler-1542.  The reported separations range from 0.04" to 0.88" and, in all cases the companion was fainter than the primary.  Using values of $\Delta$-magnitude from \citep[][mostly $\Delta$K-magnitudes]{furlan2017AJ....153...71F}, convolved with a 1" seeing-disk, it is found that the expected contaminations, in all but one target, are small, with estimated flux contaminations of 0.3\% to 4.4\%. The one primary target with a very close (0.04") companion having nearly the same brightness, is Kepler-1505 ($\Delta$ magnitude(562) = 0.00 ± 0.15 and $\Delta$ magnitude(880) = 0.16 ± 0.15; \citeauthor{furlan2017AJ....153...71F} \citeyear{furlan2017AJ....153...71F}) and, at an estimated distance of 488 pc, the projected sky separation would be $\sim$20 AU. Composite spectra that are not properly modeled can lead to uncertainties in derived stellar parameters \citep{furlanhowell2017AJ....154...66F}. Due to likely significant contamination of the spectrum of Kepler-1505 from its companion, this star was removed from the sample.

The main observational data, such as identifiers, observation dates, positions, V magnitudes (taken from NASA Exoplanet Archive), exposure time per spectrum,  
and signal-to-noise of the reduced spectra are presented in Table \ref{tab:data_star}.

\begin{figure*}
\gridline{\fig{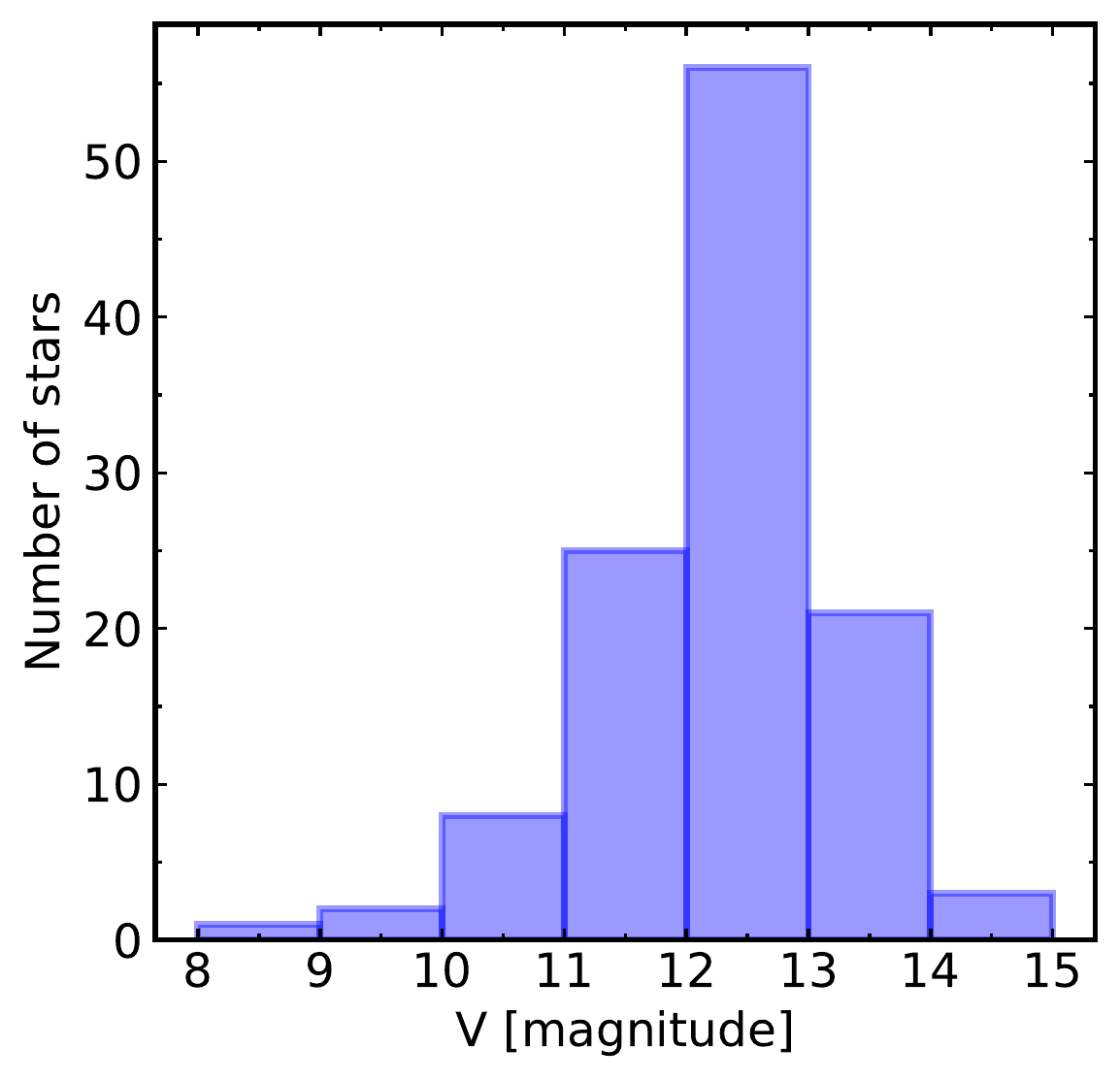}{0.508\textwidth}{}
            \hspace{-3mm}
          \fig{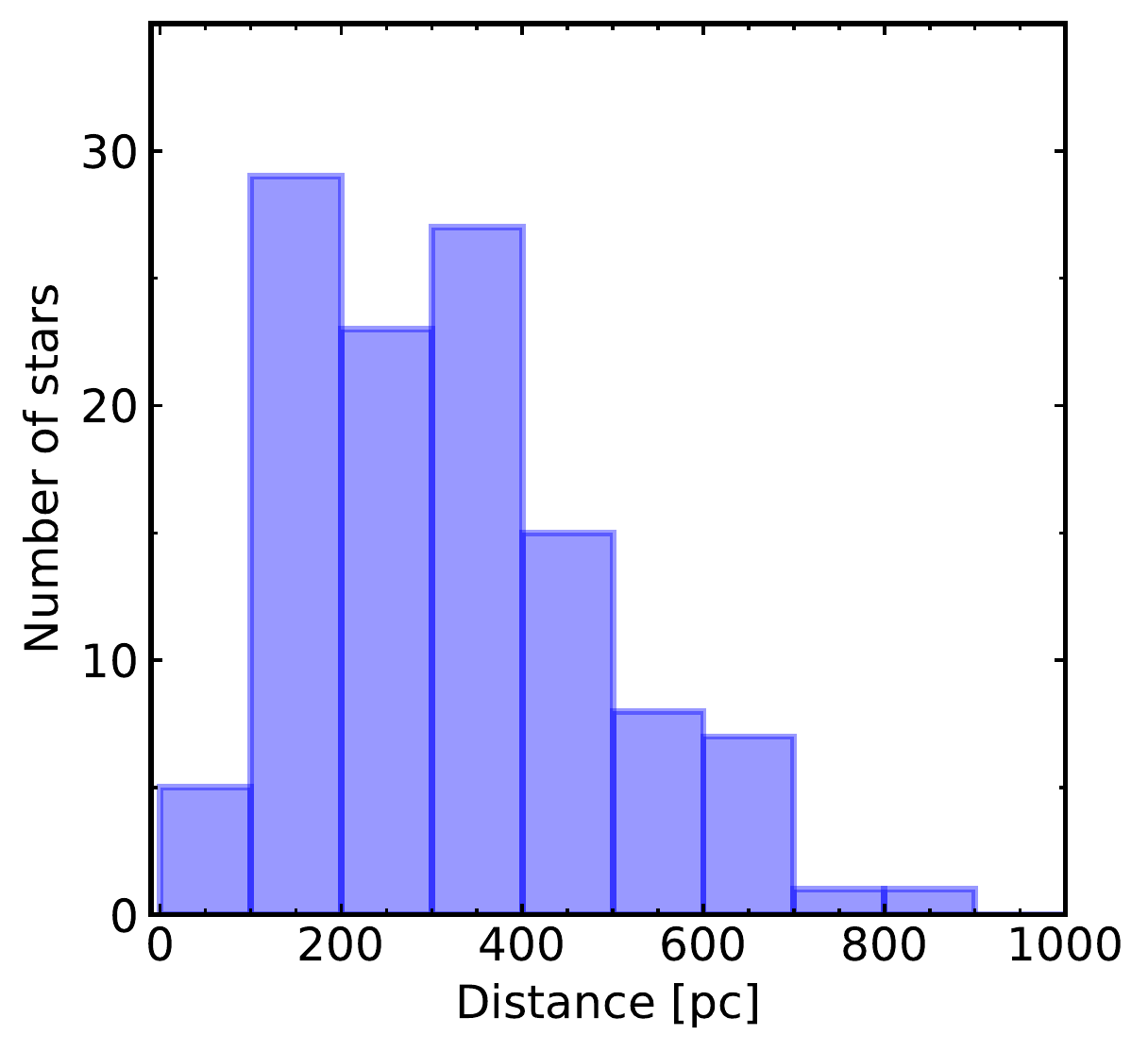}{0.53\textwidth}{}
          }
        \vspace{-7mm}
\caption{The distributions of V magnitudes and distances for the sample stars. Distances were taken from \cite{bailerjones2021}, using Gaia EDR3 data, with V magnitudes from the NASA Exoplanet Archive.}
\label{fig:dist_V_D}
\end{figure*}

Histograms with V magnitudes and distances for our sample are presented in the left and right panels of Figure \ref{fig:dist_V_D}, respectively. The V-magnitude distribution of the observed targets peaks at V$\sim$12.5 and goes as faint as V$\sim$15.
The target distances shown in the right panel of Figure \ref{fig:dist_V_D} were estimated by \cite{bailerjones2021} using the parallaxes and G-magnitude and BP - RP color from Gaia EDR3. The studied sample is dominated by stars whose distances are less than 600 pc, with the distance distribution having a peak at approximately 300 pc.  

\begin{deluxetable*}{llcccccc}
\tablecaption{Main sample data\label{tab:data_star}}
\tablecolumns{7}
\tablenum{1}
\tablewidth{0pt}
\tablehead{
\colhead{ID} & \colhead{Host name} & \colhead{UT Date} & \colhead{RA} & \colhead{DEC} & \colhead{V} & \colhead{Exposure} & \colhead{S/N} \\
\colhead{ } & \colhead{ } & \colhead{} & \colhead{} & \colhead{} & \colhead{(mag)} & \colhead{(s)} & \colhead{} 
}
\startdata
EPIC201403446 & K2-46 & 2018 Mar 30      & 11:37:03.92 & -00:54:26.10 & 12.03      & 2 x 1800              & 135          \\
EPIC211355342 & K2-181 & 2018 Mar 30      & 08:30:12.97 & 10:54:37.04  & 12.75      & 3 x 1800              & 95           \\
EPIC201736247 &	K2-15 &	2016 Mar 16 &	11:52:26.59 &	04:15:17.10 &	14.76 & 2 x 1800 &	 50 \\
\nodata  & \nodata  & \nodata & \nodata  & \nodata  & \nodata  & \nodata   \\ 
\enddata
\tablecomments{A portion is shown here for guidance regarding its form and content. This table is published in its entirety in the machine readable format.}
\end{deluxetable*}
%--------------------------------------------------
%--------------------------------------------------

%---------------------------------------
\section{Analysis} \label{sec:method}
%---------------------------------------

\subsection{Spectroscopic Stellar Parameters} \label{sec:spectroscopy}

The spectroscopic analysis employed here assumes local thermodynamic equilibrium (LTE) and uses 1-D plane parallel model atmospheres from the Kurucz ATLAS9 ODFNEW grid \citep{castelli2004a}.
Stellar spectroscopic parameters, namely effective temperature ($T_{\rm eff}$), surface gravity (log $g$), iron abundance (A(Fe)\footnote{A(X)=log(N(X)/N(H))+12.0}), and microturbulent velocity ($\xi$) were derived using a standard spectroscopic methodology which is based on measurements of equivalent widths (EWs) of selected iron lines (Fe I and Fe II lines). 

The adopted line list in this study was taken from \cite{ghezzi2018}, \cite{Melendez2014}, and \cite{Friel2003} and within the Hydra spectral window, a list of 25 Fe I lines and 5 Fe II lines was selected for analysis. The equivalent widths of the selected lines were measured using ARES code v2 \citep{sousa2015A&A...577A..67S}. In Table \ref{tab:linelist} we present the Fe I and Fe II lines, the excitation potential energies $\chi$, the oscillator strengths (log gf), and the respective references for the latter. 

Three conditions were required for obtaining a consistent solution for T$_{\rm eff}$, log $g$, and $\xi$ for the stars. To obtain  $T_{\rm eff}$, excitation equilibrium was required, or, removing trends between A(Fe I) and the excitation potential ($\chi$), of the lines. To obtain log $g$, ionization equilibrium was required, or, requiring that the average abundance of Fe I (A(Fe I)) and Fe II (A(Fe II)) lines are equal. By minimizing the slope ($<0.005$) of the relationship between A(Fe I) and the logarithm of the reduced EWs ($log(EW/\lambda$)), the microturbulent velocity, $\xi$, is obtained. Finally, the iron abundances were consistent with the input model metallicities.

%%%%%%%%%%%%%%%%%%%%%%%%%%%%%%%%%%%%%%%%%%%%
\begin{figure}[h!]
\epsscale{0.8}
\plotone{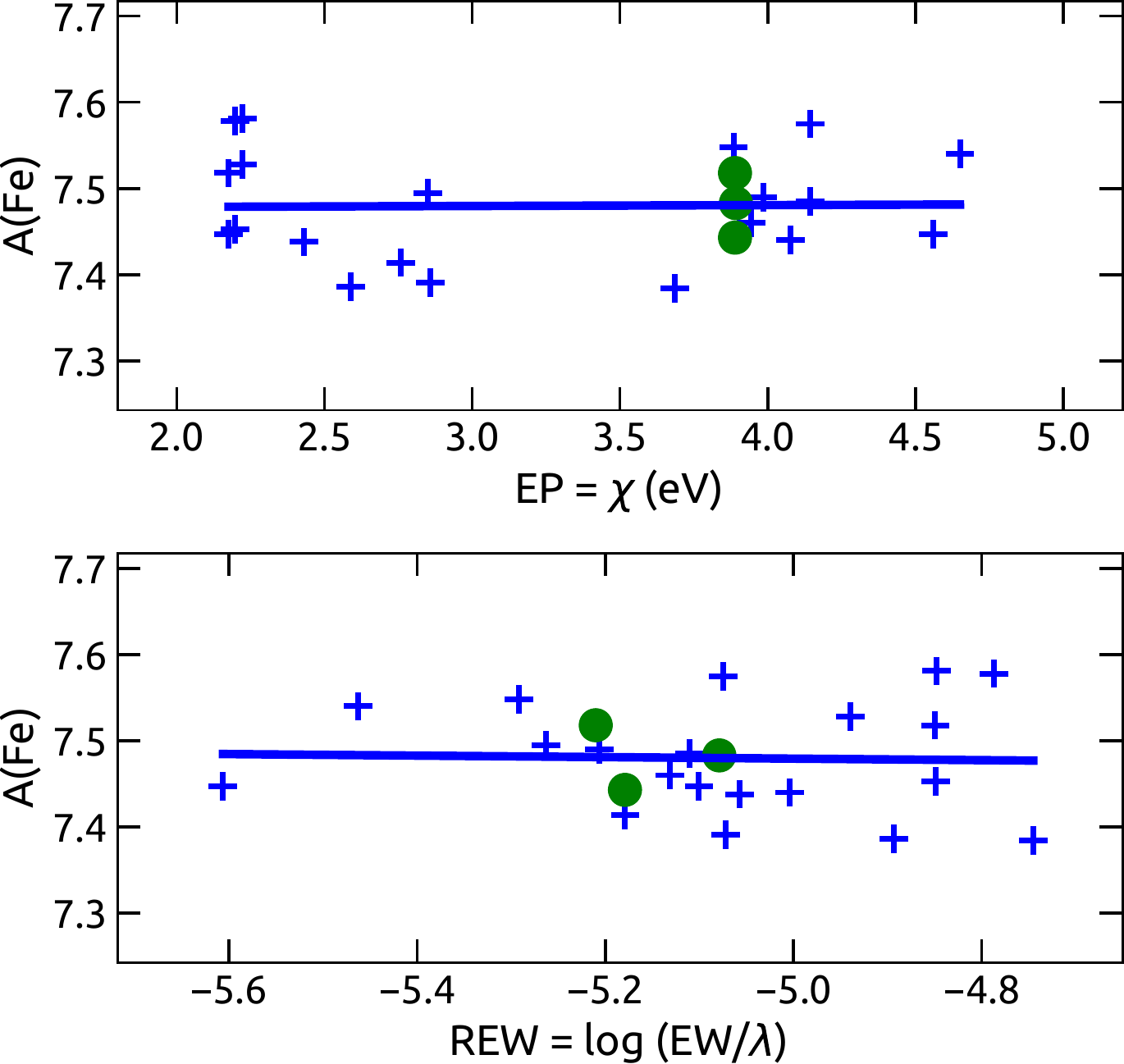}
\caption{An example of the applied methodology to the solar twin HIP 81512. The top panel shows the iron abundance of the Fe I lines (blue crosses) as a function of excitation potential $(\chi)$, which defines the effective temperature of the star. The Fe II transitions are shown as green circles. The bottom panel illustrates the Fe I abundance as a function of reduced equivalent widths (REW) of the Fe lines, which defines the microturbulent velocity parameter.}
\label{fig:method}
\end{figure}
%%%%%%%%%%%%%%%%%%%%%%%%%%%%%%%%%%%%%%%%%%%

In order to analyze a relatively large number of stars in a homogeneous and efficient way, we used the automated stellar parameter and metallicity code named \texttt{qoyllur-quipu} (or $q^2$)\footnote{\url{https://github.com/astroChasqui/q2}}. This is a Python code developed by \cite{Ramirez2014}. Briefly, \textsc{$q^2$} uses an input iron line list and measured equivalent widths, along with the 2019 version of the abundance analysis code MOOG \citep{sneden1973}, to compute the iron abundances, effective temperatures and surface gravities. 
The iterative process starts by interpolating a model atmosphere calculated assuming a given value for T$_{\rm eff}$, log $g$, and metallicity and then the values T$_{\rm eff}$ / log $g$ / A(Fe) are increased or decreased iteratively to minimize the slopes of the relationships, and until obtaining a final adjusted value for the spectroscopic parameters of each star.
Figure \ref{fig:method} shows an example of the iterated solution for the sample star  HIP\- 81512, obtained for T$_{\rm eff}=5752$ K, log $g=4.34$, and $\xi=1.06$ km s$^{-1}$, with the mean metallictity for this star represented by the solid blue line.

Table \ref{tab:parameter} presents the derived effective temperatures, surface gravities, metallicities, and microturbulent velocities for all stars in our sample, as well as the stellar radius and mass (see Section \ref{sec:radii}). %(Section 3.3).
We note that for a few stars having no measurable Fe II lines in their spectra, or having only one measurable Fe II line with large uncertainty in the equivalent width, the asteroseismic log $g$, if available, was adopted and used in the derivation of the effective temperature and microturbulence from the Fe I lines. Alternatively, when an asteroseismic log $g$ was also not available, the Fe I lines were used to estimate the parameters by varying log $g$ in steps of 0.05 dex and finding the log $g$ value that produced the most consistent solution in terms of the scatter in the Fe abundances from the individual lines.

%---------------------------------------------------
\begin{deluxetable*}{lcccc}[h!]
\tablecaption{Iron Line List\label{tab:linelist}}
\tablecolumns{11}
\tablenum{2}
\tablewidth{0pt}
\tablehead{
\colhead{$\lambda$} & \colhead{Species} & \colhead{$\chi$} & \colhead{log $gf$} & \colhead{Ref.}       \\
\colhead{(\AA)} & \colhead{} & \colhead{(eV)} & \colhead{} & \colhead{}
}
\startdata
 \\
6056.004 & Fe I & 4.733 & -0.558 & G18          \\
6085.257 & Fe I & 2.759 & -2.908 & G18          \\%& S \\
6094.374 & Fe I & 4.650 & -1.650 & F03          \\%& S \\
6096.664 & Fe I & 3.984 & -1.861 & G18          \\%& S \\
6098.243 & Fe I & 4.559 & -1.825 & G18          \\%& S \\
6100.271 & Fe I & 4.559 & -2.201 & K14         \\%& VALD \\
6127.906 & Fe I & 4.143 & -1.503 & G18          \\%& S \\
6151.617 & Fe I & 2.176 & -3.357 & G18          \\%& S \\
6157.727 & Fe I & 4.076 & -1.257 & G18          \\%& S \\
6165.359 & Fe I & 4.143 & -1.487 & G18          \\%& S \\
6173.334 & Fe I & 2.223 & -2.938 & G18          \\%& S \\
6187.989 & Fe I & 3.943 & -1.724 & G18          \\%& S \\
6200.312 & Fe I & 2.608 & -2.457 & G18          \\%& S \\
6213.429 & Fe I & 2.223 & -2.650 & G18          \\%& S \\
6219.280 & Fe I & 2.198 & -2.549 & G18          \\%& S \\
6226.734 & Fe I & 3.884 & -2.143 & G18          \\%& S \\
6229.230 & Fe I & 2.850 & -3.040 & F03          \\%& L \\
6232.640 & Fe I & 3.654 & -1.232 & G18          \\%& S \\
6252.555 & Fe I & 2.404 & -1.687 & M14          \\%& L \\
6265.132 & Fe I & 2.176 & -2.633 & G18          \\%& S \\
6270.225 & Fe I & 2.858 & -2.540 & M14          \\%& S \\
6322.690 & Fe I & 2.590 & -2.250 & F03          \\%& S \\
6335.329 & Fe I & 2.198 & -2.423 & G18          \\%& S \\
6336.824 & Fe I & 3.686 & -0.856 & BK94         \\%& L \\
6344.150 & Fe I & 2.430 & -2.970 & F03          \\%& S \\
6084.102 & Fe II & 3.199 & -3.840 & G18          \\%& S \\
6113.319 & Fe II & 3.221 & -4.155 & G18          \\%& S \\
6149.246 & Fe II & 3.889 & -2.789 & G18          \\%& S \\
6238.386 & Fe II & 3.889 & -2.634 & G18          \\%& S \\
6247.557 & Fe II & 3.892 & -2.427 & G18          \\%& S
\enddata
\tablecomments{The sources for the log gf values are given in the last column. G18: \cite{ghezzi2018}, M14: \cite{Melendez2014}, F03: \cite{Friel2003}, K14: \cite{kurucz2014dapb.book...63K}, and BK94: \cite{bard1994AA...282.1014B}.}
\end{deluxetable*}

%---------------------------------------------------

%---------------------------------------------------
\begin{deluxetable*}{lcccccccccccc}[h!]
\tablecaption{Stellar Parameters, Radii and Masses\label{tab:parameter}}
\tablecolumns{11}
\tablenum{3}
\tablewidth{0pt}
\tablehead{
\colhead{ID} & \colhead{$T_{\rm eff}$} & \colhead{$\delta T_{\rm eff}$} & 
\colhead{log $g$} & \colhead{$\delta$log $g$} & \colhead{A(Fe)} & \colhead{$\delta$A(Fe)} & \colhead{$\xi$} & 
\colhead{$\delta\xi$} & \colhead{$R_{star}$} & \colhead{$\delta R_{star}$} & \colhead{$M_{star}$} & \colhead{$\delta M_{star}$}  \\
\colhead{ } & \colhead{(K)} & \colhead{(K)} & \colhead{(dex)} & \colhead{(dex)} & \colhead{(dex)} & \colhead{(dex)} & \colhead{(km\,s$^{-1}$)} & \colhead{(km\,s$^{-1}$)} & \colhead{($R_{\odot}$)} & \colhead{($R_{\odot}$)} & \colhead{($M_{\odot}$)} & \colhead{($M_{\odot}$)}
}
\startdata
EPIC 201166680 & 6051 & 289 & 4.01 & 0.49 & 7.39 & 0.11 & 1.45 & 0.37 & 1.45 & 0.16 & 1.09 & 0.06 \\
EPIC 201211526 & 5717 & 176 & 4.13 & 0.39 & 7.21 & 0.08 & 1.07 & 0.17 & 0.92 & 0.03 & 0.89 & 0.04 \\
EPIC 201257461 & 4945 & 160 & 2.95 & 0.82 & 7.41 & 0.11 & 1.67 & 0.18 & 5.48 & 0.40 & 1.25 & 0.22 \\
\nodata  & \nodata  & \nodata & \nodata  & \nodata  & \nodata  & \nodata  & \nodata & \nodata & \nodata &  \nodata  & \nodata & \nodata \\ 
\enddata
\tablecomments{(*) Asteroseismic log $g$; (**) no measurable Fe II lines; ($a$) R$_{star}$ computed using log $g$ values. This table is published in its entirety in the machine readable format. A portion is shown here for guidance regarding its form and content.}
\end{deluxetable*}
%---------------------------------------------------

\subsection{Solar Proxies and a Solar twin as Benchmarks} \label{sec:proxy}

As benchmarks to our methodology and analysis techniques, we also analyzed solar proxy spectra obtained with the Hydra spectrograph for reflected solar light
from two asteroids, Astraea and Parthenope,  as well as the well-studied solar twin HIP 81512 as a benchmark. Results are presented in Table \ref{tab:proxies}. The parameters and metallicities obtained for the solar proxy Astraea (T$_{\rm eff}=5778$ K, log $g=4.34$, A(Fe) $=7.51$, $\xi =1.16$ km-s$^{-1}$) and for Parthenope (T$_{\rm eff}=5770$ K, log $g=4.40$, A(Fe) $=7.54$, $\xi =1.06$ km-s$^{-1}$) are in excellent agreement with the solar parameters, indicating that our methodology does not harbor strong biases for solar type stars. We note, however, that the mean metallicity for the solar proxies of A(Fe) $=7.52$ is slightly more metal rich than the \citet[][A(Fe)$_{\odot}$=7.46]{asplund2021AA...653A.141A} scale, although it is in good agreement with the \citet[][A(Fe)$_{\odot}$=7.50]{magg2022AA...661A.140M} scale. 
We also note excellent agreement with the stellar parameters derived in the high-precision analysis of the solar twin HIP 81512 by \cite{Ramirez2009} and \cite{ramirez2013}; the latter studies are based on the analysis of high-resolution spectra ($R=\lambda/\Delta \lambda \simeq 60 000$) obtained with the Robert G. Tull coudé spectrograph on 2.7 m Harlan J. Smith telescope and measurements of 128 Fe I and 16 Fe II lines, obtaining T$_{\rm eff}=5755 \pm 32$ K, log $g= 4.43 \pm 0.04$ dex, A(Fe) $= 7.40 \pm 0.04$ dex.  These comparisons with benchmark spectra can serve as validations for the technique adopted in this study for the analysis of Hydra spectra covering $\lambda$6050$-$6350\AA\ and the Fe I/Fe II line list from Table \ref{tab:linelist}.
  
%------------------------------------------
\begin{deluxetable*}{lcccc}[h!]
\tablecaption{Stellar Parameters of Solar Proxies\label{tab:proxies}}
\tablecolumns{11}
\tablenum{4}
\tablewidth{0pt}
\tablehead{
\colhead{ID} & \colhead{$T_{\rm eff}$} & 
\colhead{log $g$} & \colhead{A(Fe)} & \colhead{$\xi$} \\
\colhead{ } & \colhead{(K)} & \colhead{(dex)} & \colhead{(dex)} & \colhead{(km\,s$^{-1}$)}
}
\startdata
Astraea     & 5778 $\pm$ 160 & 4.34 $\pm$ 0.33 & 7.51 $\pm$ 0.09 & 1.16 $\pm$ 0.14 \\
Parthenope & 5770 $\pm$ 158 & 4.40 $\pm$ 0.36 & 7.54 $\pm$ 0.09 & 1.06 $\pm$ 0.24 \\
HIP 81512   & 5752 $\pm$ 113 & 4.34 $\pm$ 0.33 & 7.48 $\pm$ 0.07 & 1.06 $\pm$ 0.10 \\
\enddata
\tablecomments{A(Fe)$_\odot$ = $7.46 \pm 0.04$, \cite{asplund2021AA...653A.141A}; A(Fe)$_\odot$ = $7.52 \pm 0.06$ dex, \cite{magg2022AA...661A.140M}}
\end{deluxetable*}
%------------------------------------------

%=======================================================
\subsection{Stellar Masses and Radii and Planetary Radii} \label{sec:radii}
%=======================================================

Fundamental stellar properties, such as stellar radius, age, or mass can be estimated by comparing the positions of stars in a color-magnitude diagram with theoretical isochrones. Gaia DR3 \citep{gaia2021A&A...649A...1G} currently provides the high precision parallaxes  that can be used to determine absolute magnitudes of large numbers of stars.

Different codes that are available to the community, such as \texttt{PARAM} \citep{girardi2000A&AS..141..371G,dasilva2006,rodrigues2014MNRAS.445.2758R,rodrigues2017MNRAS.467.1433R}, or \texttt{isochrones} \citep{morton2015}, have been developed as interfaces to find best fits to various isochrones, such as MESA Isochrones \& Stellar Tracks \citep[MIST,][]{dotter2016ApJS..222....8D,choi2016}, and several works have adopted similar methodologies to obtain stellar masses and radii \citep[e.g.,][]{johnson2017,mayo2018,wittenmyer2020}. 

In this work, stellar masses and radii were computed using the isochrone method via the $q^2$ code \citep[\texttt{qoyllur-quipu};][]{Ramirez2014}, which determines stellar mass, age, luminosity, and radius, using the grid of Yonsei-Yale isochrones \citep{yi2001}. Briefly, to determine which isochrones best represent a particular set of observed stellar parameters (spectroscopic T$_{\rm eff}$ and [Fe/H] from this work, along with M$_V$ absolute magnitude derived from parallax), probability distributions of those parameters are determined and matched to the isochrones, assuming that the errors in the observed stellar parameters ($\delta T_{\rm eff}$, $\delta$M$_V$ and $\delta \rm [Fe/H]$) have Gaussian probability distributions. 
For more details see \cite{ramirez2013} and \cite{Ramirez2014}. Isochrone-derived masses and radii of the sample stars are shown in Table \ref{tab:parameter}.
For five stars, instead of using M$_V$, we used the log $g$ values to compute the stellar radii and the latter are flagged with `$a$' in Table \ref{tab:parameter}. In particular, we opted for using log $g$ when there was a range of V magnitudes reported for a star, which was the case, for example, for two of the eclipsing binaries in our sample. We note that the stellar masses in these cases were estimated in the same way as the other stars using $Y^2$ isochrones.

Planetary radii were then obtained using the derived stellar radii and the values of transit depth ($\Delta F$), which are the fraction of stellar flux lost at the minimum of the planetary transit, and the equation from \cite{seager2003}:

\begin{equation}\label{eq:Rpl}
	R_{pl} = 109.1979 \times \sqrt{\Delta F \times 10^{-6}} \times R_{star}
\end{equation}
where the radius of the planet is in Earth-radii.

We note that in this study, only values of $\Delta F$ from confirmed planets were used; we did not consider planets classified as planet candidates and false positives (according to the Kepler 1 and K2 notes in NASA Exoplanet Archive).
Most of the planetary transit depths of K2 stars were from \cite{kruse2019} (for 66 confirmed planets); for planets not catalogued in \cite{kruse2019} we used values from  \cite{vanderburg2016}, \cite{pope2016MNRAS.461.3399P}, \cite{barros2016AA...594A.100B}, \cite{rizzuto2017AJ....154..224R}, and \cite{livingston2018}. We note that for K2-100 we noticed that the $\Delta F$ from \cite{kruse2019} was discrepant when compared to other literature sources \citep[e.g.,][]{livingston2018}, \cite{stefansson2018AJ....156..266S}, \cite{mann2017AJ....153...64M}, \cite{librato2016MNRAS.463.1780L}, \cite{pope2016MNRAS.461.3399P} and in this study we adopted the transit depth from \cite{livingston2018}.
The transit depth of planets of Kepler 1 stars were from \cite{thompson2018} (for 56 confirmed planets). The $\Delta F$ values are provided in the Table \ref{tab:Rpl}.

%---------------------------------------------------
\subsection{Uncertainties in the Derived Parameters} \label{sec:errors}
%---------------------------------------------------
The formal errors adopted for the stellar parameters T$_{\rm eff}$, log $g$, and $\xi$ were computed using $q^2$, which follows the error analysis discussed in \cite{epstein2010} and \cite{bensby2014}.
Errors in the iron abundances, A(Fe I) and A(Fe II), were obtained by combining errors estimated from the equivalent-width measurements with stellar parameter uncertainties. 
The individual errors of these parameters are presented in Table \ref{tab:parameter} and \ref{tab:proxies}.
 
The median errors in the stellar parameters derived in this study are reported in Table \ref{tab:error_budget}: $\delta$T$_{\rm eff}$ = 154 K, $\delta$log $g$ = 0.36 dex,  $\delta$A(Fe) = 0.09 dex, $\delta$$\xi$ = 0.24 km-s$^{-1}$. We note that these are somewhat larger than typical values  from the analysis of high-resolution spectra (R $\sim$ 60,000) in the literature \citep[e.g.,][]{Ramirez2014,martinez2019,Ghezzi2021}. Larger errors here are to be expected given that the Hydra spectra have lower resolution (R $\sim$ 18,500) and smaller wavelength coverage, which results in having a smaller number measurable Fe I lines (25) and  Fe II lines (5).

Since stellar masses and radii and planetary radii of our sample were derived from other parameters, we considered  the individual contributions of the errors in each one of the parameters to estimate the error budget \citep[similar to the discussions in][]{martinez2019,fulton2018}. 
The internal precisions (median errors) in the derived effective temperatures and metallicities were discussed above.
The error in the V magnitude contributes $\sim 1 \%$ to the stellar radius error, when taking 0.07 mag to be the median error in V magnitude for our stars. The contribution due to errors in the parallaxes corresponds to a median error of 0.02 mas and represents a 0.45 $\%$ error in stellar mass and radius.

The stellar radii uncertainty and the transit depth ($\Delta F$) errors have a direct impact on the determination of planetary radii errors. The median internal uncertainty in our derived stellar radii distribution is 4.2 $\%$; we adopted the transit depth values $\Delta F$ and respective errors from \cite{kruse2019} and \cite{thompson2018}, which, for the planets in our sample, result in 3.4 $\%$ internal precision in $\Delta F$. Finally, these uncertainties lead to a 4.4 $\%$ internal precision for the R$_{pl}$ error budget. A summary of the contributions to the error budgets in the R$_{star}$ and R$_{pl}$ determinations is presented in Table \ref{tab:error_budget}.
To assess possible differences in the choice of isochrones we adopted the Darmouth isochrones instead of the Yonsei-Yale ones and found no signicant differences in the derived stellar radii and masses. As an example, for the star Kepler-62 we find R$_{star}$ = 0.659 $\pm$ 0.018 R$_\odot$, M$_{star}$ = 0.711 $\pm$ 0.024 M$_\odot$ (using Dartmouth isochrones), while the result in this study is the same but just with a slightly higher uncertainty in the mass: R$_{star}$ = 0.659$\pm$0.018 R$_\odot$, M$_{star}$ = 0.711 $\pm$ 0.026 M$_\odot$ (using Yonsei-Yale isochrones).

\begin{deluxetable*}{ll}%[b!]
\tablecaption{Error budget\label{tab:error_budget}}
\tablecolumns{2}
\tablenum{5}
\tablewidth{0pt}
\tablehead{
\colhead{Parameter} & \colhead{Median Uncertainty}
}
\startdata
T$_{\rm eff}$  & 154 K \\
log $g$ & 0.36 dex \\
A(Fe)         & 0.09 dex \\
V             & 0.07 mag \\
plx           & 0.02 mas \\
M$_{star}$ & 0.04 M$_\odot$ \\
R$_{star}$     & 4.15 \% \\
$\Delta$F     & 3.35 \% \\
R$_{pl}$      & 4.44 \% \\
\enddata
\end{deluxetable*}

%------------------------------------------------
\subsection{Possible Impact of Magnetic Activity on Stellar Parameter Determination}
%------------------------------------------------

In addition to the various sources of uncertainty discussed in the previous subsection, stellar magnetic activity can also affect the determinations of T$_{\rm eff}$, $\xi$, and [Fe/H], as shown by \cite{flores2016A&A...589A.135F}, \cite{yana2019MNRAS.490L..86Y}, or \cite{spina2020ApJ...895...52S}.

Three stars in our sample had activity indices reported in \cite{brown2022MNRAS.514.4300B}, who compiled a database of chromospheric activity measurements and surface-averaged large-scale magnetic-field measurements for a sample of FGK main-sequence stars. These stars are K2-229, EPIC 202089657, and Kepler-409, with log R'$_{HK}$= --4.73, --4.75, and --4.82, respectively, keeping in mind that the Sun has a value of log R'$_{HK}$= --5.02, which varies by about $\pm$0.01 dex over the Solar activity cycle \citep{lorenzo2018AA...619A..73L}.

Stellar magnetic activity effects have been quantified by \cite{spina2020ApJ...895...52S} via the log R'$_{HK}$ index, based on a stellar parameter analysis that relies on relations between the Fe I excitation equilibria and the Fe I reduced equivalent widths (EW/$\lambda$), both as functions of the Fe I abundance, in addition to the ionization balance of Fe I and Fe II. These constraints provide the values of T$_{\rm eff}$, log $g$, $\xi$, and [Fe/H], and this is the technique used in our study. The impact that stellar activity has on the derivation of specific stellar parameters using Fe I and Fe II lines is found to be the greatest for T$_{\rm eff}$, $\xi$, and [Fe/H], with almost no effect on log $g$ (as illustrated in Figure 3 of \citeauthor{spina2020ApJ...895...52S} \citeyear{spina2020ApJ...895...52S}). 
In rough numbers \citep[from Figure 4 in][]{spina2020ApJ...895...52S}, R'$_{HK}\sim$ --4.25 leads to a lower value of T$_{\rm eff}$ by $\sim$100 K, while log R'$_{HK}\sim$--4.40 causes a metallicity change of [Fe/H]=$-0.05$, and a value of log R'$_{HK}\sim$ --4.35 would lead to a derived microturbulent velocity that is too large by $\sim$+0.3 km-s$^{-1}$.
A much lower activity level of log R'$_{HK}\sim$ --4.7 - --4.8, as measured, for example, for the three target stars mentioned above, would cause an unmeasurable change in the effective temperature, a metallicity change of -0.02 dex, and would lead to a microturbulent velocity change of $\sim$+0.05 km-s$^{-1}$. All of these variations are well within the uncertainties in our analysis. These are also in line with the results by \cite{lorenzo2018AA...619A..73L} that, for the young solar twin HIP 36515 (log R'$_{HK}\sim$ --4.70), stellar parameters derived over its six-year activity cycle find a scatter in T$_{\rm eff}$ of $\pm$10 K, $\pm$0.01 in [Fe/H], and $\pm$0.07 km-s$^{-1}$ in microturbulent velocity.
It is possible, however, that other stars in our sample may have higher levels of activity, although such high activity levels would suggest ages $<$ 1 Gyr \citep{lorenzo2018AA...619A..73L} or that stars are members of close binary systems \citep{olah2007IAUS..240..442O}.

As part of an analysis of the young active solar twin HIP 36515 (age $\sim$ 0.4 Gyr), \cite{yana2019MNRAS.490L..86Y} established a list of Fe I and Fe II lines that are sensitive to stellar magnetic activity. Among the lines in our line list that were identified as sensitive to magnetic fields are Fe I $\lambda$ 6173 \AA, 6200 \AA, 6213 \AA, 6219 \AA, 6252 \AA, 6265 \AA, and 6270 \AA\ (Table \ref{tab:linelist}) and these have both large Landé factors, coupled with large equivalent widths (through a combination of log gf values and excitation potentials). None of our Fe II lines were identified as sensitive to stellar magnetic activity.

We investigated the impact that the inclusion of ``sensitive" Fe I lines has on the derivation of stellar parameters by removing these sensitive lines and rederiving the parameters for a selected sample of 15 stars. This sub-sample contains the three stars with measured values of log R'$_{HK}$ mentioned above, plus five stars in our sample with approximate ages from $q^2$ of $<$ 4 Gyr (K2-223, K2-44, Kepler-1339, KOI-6, and KOI-293), and two stars (K2-100; K2-101) which are members of the young \citep[650 $\pm$ 70 Myr;][]{martin2018ApJ...856...40M} open cluster M 44. We also added the stars K2-186, K2-34, Kepler-139, Kepler-1445, Kepler-396, as these have measured flares with energies $>$ 10$^{34}$ ergs along with log R'$_{H\alpha}$ $>$ -4.4 in \cite{su2022ApJS..261...26S}, who determined the R'$_{H\alpha}$ activity index and flare energy using LAMOST spectra and light curves of stars observed by Kepler 1 and K2.

%-----------------------------------
\begin{figure*}
\epsscale{0.9}
\plotone{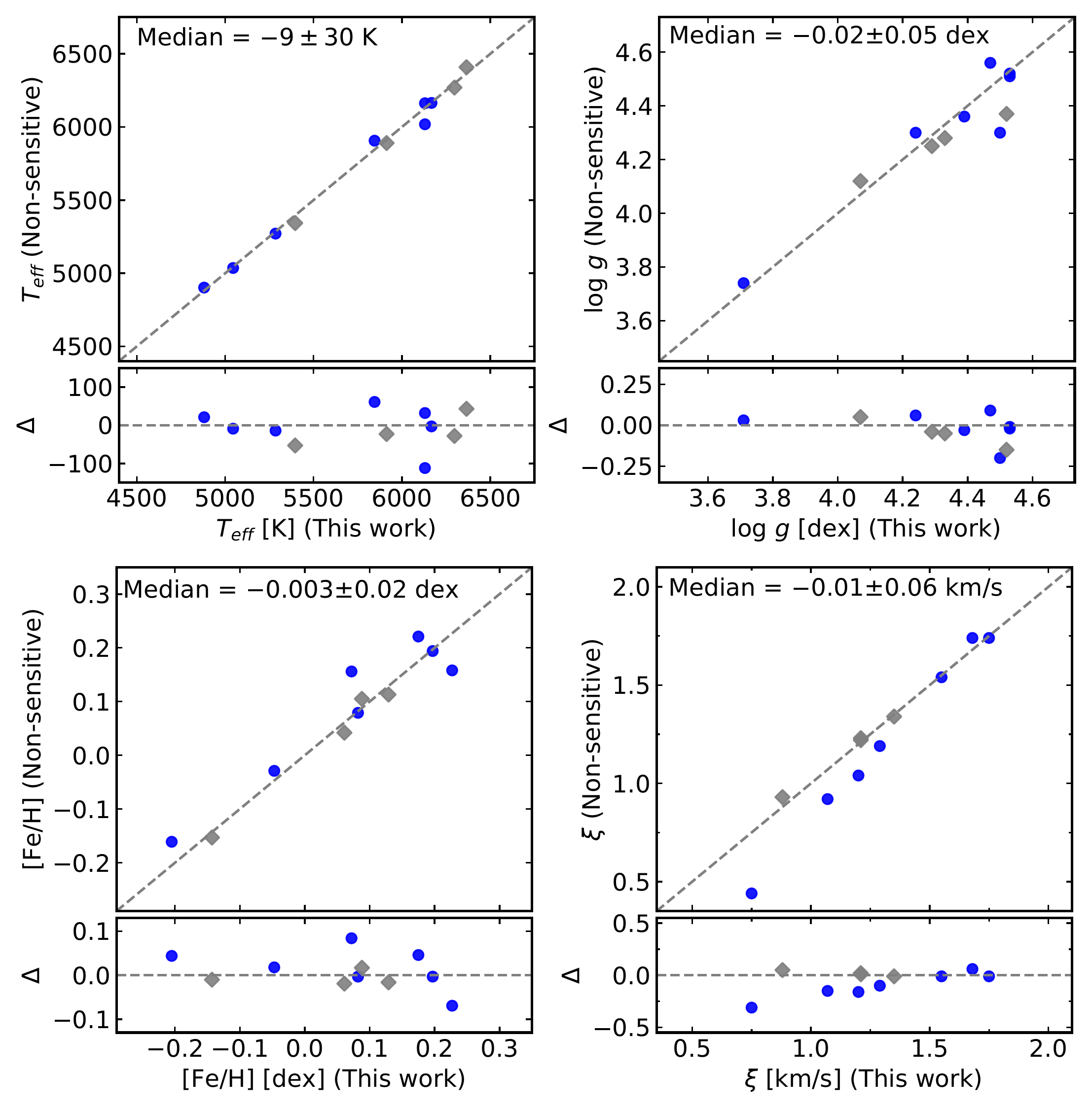}
\caption{Comparison between the effective temperatures, surface gravities, metallicities, and microturbulent velocities derived in this work with stellar parameters that were rederived using only the subset of the Fe I lines that are deemed as not sensitive to effects of stellar activity (`Non-sensitive'). The bottom panels present the difference, $\Delta$, between the stellar parameters for `Non-sensitive - This work'. Blue circles respresent K2 stars and grey diamonds Kepler 1 stars. The selected targets include host stars that exhibit flare activity (flares with energies $>$ 10$^{34}$ ergs) along with log R'$_{H\alpha}$ $>$ --4.4, three stars with measured chromospheric Ca II index log R'$_{HK}$ $\sim$ --4.7 - --4.8, five stars that have estimated ages $<$ 4 Gyr, and two stars that are members of the young open cluster M 44.}
\label{fig:stel_act}
\end{figure*}
%-----------------------------------

Figure \ref{fig:stel_act} provides the comparisons between stellar parameters derived in this study (without taking into consideration possible effects of magnetic fields in the measured Fe I lines) and those obtained including only those Fe I lines that were deemed as ``non-sensitive" to magnetic fields. The median and MAD values of the differences for each stellar parameter are included in each panel of the figure.  
The upper left panel shows that the differences in effective temperatures are less than $\sim$50 K for most stars, with only two stars, K2-44 and Kepler-1445, having  larger differences of $\Delta$T$_{\rm eff}$ (`Non-sensitive' - `This Work')=$-112$ K and $-142$ K, respectively.
This result suggests that the differences in the effective temperatures are not significant and, in general, fall within the range of our uncertainties (see Table \ref{tab:error_budget}).
In the case of log $g$ (top right panel), all stars have differences less than 0.09 dex, with the exception of K2-44 and Kepler-409 having differences of -0.2 and -0.15 dex, respectively. Given the uncertainties in log $g$, all differences are well within the estimated median uncertainty of 0.36 dex.
A similar result is found for values of [Fe/H] (lower left panel), as the median uncertainty in [Fe/H] in this study is 0.09 dex (Table \ref{tab:error_budget}), with the largest metallicity difference found being $-0.118$ for Kepler-1445.
According to \cite{spina2020ApJ...895...52S}, magnetic activity is expected to result in an increase in microturbulent velocities for log R'$_{HK}$ $>$ $ -5.0$, or in active young stars (ages $<$ 4 - 5 Gyr). The median difference found for the microturbulent velocity parameter is -0.01 and MAD of 0.06 km-s$^{-1}$ (Figure \ref{fig:stel_act}), indicating that there is no significant evidence that magnetic activity is measurably affecting the microturbulent velocities in this analysis.

In summary, the stellar parameters derived without the use of magnetically sensitive lines are within the uncertainties when compared to parameters from this study, that include some Fe I lines that are deemed as sensitive lines to the effects of magnetic fields and activity. This result suggests that the final stellar parameters derived for this sample of stars have not been perturbed significantly by strong magnetic activity.

%------------------------------------------------
\section{Results and Comparisons with the Literature} \label{sec:result}
%------------------------------------------------

\subsection{Effective Temperatures and Surface Gravities} \label{subsec:tef_log}

In general, there are more results available for Kepler 1 mission targets than for K2. 
Several studies in the literature \citep{petigura2017,Brewer2018,martinez2019,Ghezzi2021} analyzed the California Kepler Survey sample \citep[CKS,][]{petigura2017} but used different analysis techniques.  The stellar parameters obtained by \cite{petigura2017} were derived using synthetic spectra and the codes \texttt{SpecMatch} and \texttt{SME@XSEDE}; \cite{Brewer2018} derived the stellar parameters also using spectral synthesis and the Spectroscopy Made Easy \citep[SME;][]{piskunov2017} code, while \cite{martinez2019} and \cite{Ghezzi2021} adopted a methodology that was based on the classical spectroscopic equivalent-width method and used the code MOOG \citep{sneden1973}. 
Concerning results for K2 targets, in particular, \citet[][using the same methodology of \cite{petigura2017}]{petigura2018} analyzed a sample of 141 K2 candidate planet host stars, while \cite{wittenmyer2020} analyzed a sample of 129 K2 planet candidate host stars whose spectra were observed in the K2-HERMES program \citep{wittenmyer2018AJ....155...84W,sharma2019MNRAS.490.5335S,clark2022MNRAS.510.2041C} from the Galactic Archaeology with HERMES \citep[GALAH,][]{Buder2021} survey. 
In addition, results for a large number of both Kepler 1 and K2 targets were obtained by the optical low-resolution spectroscopic LAMOST survey \citep{cui2012,zong2018}, as well as the high-resolution spectroscopic near-infrared APOGEE survey \citep[][see also \citeauthor{wilson2018} \citeyear{wilson2018}]{majewski2017}. 

\begin{figure*}
\gridline{\fig{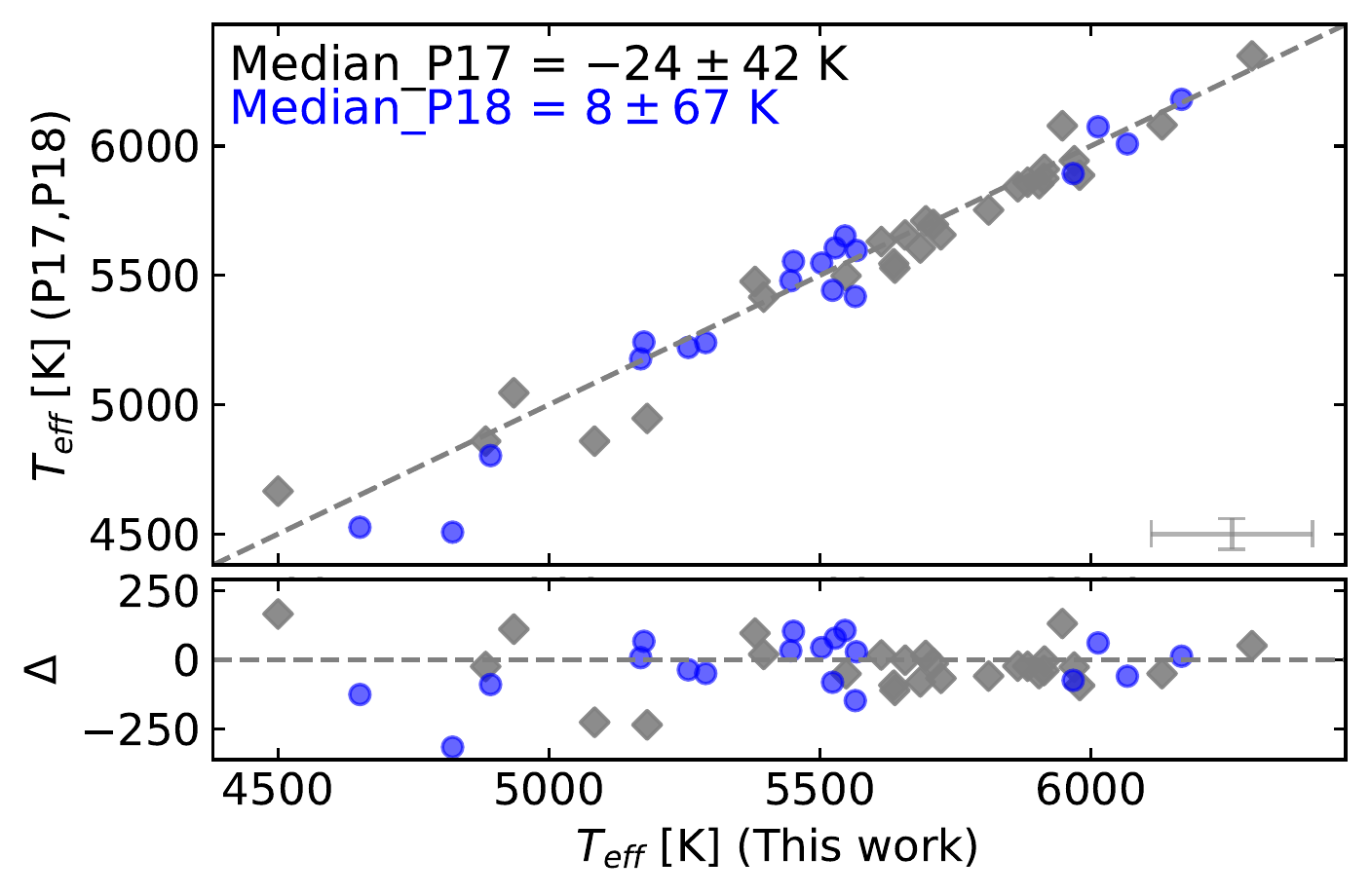}{0.48\textwidth}{}
            \hspace{-3mm}
          \fig{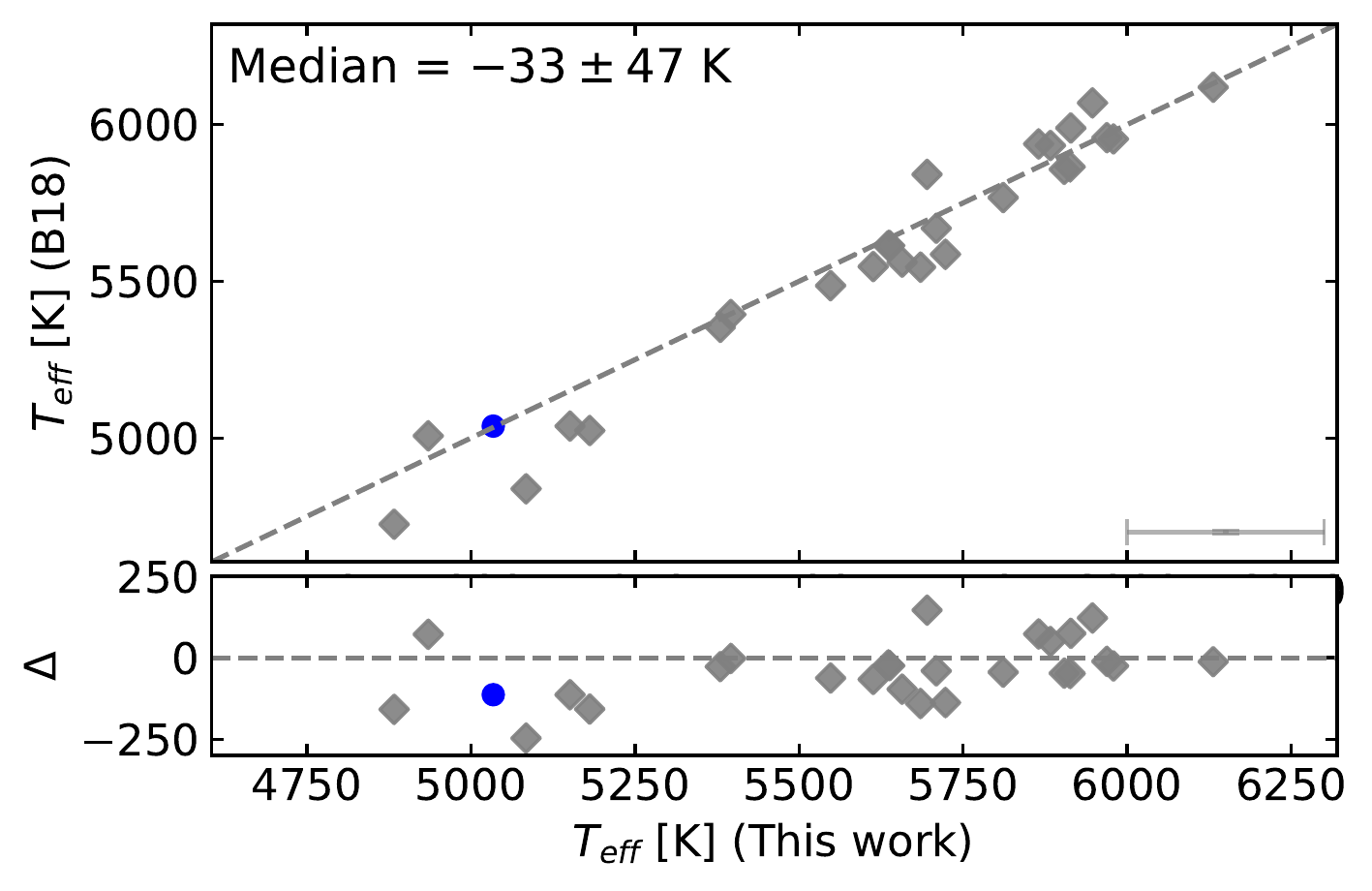}{0.48\textwidth}{}
          }
          \vspace{-11mm}
\gridline{\fig{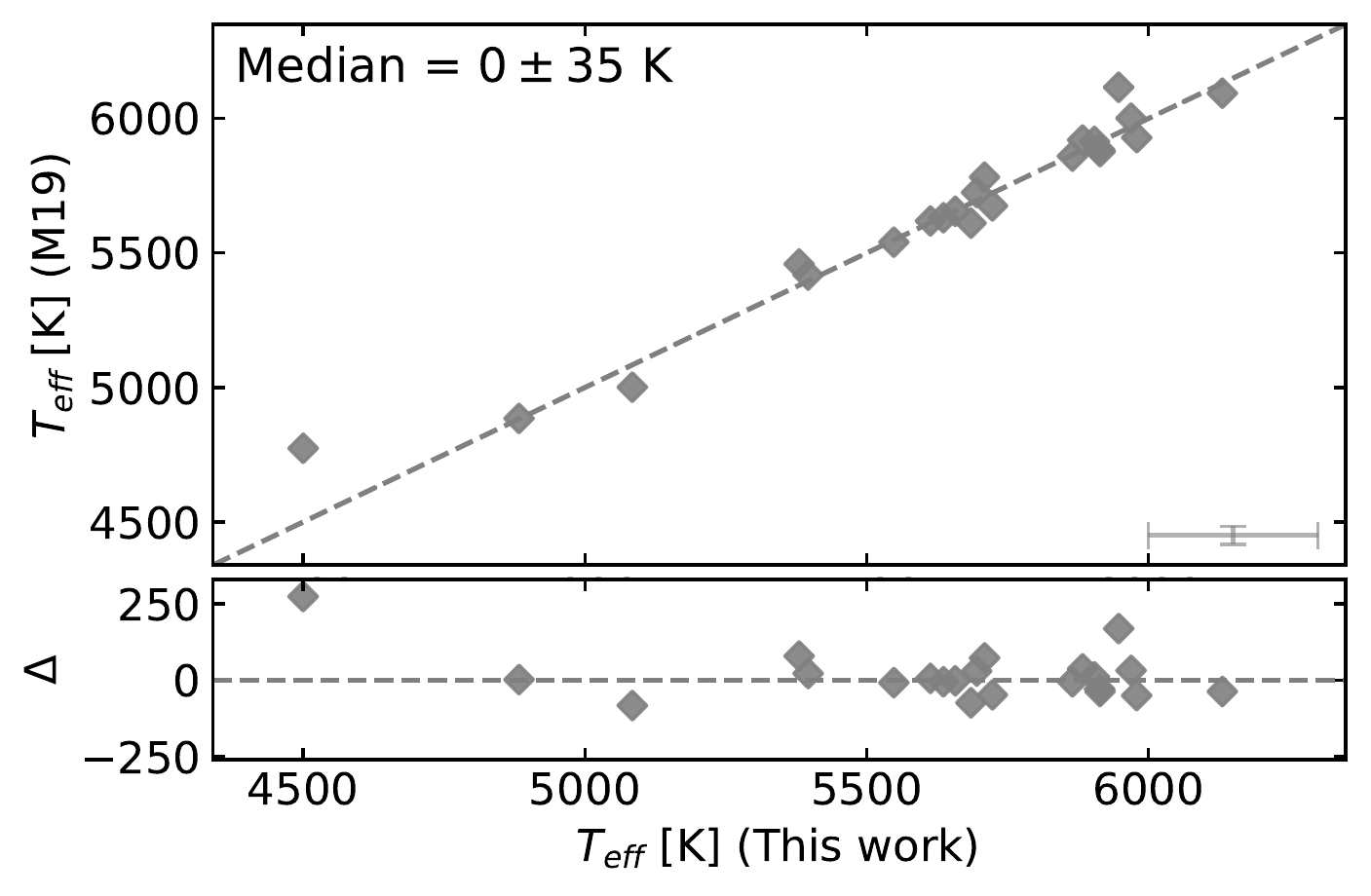}{0.48\textwidth}{}
            \hspace{-3mm}
          \fig{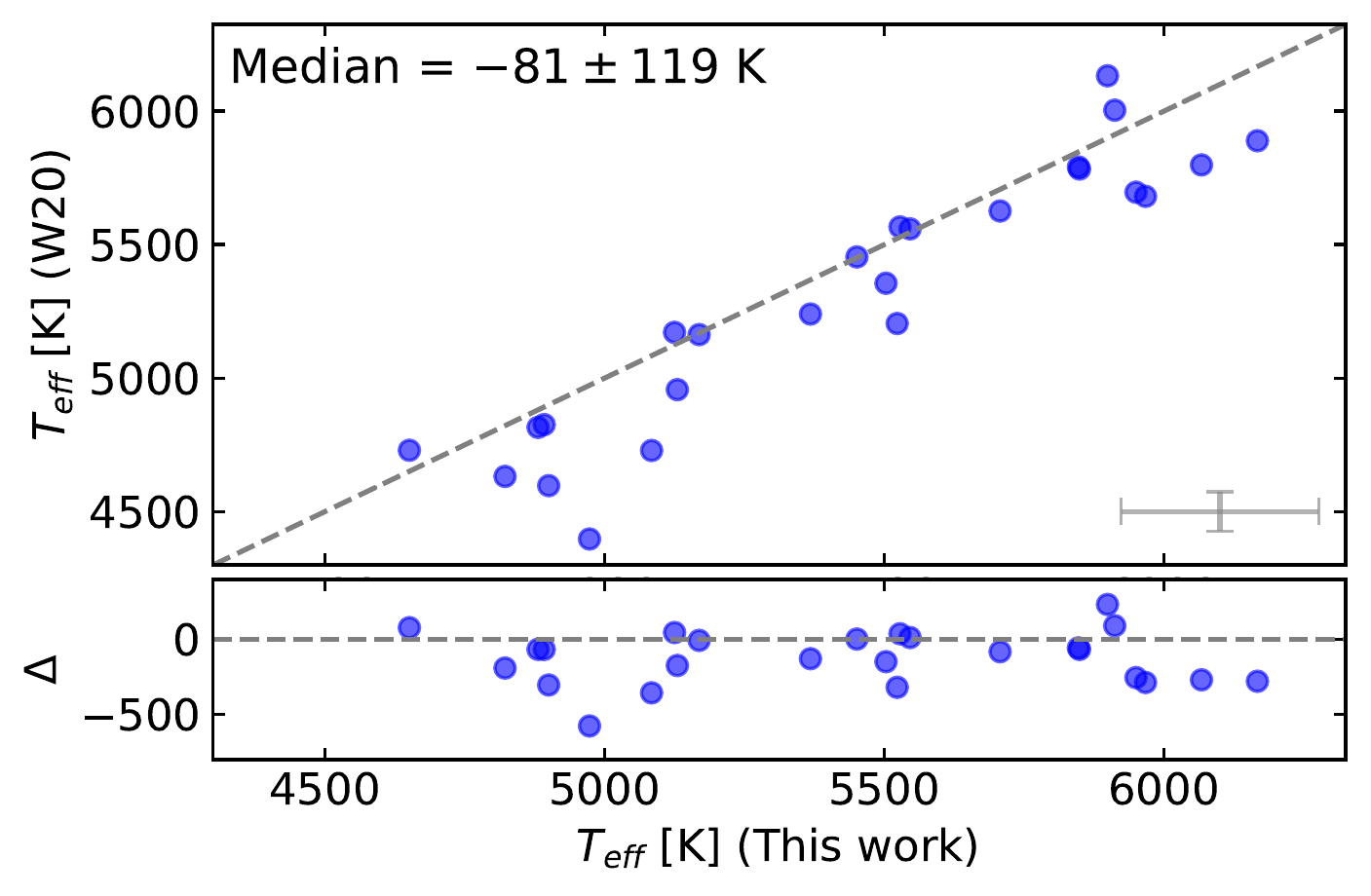}{0.48\textwidth}{}
           }
          \vspace{-11mm}
\gridline{\fig{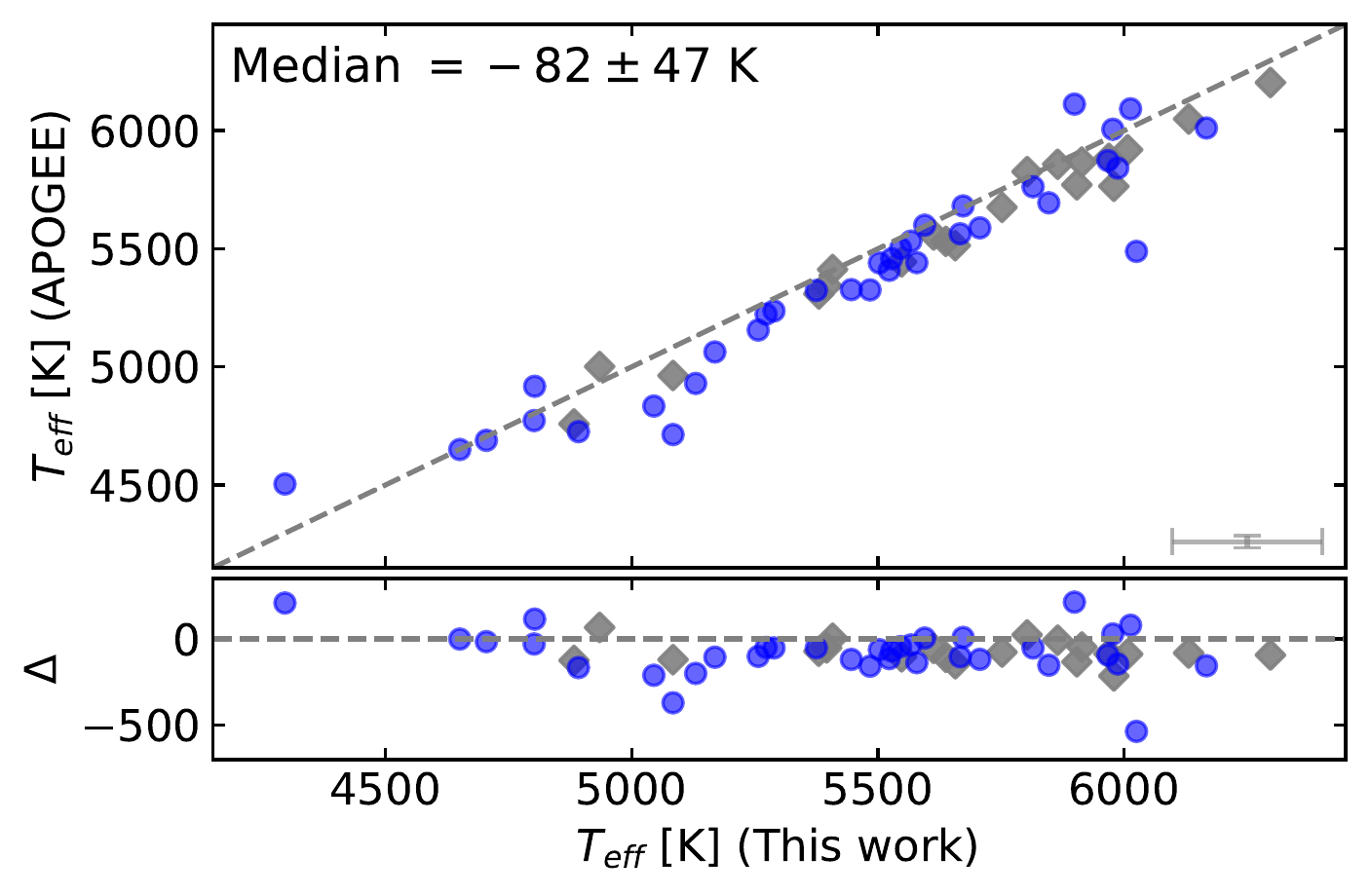}{0.48\textwidth}{}
            \hspace{-3mm}
          \fig{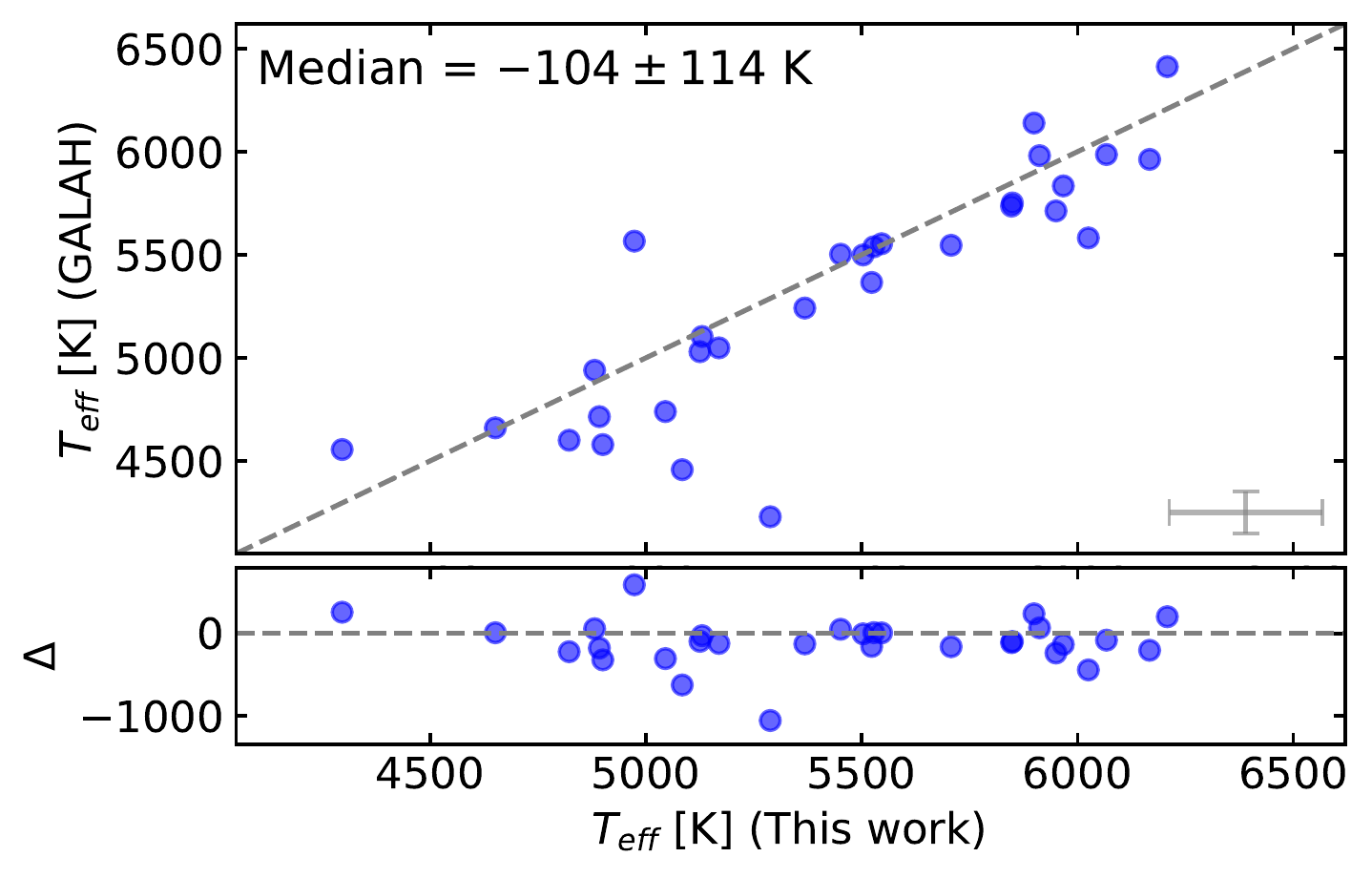}{0.48\textwidth}{}
           }
            \vspace{-11mm}
\gridline{\fig{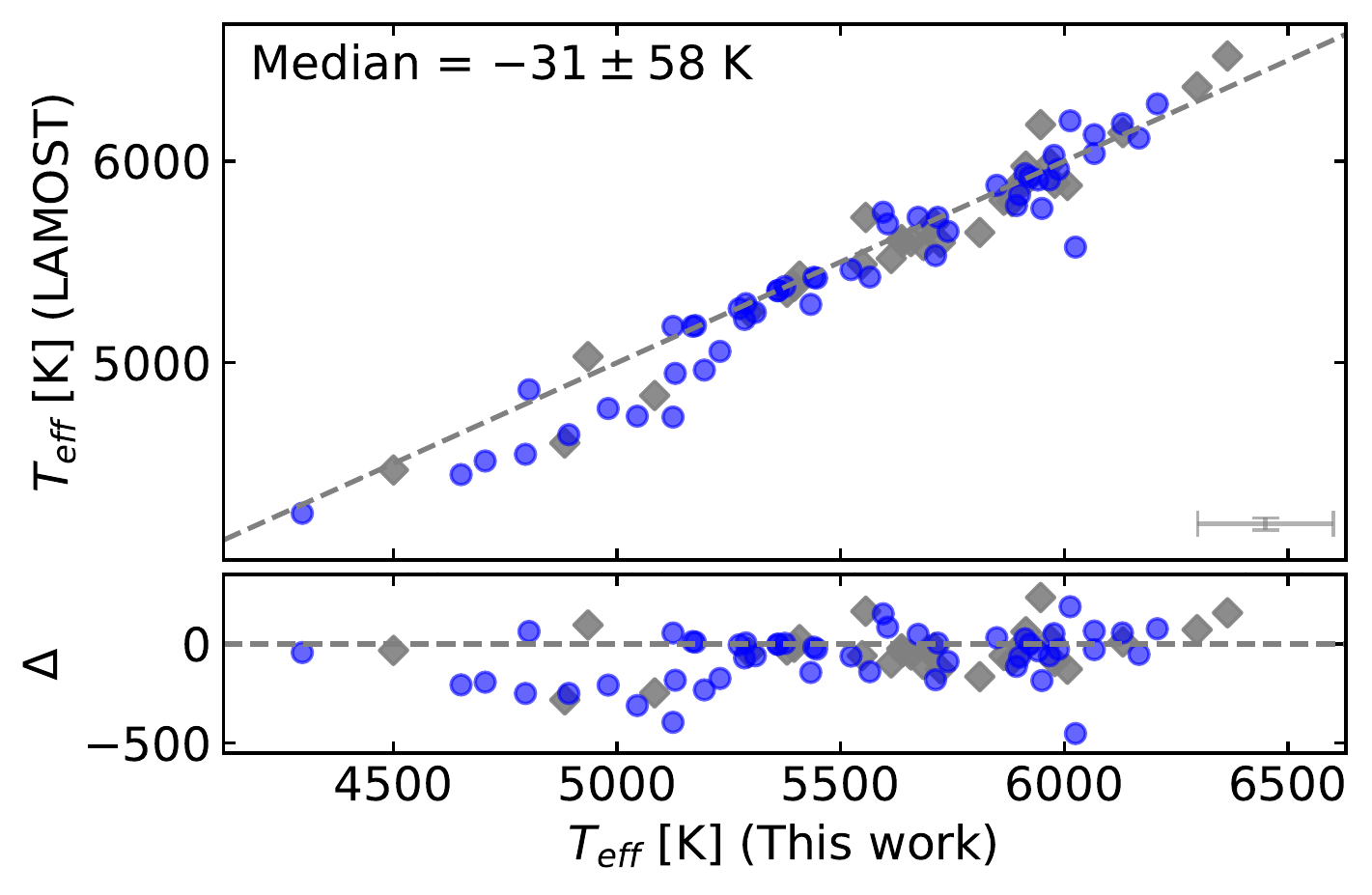}{0.48\textwidth}{}
          }
        \vspace{-8mm}
\caption{Comparisons of the effective temperatures in this study with those from  \citet[P17,][P18]{petigura2017,petigura2018},  \citet[B18,][]{Brewer2018}, \citet[M19,][]{martinez2019}, \citet[W20,][]{wittenmyer2020}, APOGEE DR17, GALAH DR3, and LAMOST DR5. K2 stars are blue circles and Kepler 1 stars are grey diamonds. The bottom sub-panels show the difference between 
`Other Work - This Work' ($\Delta$). The median differences between the parameters and the corresponding MAD are indicated in each case. The black dashed lines represent equality.}
\label{fig:teff}
\end{figure*}

Comparisons of our T$_{\rm eff}$s with the studies mentioned above are shown in the different panels of Figure \ref{fig:teff}; for each case, the median differences ``Other Work" - ``This Work" are given in the top left of each panel, along with the corresponding median absolute deviation (MAD); in all panels the grey diamonds correspond to Kepler 1 stars, while the blue circles correspond to K2 stars. From the medians and MADs in Figure \ref{fig:teff} we can conclude that there is an overall good agreement between our effective temperatures and those from the APOGEE and LAMOST surveys, as well as those available from the literature. The mean of the median differences is small, $\Delta$T$_{\rm eff}\sim-45$ K and the MADs are all below 70 K, except for the comparison with the GALAH results, where MAD = 119 K \citep{wittenmyer2020} and 114 K (GALAH). Overall our T$_{\rm eff}$ scale is just slightly hotter than the other scales, except for the comparison with \citet[][K2 targets median difference = +8 K]{petigura2018} and \citet[][median difference = 0 K]{martinez2019}. 

%%%%%%%%%%%%%%%%%%%%%%%%%%%%%%%%%%%%%%%%%%%%%%%%%%%%
%%%%%%%%%%%%%%%%%%%%%%%%%%%%%%%%%%%%%%%%%%%%%%%%%

\begin{figure*}
\gridline{\fig{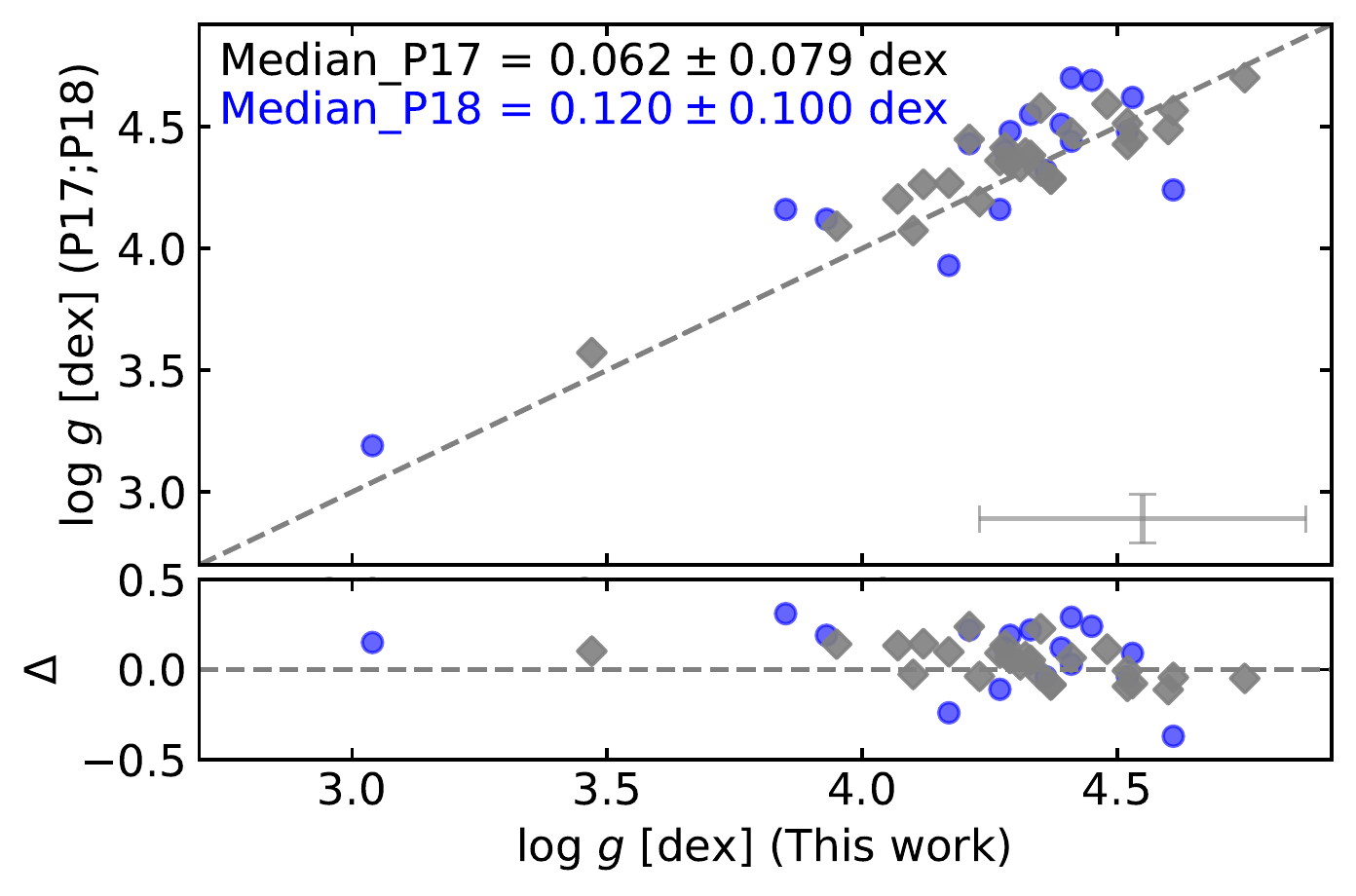}{0.48\textwidth}{}
            \hspace{-3mm}
          \fig{logg_brewer+2018_hydra}{0.48\textwidth}{}
          }
          \vspace{-11mm}
\gridline{\fig{logg_martinez+2019_hydra.pdf}{0.48\textwidth}{}
            \hspace{-3mm}
          \fig{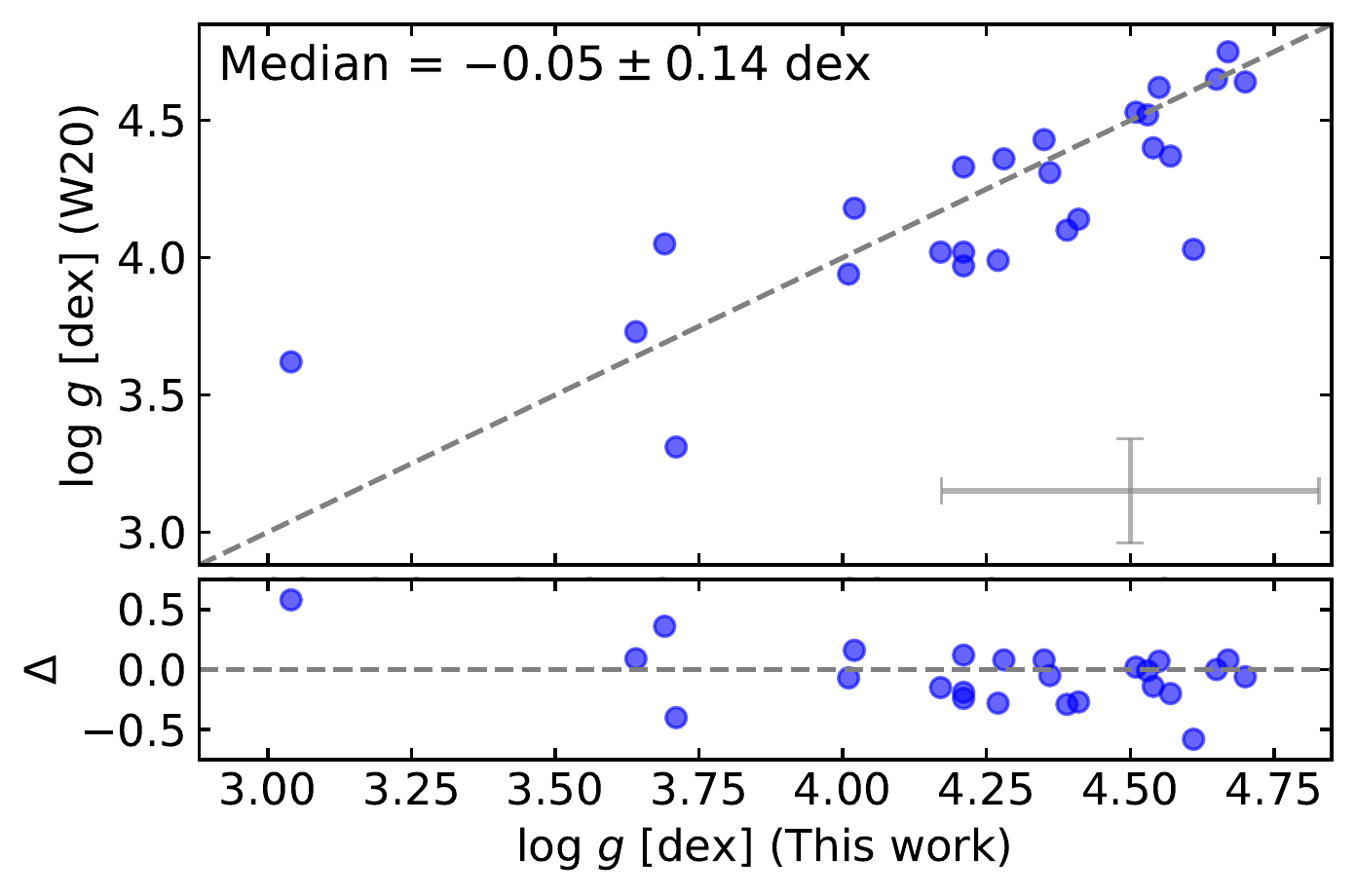}{0.48\textwidth}{}
          }
          \vspace{-11mm}
\gridline{\fig{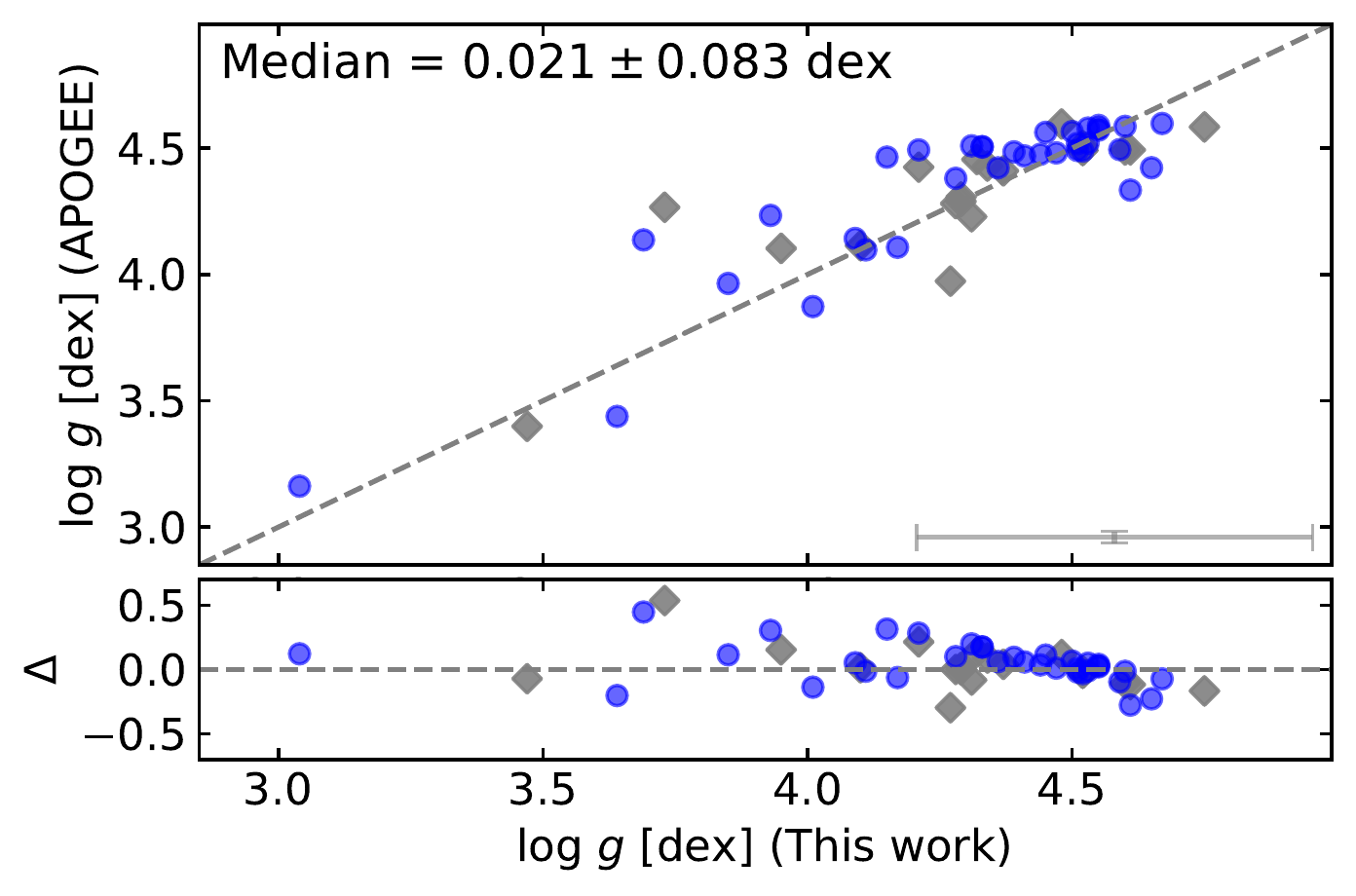}{0.48\textwidth}{}
            \hspace{-3mm}
          \fig{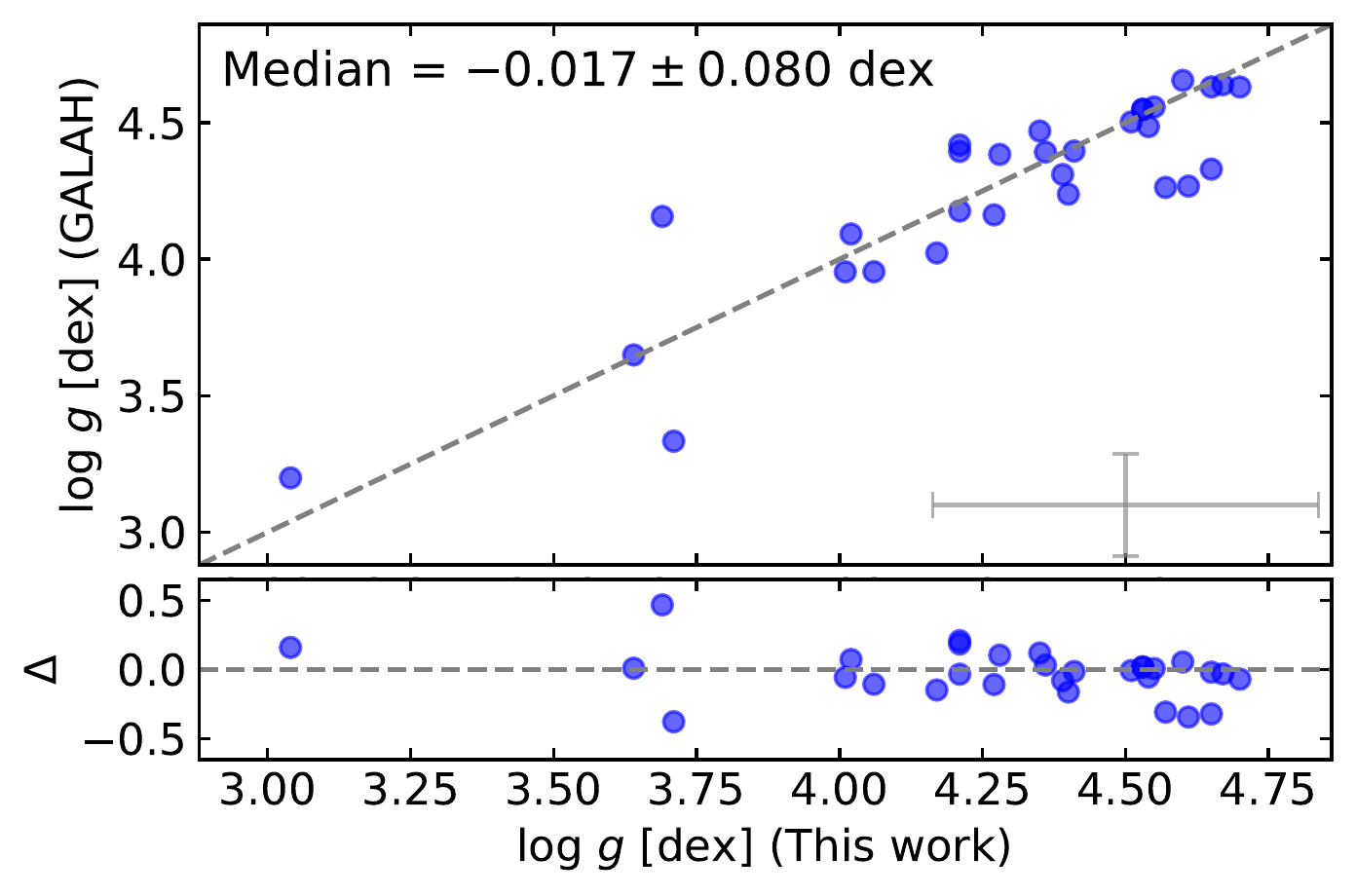}{0.48\textwidth}{}
          }
          \vspace{-11mm}
\gridline{\fig{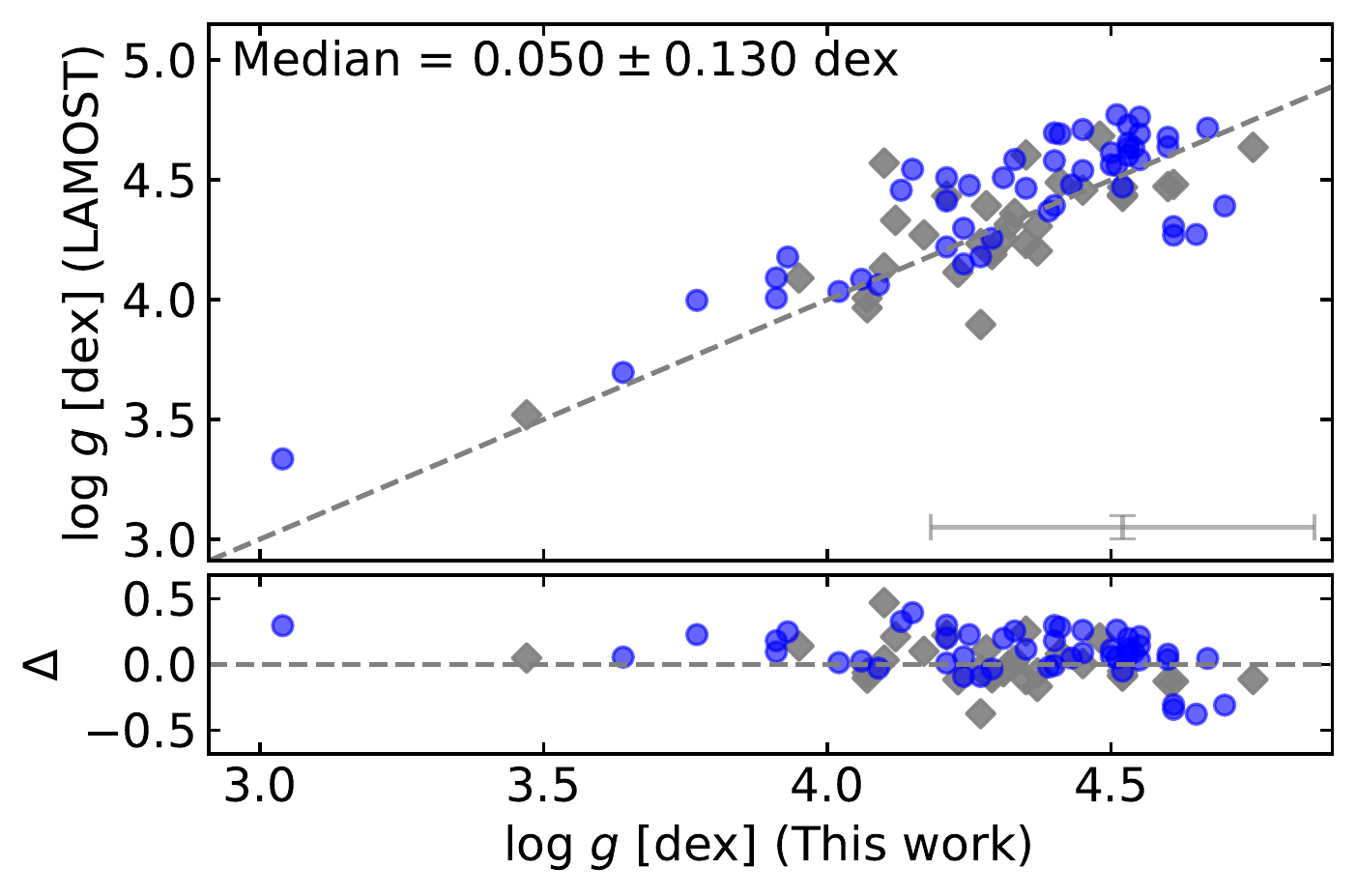}{0.48\textwidth}{}
          }
          \vspace{-8mm}
\caption{Comparison of the log $g$ values derived in this work and \citet[P17,][P18]{petigura2017,petigura2018},  \citet[B18,][]{Brewer2018}, \citet[M19,][]{martinez2019}, \citet[W20,][]{wittenmyer2020}, APOGEE DR17, GALAH DR3, and LAMOST DR5 for K2 stars (blue circle) and Kepler 1 stars (grey diamond). The median differences between the parameters and the corresponding MAD are indicated in each case. The black dashed lines represent equality.  Each bottom sub-panel plots the difference, $\Delta$, between `Other Work - This Work'.}
\label{fig:logg}
\end{figure*}

Comparisons of the surface gravities derived here - from Fe I and Fe II ionization balance -  with those from the works discussed above are shown in Figure \ref{fig:logg} (as in Figure \ref{fig:teff}, the median differences (``Other Work" - ``This Work") $\pm$ MAD are given in each panel). The median log $g$ differences for all studies and the surveys are surprisingly small (see Section \ref{sec:astero}) given the different analysis methodologies and line lists, all $<$ 0.06 dex, except for \cite{petigura2018} that has a mean log $g$ difference of 0.12 dex. In all comparisons the MAD values are smaller or equal to 0.14 dex, which is smaller than typical uncetainties in log $g$ values.

%##############################################
%###########################################################

\subsubsection{Asteroseismic vs. Spectroscopic Surface Gravities} \label{sec:astero}
%----------------------------------------------------------------

The stellar parameters derived here (T$_{\rm eff}$, log $g$, $\xi$, [Fe/H]) are based on the analysis of a sample of Fe I and Fe II lines, where correlations between parameters in such a spectroscopic analysis can lead to systematic errors, especially in the derived values for log $g$; for example, Fe I lines are typically stronger than Fe II lines, leading to potential correlations between the microturbulent velocity and log $g$.  Accurate surface gravities can thus be one of the more difficult parameters to constrain via spectroscopy, especially if the Fe II lines are few in number and weak.  In order to investigate possible systematic offsets in the log $g$ values derived in this study, we compare our results with those computed via asteroseismology, where the surface gravity can be derived with quite good precision \citep{pinsonneault2018}. 

\begin{figure*}
\epsscale{0.75}
\plotone{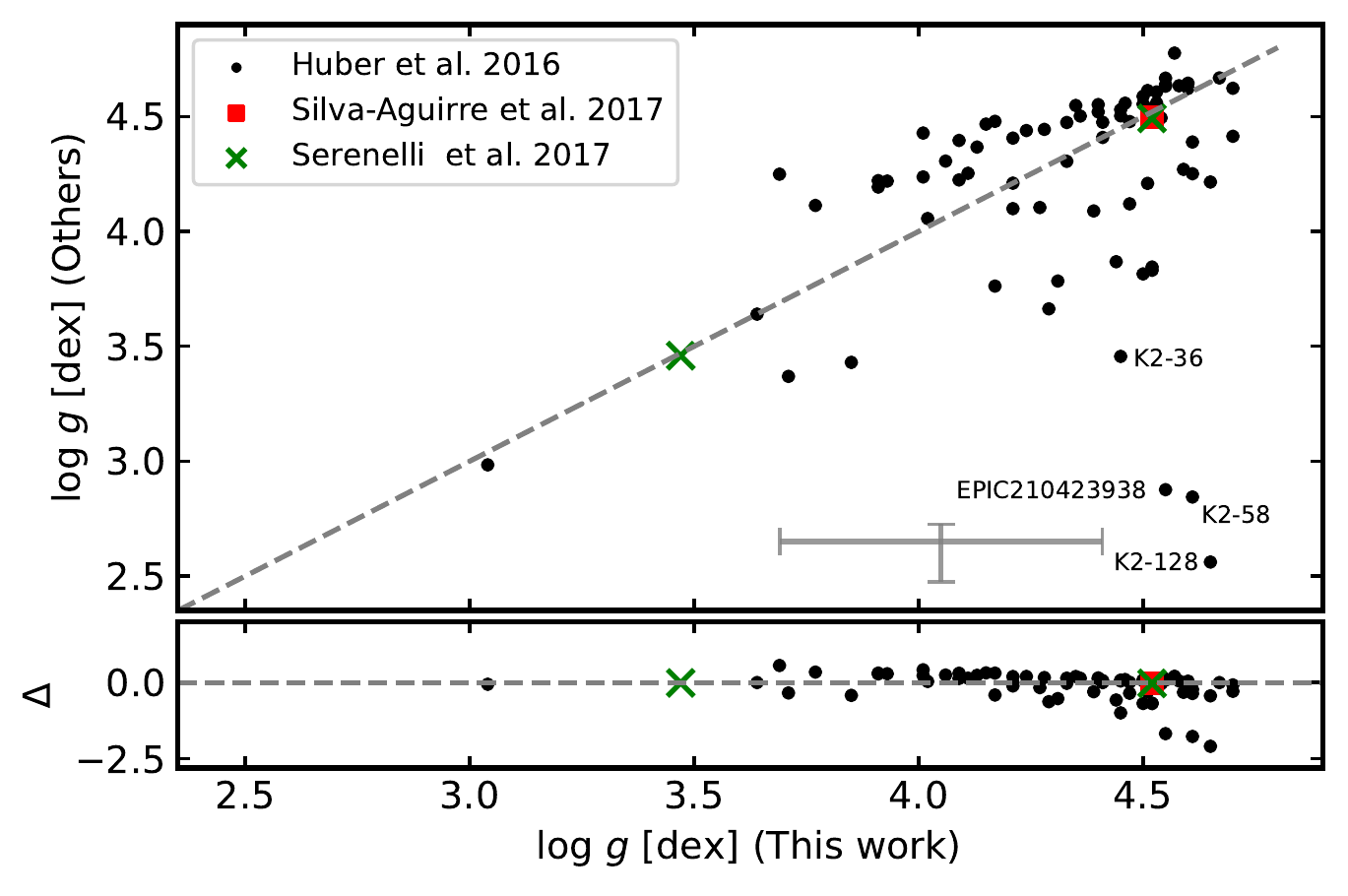}
\caption{Comparisons of spectroscopic surface gravities derived in this work with asteroseismic surface gravities from \cite{huber2016ApJS..224....2H}, \cite{silva-aguirre2017} and \cite{serenelli2017}.  The four most discrepant values of stellar log $g$ between this work and \cite{huber2016ApJS..224....2H} are labeled and these are discussed in the text.  The bottom panel shows the differences between `Other Work - This Work'. Median uncertainty error bars are presented.}
\label{fig:asteroseismic}
\end{figure*}

For such a comparison, we collected in Figure \ref{fig:asteroseismic} asteroseismology results for 83 stars from \cite{huber2016ApJS..224....2H}, \cite{silva-aguirre2017}, and \cite{serenelli2017}. In order to compute surface gravities via asteroseismology, \cite{huber2016ApJS..224....2H} and  \cite{serenelli2017} used the T$_{\rm eff}$ and [Fe/H] from the APOGEE survey (DR13), while \cite{silva-aguirre2017} adopted T$_{\rm eff}$ and [Fe/H] from several literature sources.  Most of the stars in common with this study are from \cite{huber2016ApJS..224....2H}; the log $g$s from that study show no significant offsets and are in good agreement when compared to ours, with the median log $g$ difference (``Huber et al. 2016" - ``This Work") = 0.033 $\pm$ 0.164 dex. However, there are four stars for which the log $g$ differences are $>1$ dex. These four discrepent stars (which are labeled in Figure \ref{fig:asteroseismic}) have asteroseismic gravities that would indicate that they are giant or subgiant stars, whereas other studies have derived surface gravities that suggest that these stars are dwarfs as discussed below:

\begin{itemize}
 \item K2-36: this star was analyzed individually in \cite{damasso2019} using high-resolution spectra from HARPS-N, with stellar parameters derived using the spectroscopic technique and finding log $g$ values of 4.73, 4.60, and 4.57 with Equivalent Widths, Atmospheric Stellar Parameters from Cross-Correlation Functions (CCFpams), and Stellar Parameter Classification \citep[SPC;][]{buchhave2014}, respectively. Using \texttt{SME} and \texttt{SpecMatch} with HIRES spectra, \cite{sinukoff2016}, \cite{brewer2016a}, and \cite{crossfield2016} derived log $g$ =4.65, 4.55, and 4.60, respectively. Finally, using SPC in Tillinghast Reflector Echelle Spectrograph (TRES) spectra, \cite{vanderburg2016} derived log $g$ =4.70.  Our result (log $g$ = 4.45) is $\sim$0.1-0.3 dex lower than these studies, although all of these results indicate that K2-36 is a dwarf.

 \item K2-58: using \texttt{SpecMatch} and \texttt{SME} with HIRES spectra, \cite{crossfield2016} and \cite{Brewer2018} derived log $g$ =4.52 and 4.50. Using \texttt{SPC} with TRES spectra, \cite{vanderburg2016} derived log $g=4.54$, with these values being consistent with our log $g=4.50$ dex, pointing to K2-58 as being a dwarf.

\item K2-128: using \texttt{SPC} with TRES spectra, \cite{crossfield2018} and \cite{mayo2018} derived log $g = 4.70$, while this result is 0.20 dex larger than our log $g$; taken together, these studies point to K2-128 as being a dwarf. 

\item EPIC 210423938: for this star, the log $g$ value obtained in \cite{mayo2018} is 4.69. \cite{stassun2019} determined that log $g$ = 4.55 from stellar radius and mass. Considering our result (log $g = 4.55$ dex), we identify EPIC 210423938 as a dwarf. 
\end{itemize}

In summary, spectroscopic studies have found the four stars discussed above to have dwarf-star surface gravities.  In addition, this status is confirmed from their DR3 Gaia parallaxes and distances, with distances to K2-36 of 109 pc, K2-58 of 182 pc, K2-128 of 114 pc, and  EPIC 210423938 of 149 pc, resulting in absolute Gaia or V-magnitudes of M$_{V}$=6.5, M$_{G}$=5.8, M$_{V}$=7.3, and M$_{V}$=7.1, respectively.  These absolute magnitudes confirm them as K-dwarfs. 
Without considering the four discrepant stars discussed above, we find median log $g$ differences with \cite{huber2016ApJS..224....2H} = $-$0.049 $\pm$ 0.136. We note that for the few stars in common with \cite{silva-aguirre2017} and \cite{serenelli2017}, there is excellent agreement. 
%---------------------------------------
%---------------------------------------

%-----------------------------
\subsection{Metallicities}
%-----------------------------

\begin{figure*}
\gridline{\fig{FeH_Petigura2017_2018_hydra.pdf}{0.48\textwidth}{}
            \hspace{-3mm}
          \fig{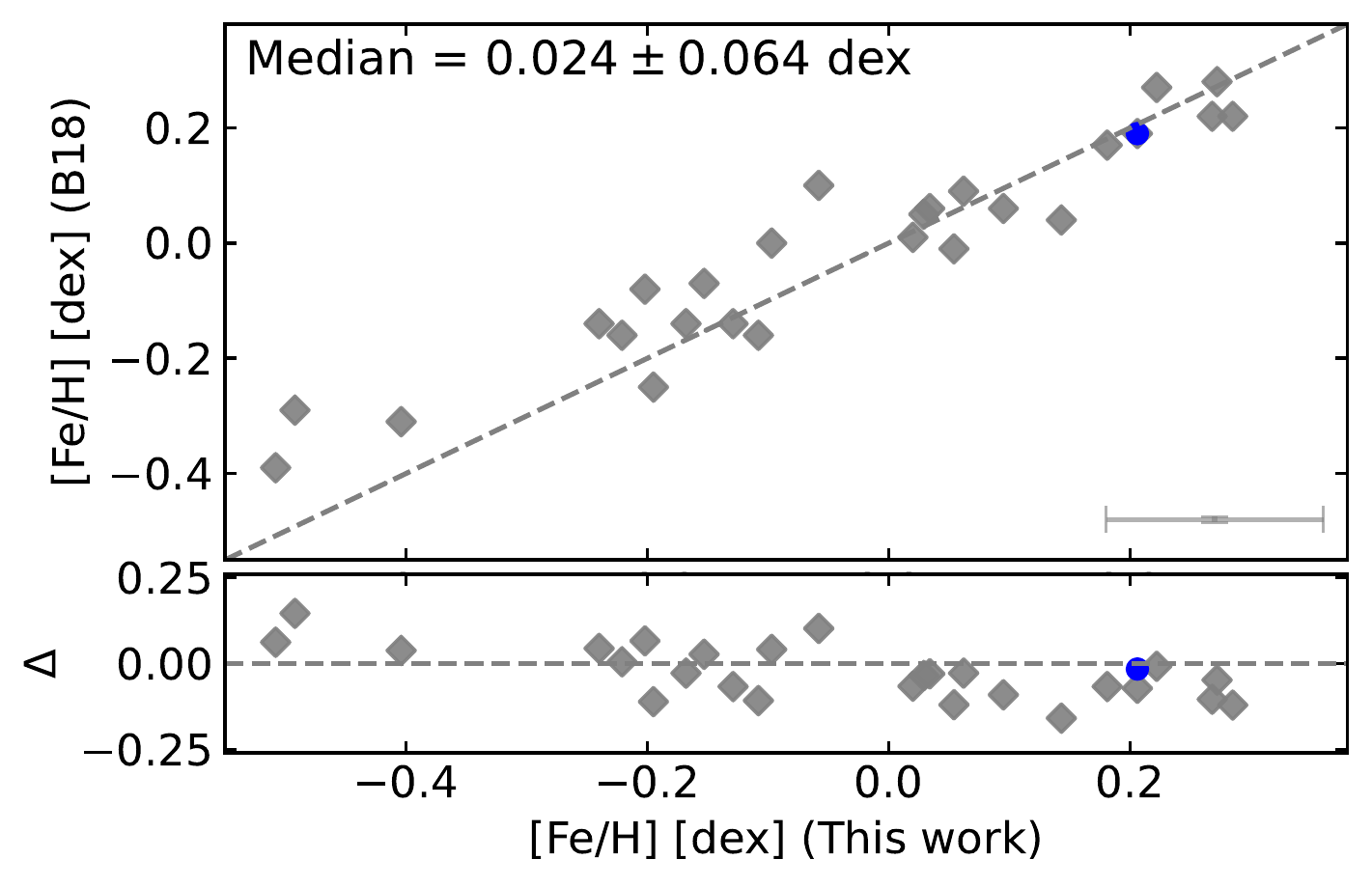}{0.48\textwidth}{}
          }
            \vspace{-11mm}
\gridline{\fig{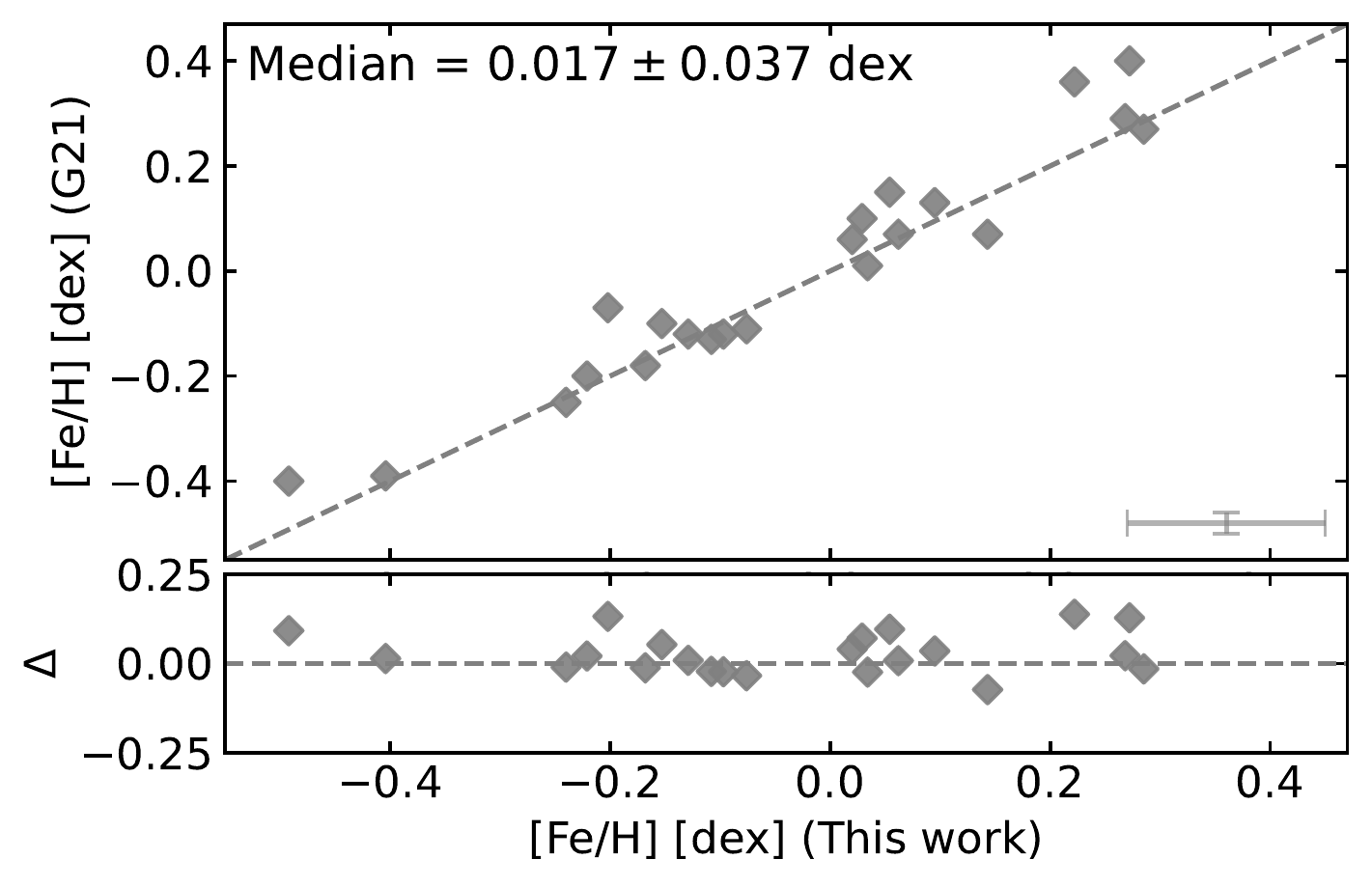}{0.48\textwidth}{}
            \hspace{-3.1mm}
          \fig{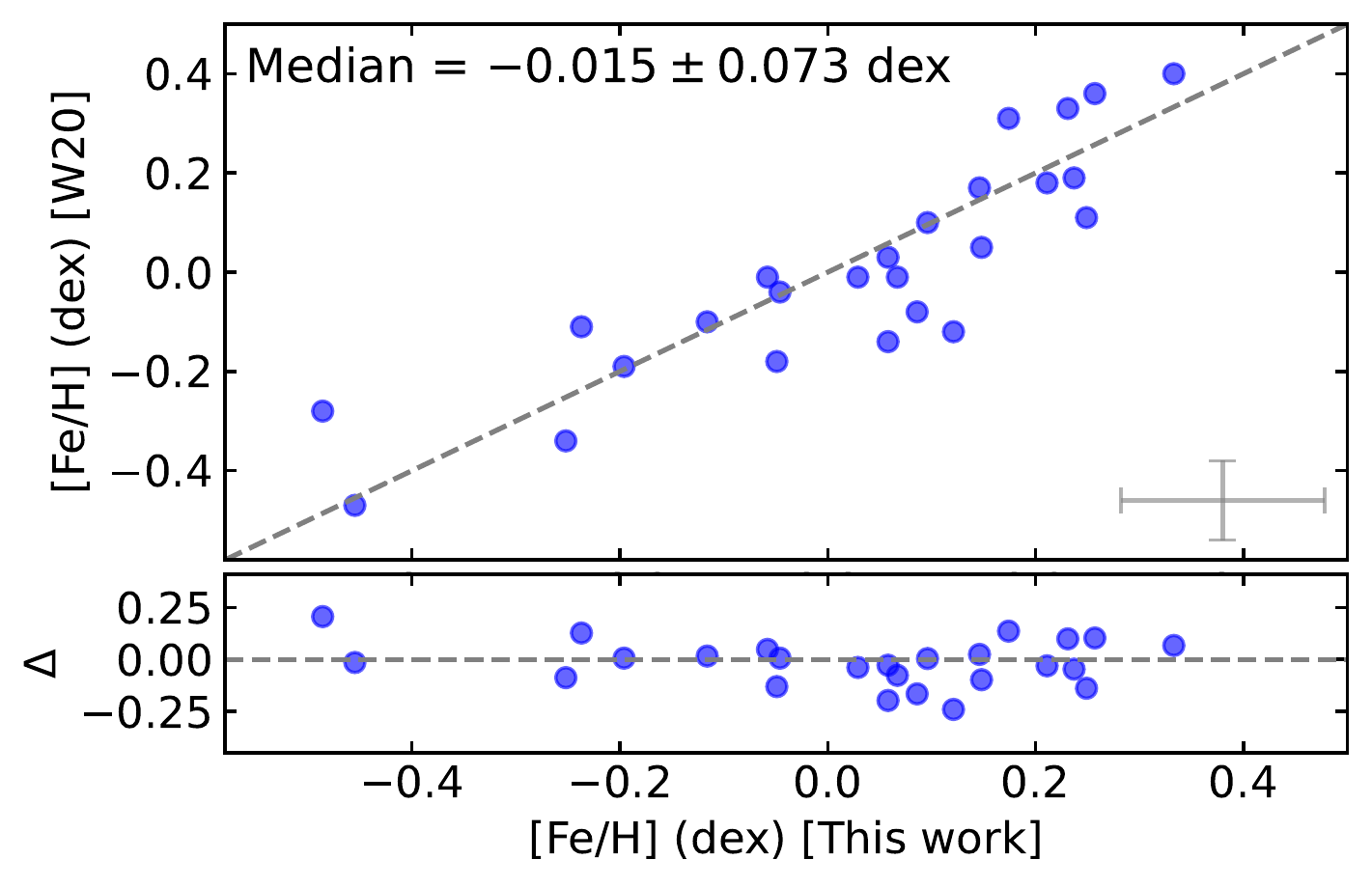}{0.48\textwidth}{}
          }
          \vspace{-11mm}
\gridline{\fig{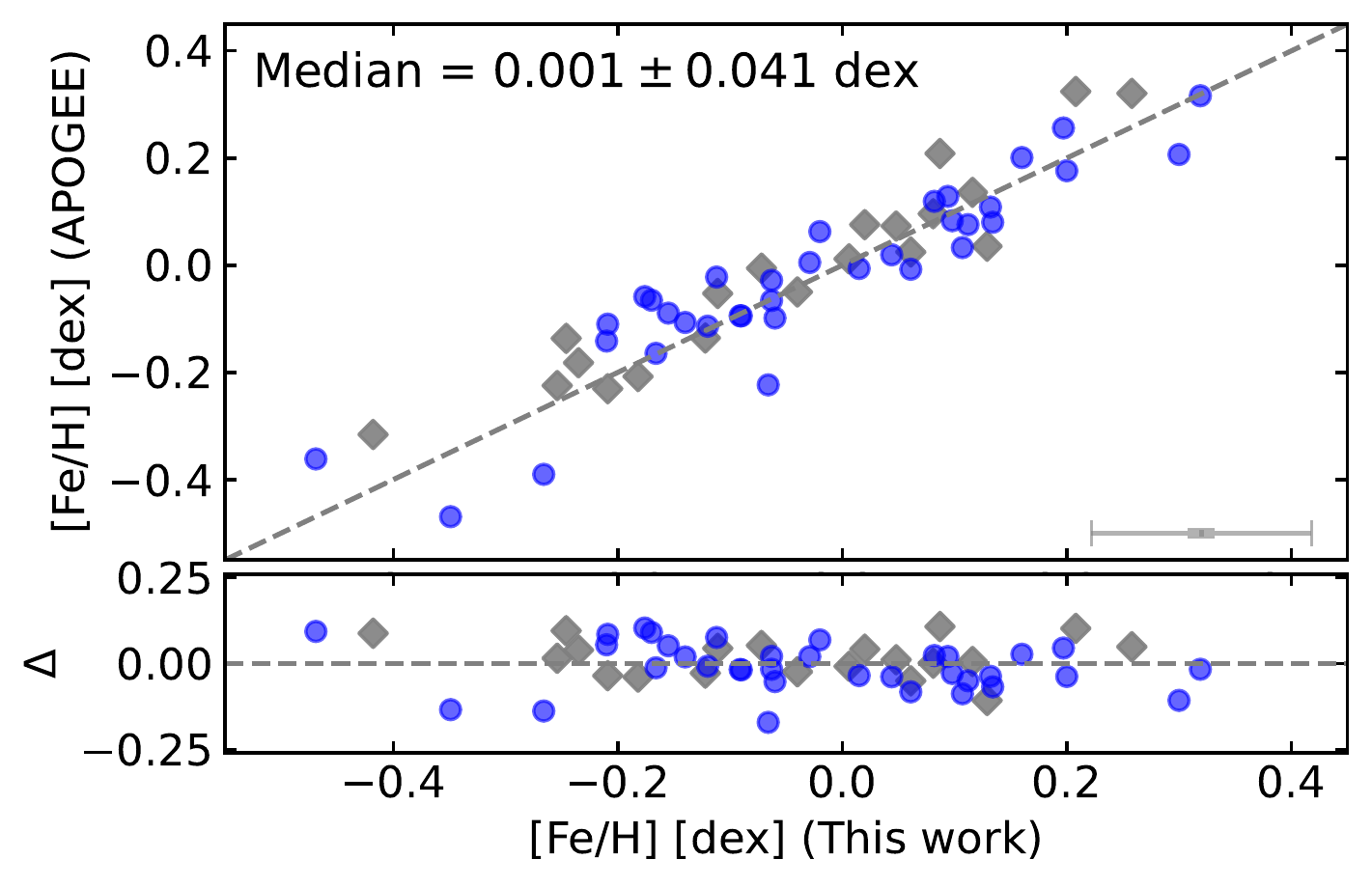}{0.48\textwidth}{}
            \hspace{-3mm}
          \fig{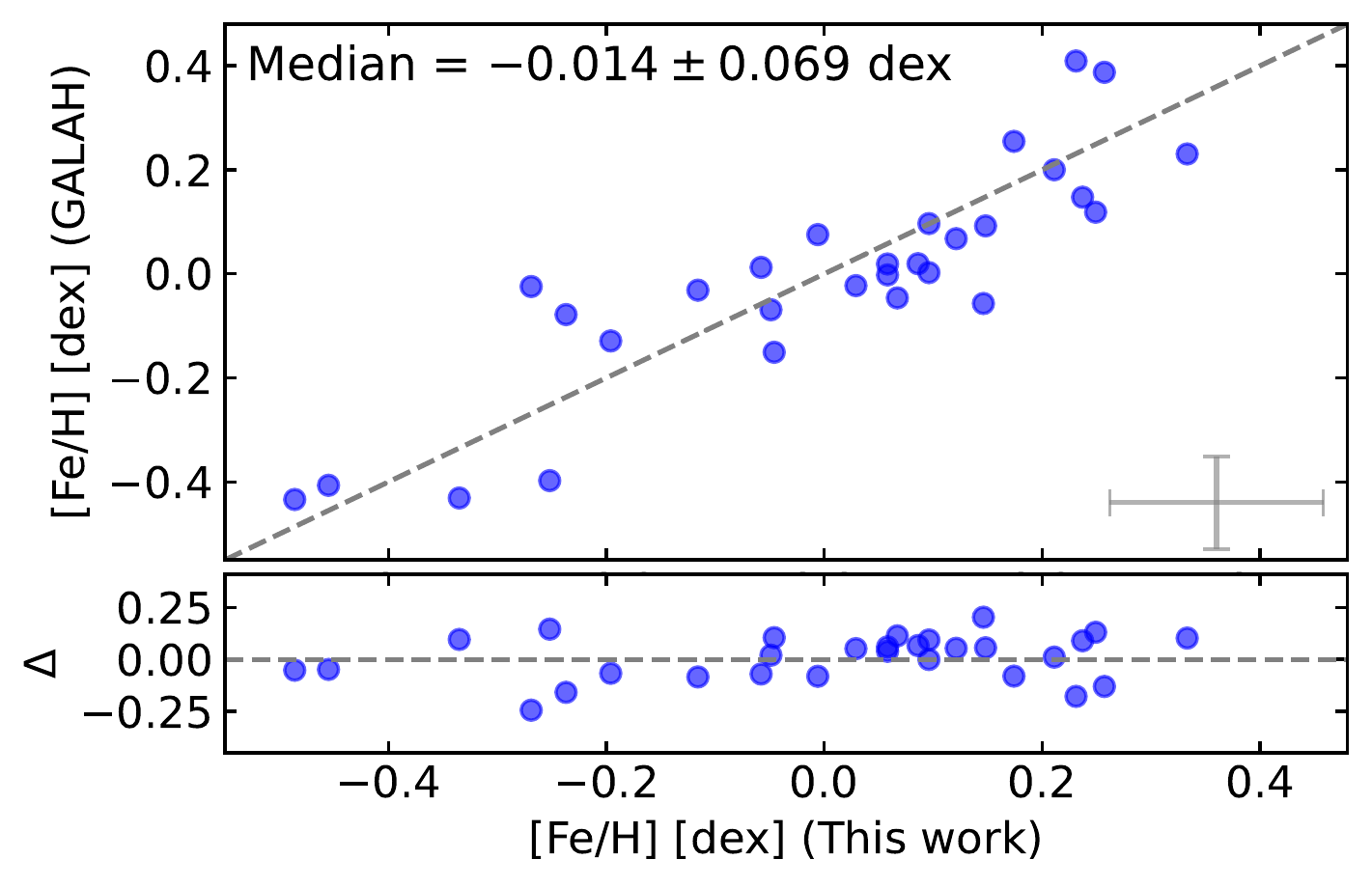}{0.48\textwidth}{}
          }
            \vspace{-11mm}
\gridline{\fig{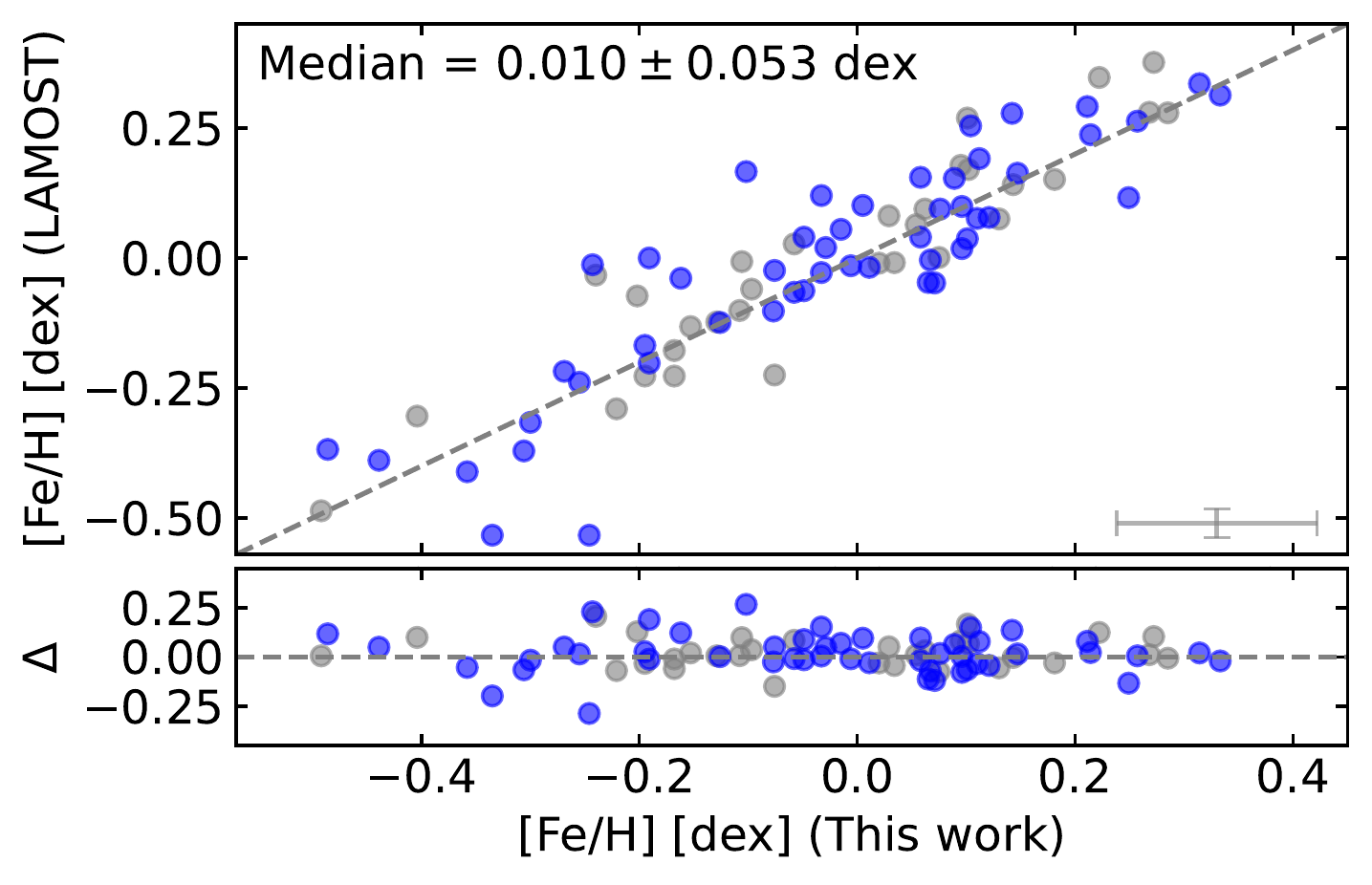}{0.48\textwidth}{}
          }
          \vspace{-8mm}
\caption{Comparisons of metallicities derived in this work with results from \citet[P17,][P18]{petigura2017,petigura2018},  \citet[B18,][]{Brewer2018}, \citet[G21,][]{Ghezzi2021}, \citet[W20][]{wittenmyer2020}, APOGEE DR17, GALAH DR3, and LAMOST DR5, for K2 stars (blue circles) and Kepler 1 stars (grey diamonds). The median differences between the parameters and the corresponding MAD scatters are indicated in each case. The black dashed lines represents equality. The bottom sub-panels show the differences between `Other Work - This Work'.}
\label{fig:Afe}
\end{figure*}

Figure \ref{fig:Afe} summarizes the comparison of the metallicities for K2 stars (blue circles) and Kepler 1 stars (grey diamonds) derived in this work with those derived in other studies and surveys. The metallicities from the literature were obtained either via spectrum synthesis method \citep{petigura2017, petigura2018, Brewer2018, wittenmyer2020}, \citet[][GALAH]{Buder2021}, \citet[][APOGEE]{jonsson2020AJ....160..120J}, \citet[][LAMOST]{zong2018} or were based on equivalent-width measurements of Fe I and Fe II lines \citep{Ghezzi2021}. For Kepler 1 stars in common with \citet{petigura2017}, the median metallicity difference is $+0.040 \pm  0.034$ dex, while with \cite{Brewer2018} it is $0.024\pm  0.064$ dex. There are 22 stars in common with \cite{Ghezzi2021}, with a median difference of $0.017 \pm 0.037$ dex. The K2 stars in common with \cite{petigura2018} have a median difference of $+0.021 \pm 0.052$ dex, and for those in common with \cite{wittenmyer2020} the median [Fe/H] difference is $-0.015 \pm 0.073$ dex. Comparisons of metallicities with the spectroscopic surveys also find excellent agreement: $0.001 \pm 0.041$ dex for APOGEE DR17, while with GALAH DR3 it is $-0.014\pm  0.069$ dex.
Finally, the median difference for stars in common with LAMOST DR5 is $+0.010 \pm 0.053$ dex.

Overall, the metallicity comparisons in Figure \ref{fig:Afe} indicate good consistency between our results and the other metallicity scales, with median metallicity differences in all cases being smaller than $\sim$0.04 dex and MAD less than 0.07 dex. 
Save for a few outliers, the scatter in [Fe/H] differences are well within the expected uncertainties.

%##############################################
%##############################################

\subsection{Stellar Radii and Masses} \label{sec:rad_mass}

A number of studies in the literature obtained stellar radii and, in some instances, also masses for Kepler 1 targets using different combinations of codes, models and parameters. \cite{johnson2017} used the isochrone method, stellar parameters from \cite{petigura2017}, and the Dartmouth Stellar Evolution Program (DSEP) models \citep{dotter2008}; \cite{fulton2018}, computed stellar radii using Gaia DR2 inverted parallaxes, and \cite{martinez2019} used their stellar parameters, combined with Gaia DR2 distances from \cite{BailerJones2018}. The large Gaia-Kepler stellar properties catalog by \cite{berger2020a} combined isochrones, Gaia DR2 parallaxes, and spectroscopic metallicities using the \texttt{isoclassify} \citep{huber2017} package, while \cite{Hardegree2020} used stellar parameters from LAMOST and derived K2 stellar radii from the Stefan-Boltzmann law and stellar masses from stellar radii. Via asteroseismology, \cite{huber2016ApJS..224....2H} also derived stellar masses and radii of K2 stars using T$_{\rm eff}$ and [Fe/H] from the APOGEE DR14, while \cite{mayo2018} used the \texttt{isochrone} package \citep{morton2015}, which requires effective temperature, surface gravity and metallicity as input parameters \citep[these parameters were derived using the spectral synthesis method described in][]{mayo2018}.

\begin{figure*}

\gridline{\fig{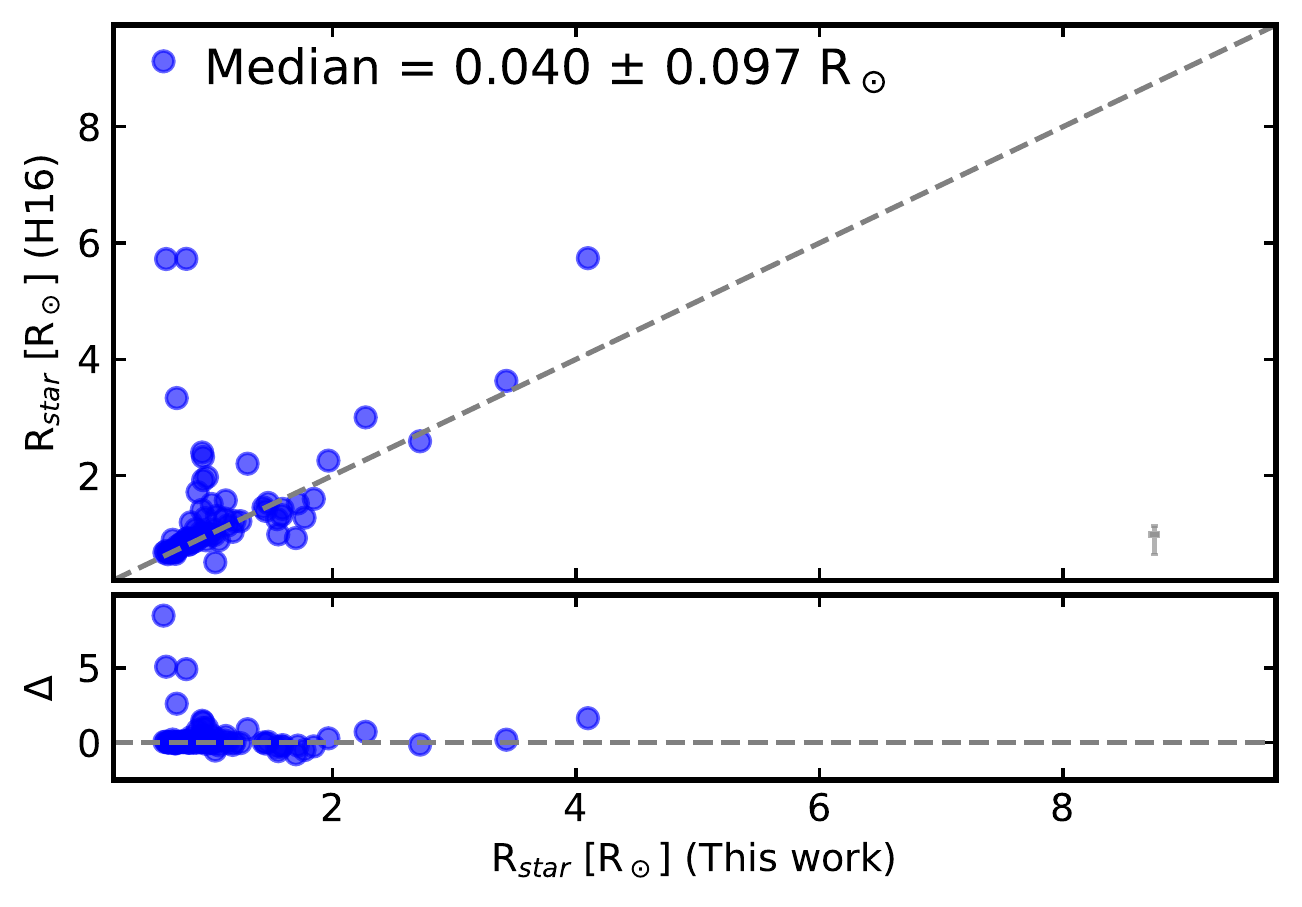}{0.5\textwidth}{}
            \hspace{-3mm}
          \fig{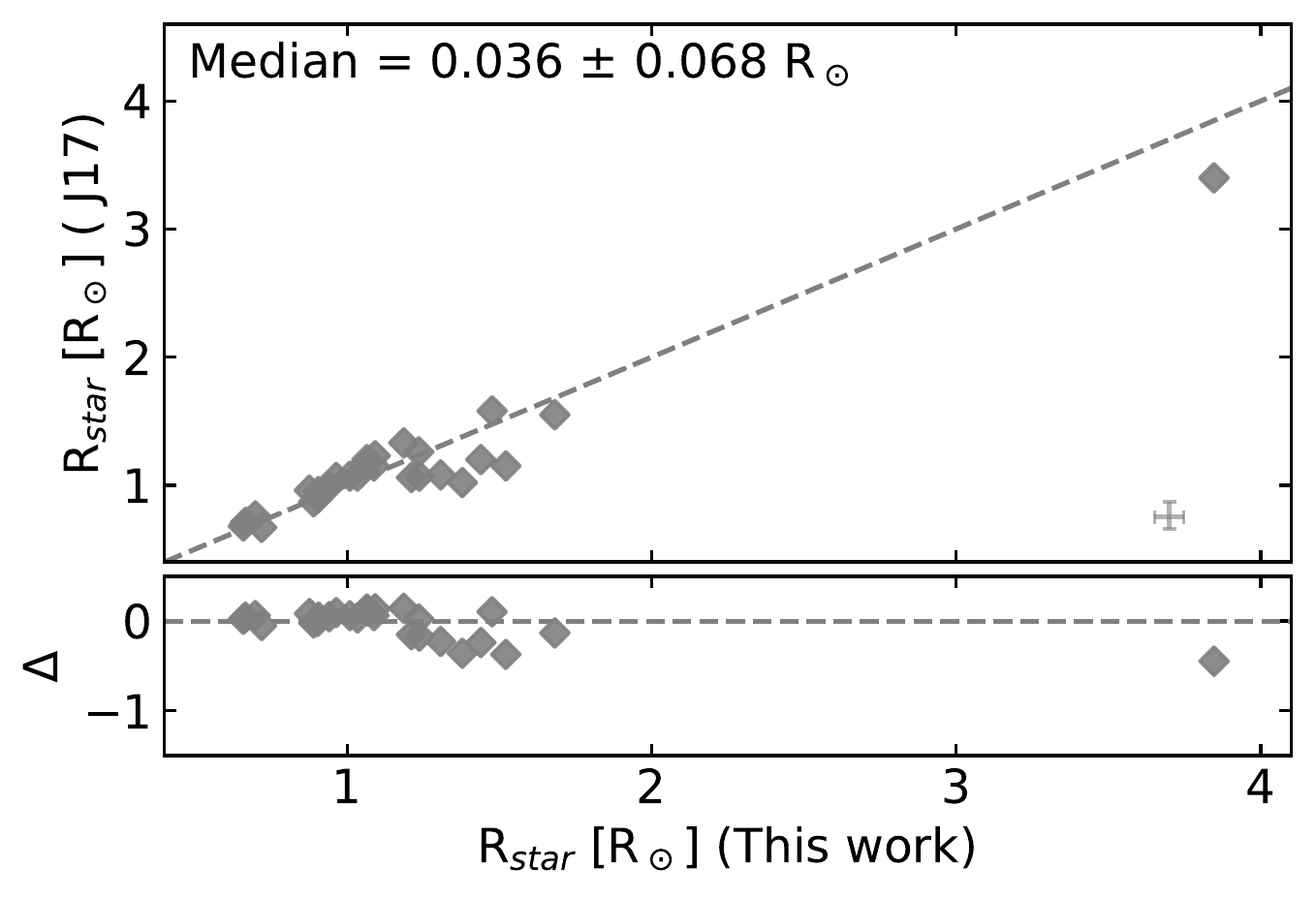}{0.5\textwidth}{}
          }
          \vspace{-10mm}
\gridline{\fig{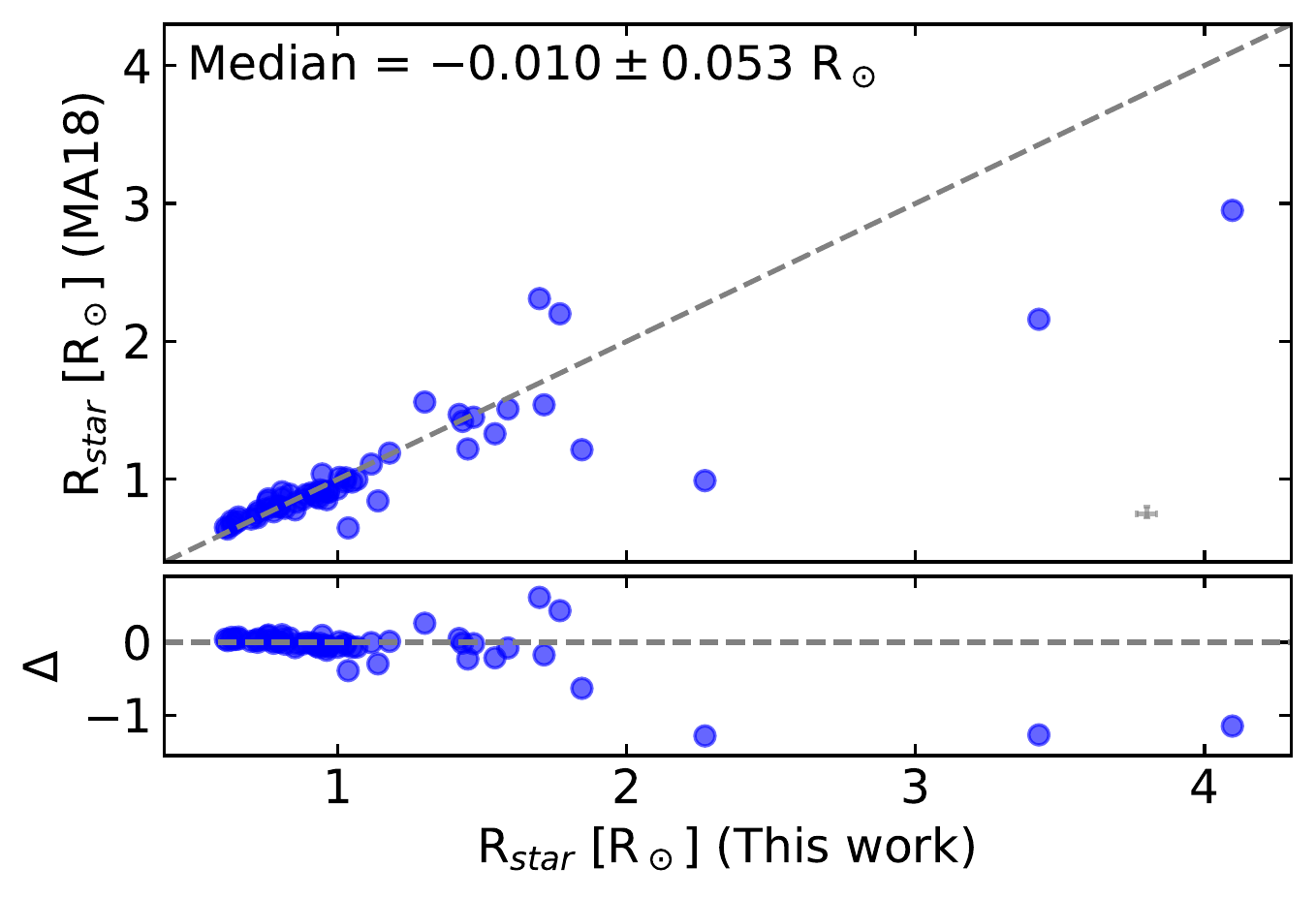}{0.5\textwidth}{}
            \hspace{-3mm}
          \fig{Radii_martinez+2019_HYD_GZ21}{0.5\textwidth}{}
          }
          \vspace{-10mm}
\gridline{\fig{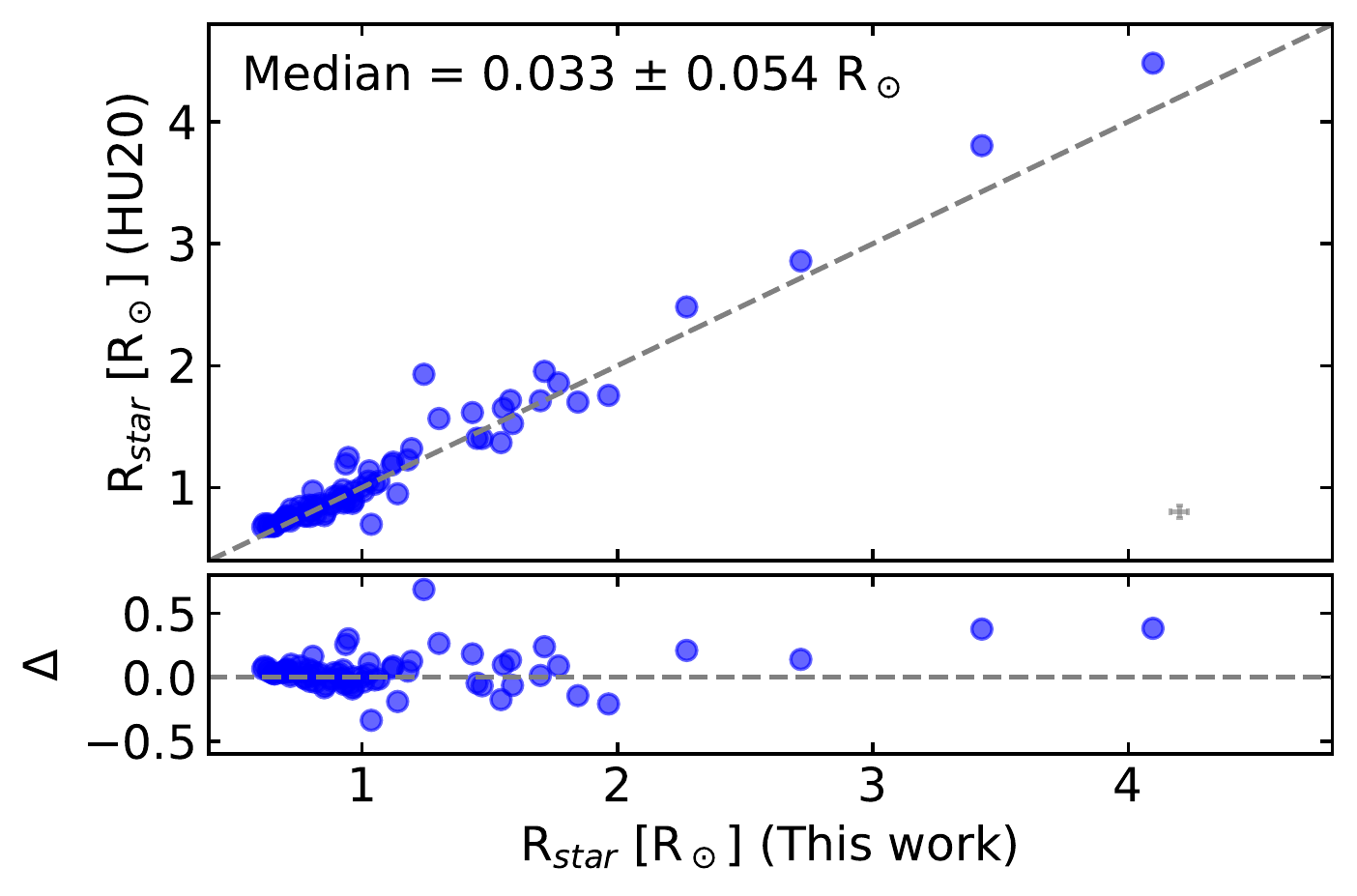}{0.5\textwidth}{}
            \hspace{-3mm}
          \fig{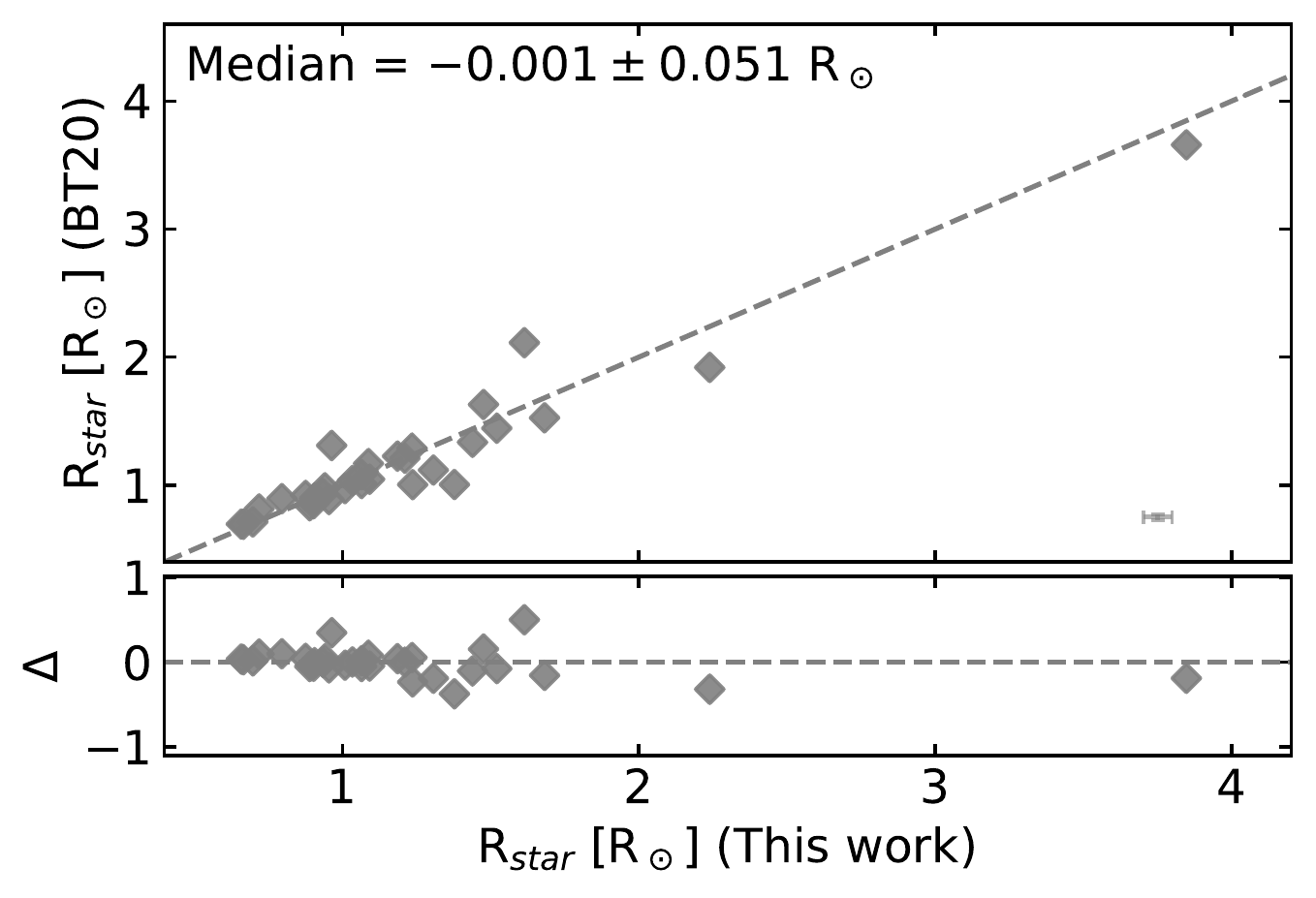}{0.5\textwidth}{}
          }
          \vspace{-8mm}
\caption{Comparisons of the stellar radii derived in this work with stellar radii from \citet[H16,][]{huber2016ApJS..224....2H}, \citet[MA18,][]{mayo2018} and \citet[HU20,][]{Hardegree2020} for K2 stars (left panels), and \cite[J17,][]{johnson2017}, \citet[M19,][]{martinez2019} and \citet[BT20,][]{berger2020b} for Kepler 1 stars (right panels).  The four most discrepant radii between this study and \cite{huber2016ApJS..224....2H} are discussed in the text.  Each bottom sub-panel shows the difference, $\Delta$=`Other Work - This Work'.}
\label{fig:kep_k2_R}
\end{figure*}

Comparisons of our derived radii (Section \ref{sec:radii}) with those in six studies mentioned above are shown in the left and right panels of Figure \ref{fig:kep_k2_R}, respectively for K2 and Kepler 1 targets. Please note the different scales for the x- and y- axis in the different panels. Firstly, when considering the median radius differences ``Other works - This work", overall, there is not a clear bias in any direction.
Overall, all results from the literature present similar levels of consistency relative to ours. We note, however, the presence of the four significant outliers in \cite{huber2016ApJS..224....2H} for which their radii are much larger than ours (the log $g$ values for these cases are also very discrepant as discussed in Section \ref{sec:astero} and shown in Figure \ref{fig:asteroseismic}), which when removed improve the consistency relative to our results ($+0.029 \pm 0.082$ R$_\odot$).

For example, for the most discrepant case (K2-128), the stellar radius in \cite{huber2016ApJS..224....2H} would be roughly 15 times larger than ours, which would imply a correspondingly larger planet radius.
For the comparisons with \cite{berger2020a}, \cite{mayo2018}, and \cite{martinez2019}, the median differences are $\Delta$R$_{star} < -0.01$ R$_\odot$, noting again the presence of outliers.
In the comparison with \cite{mayo2018}, in particular, the outliers are mostly for radii larger than $\sim$1.5 R$_\odot$,
with a tendency that our radii are larger.
For the comparisons with \cite{Hardegree2020} and \cite{johnson2017}, the systematics go in the opposite direction, with median $\Delta$R$_{star} \sim$ $+0.03$ R$_\odot$. In all comparisons, the MADs are less than 0.05 R$_\odot$, except for \cite{johnson2017} that is slightly larger (0.07 R$_\odot$), while \cite{huber2016ApJS..224....2H} has a much larger MAD of 0.08 R$_\odot$.

%----------------------------------------
%----------------------------------------

\begin{figure*}
\gridline{\fig{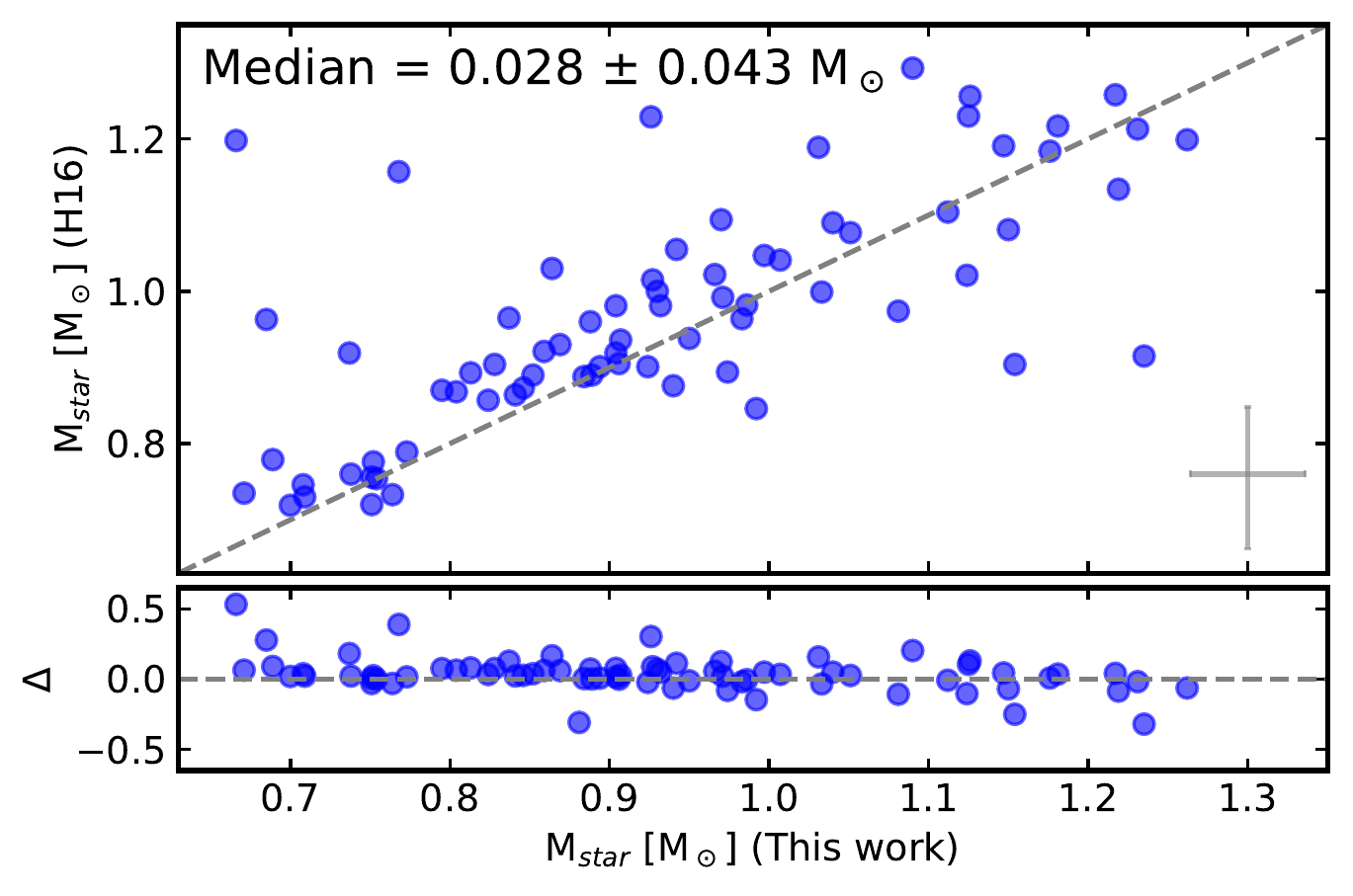}{0.5\textwidth}{}
            %\hspace{-3mm}
          \fig{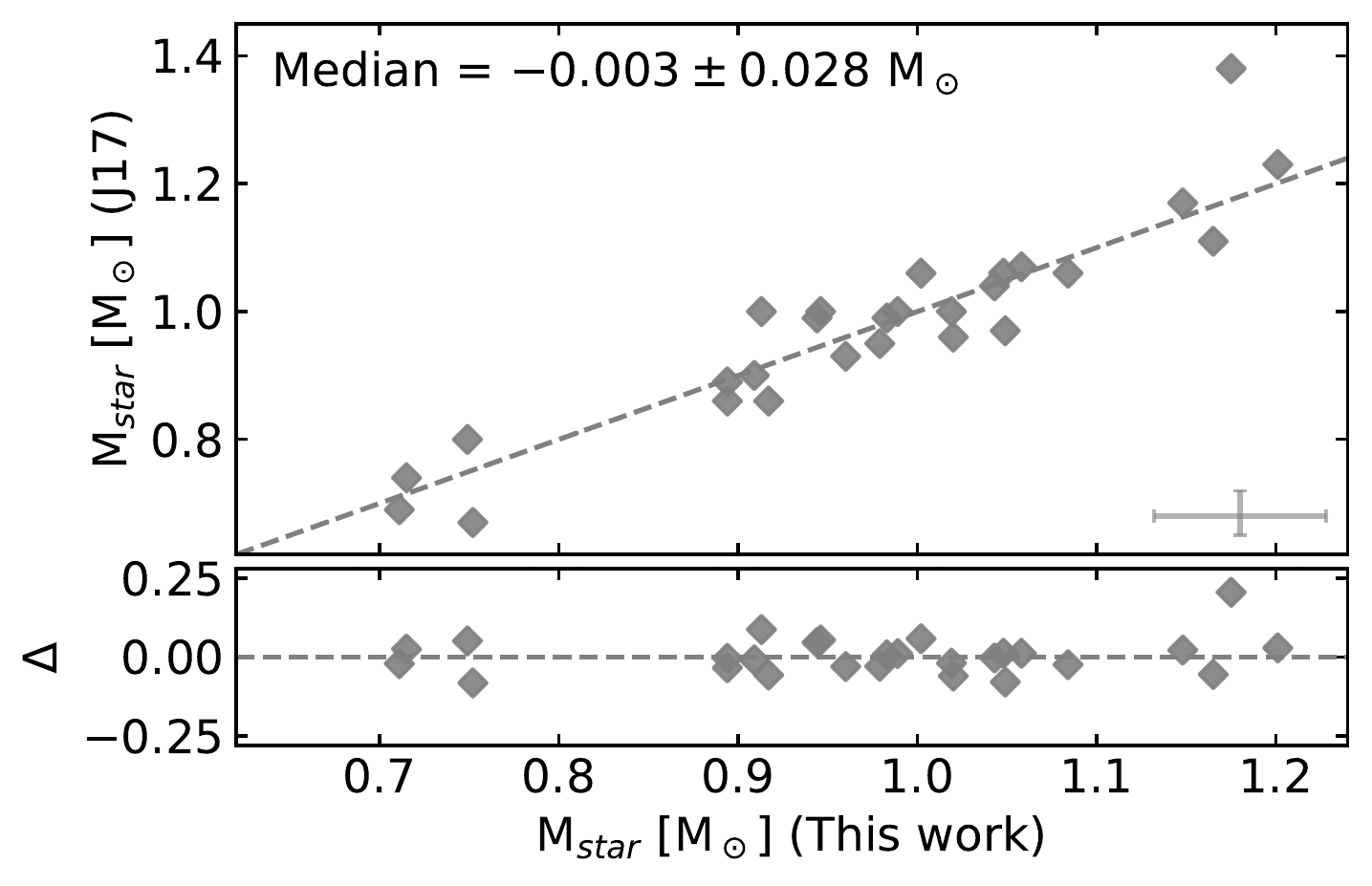}{0.525\textwidth}{}
          }
          \vspace{-10mm}
\gridline{\fig{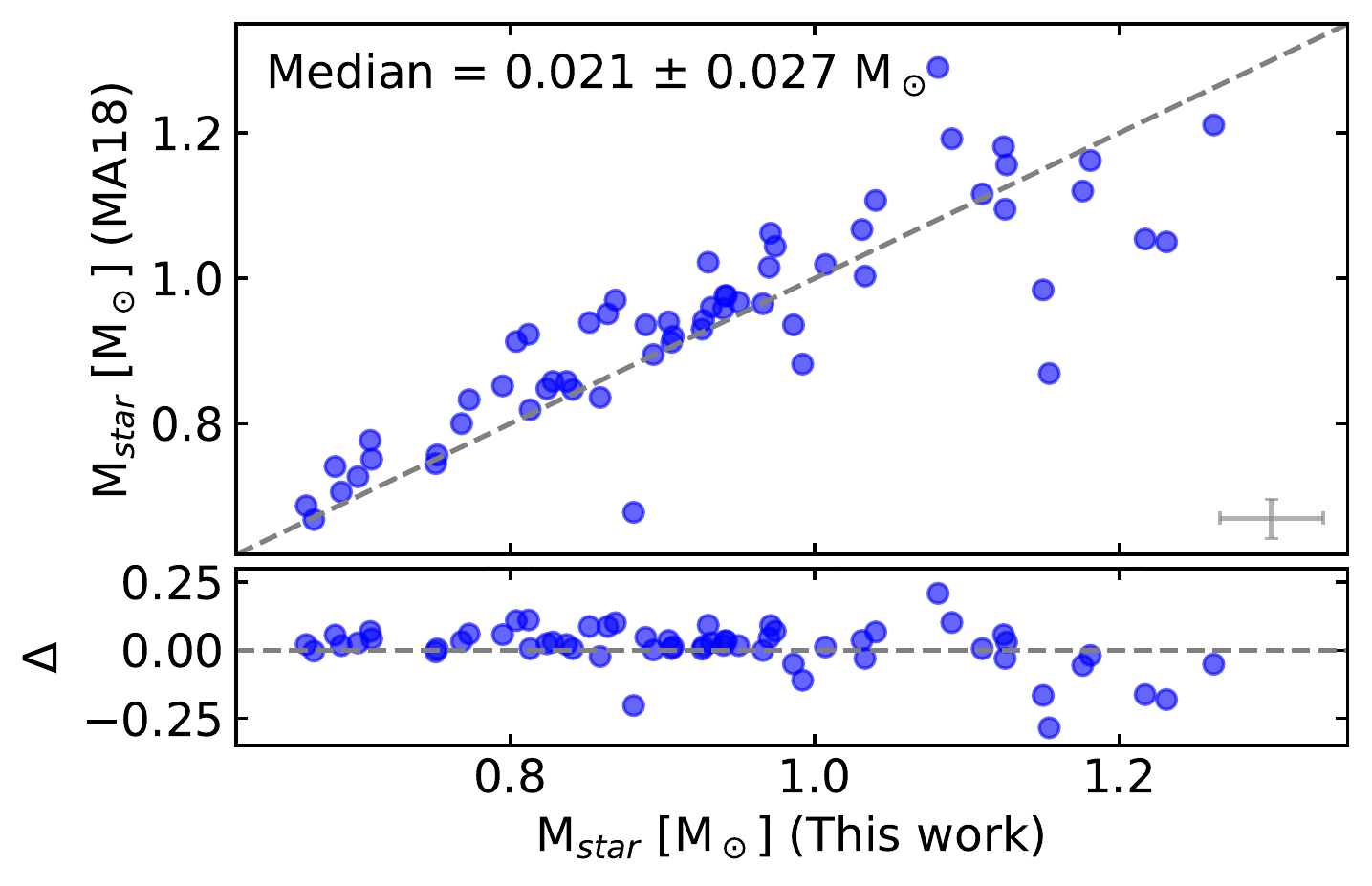}{0.51\textwidth}{}
            %\hspace{-3mm}
          \fig{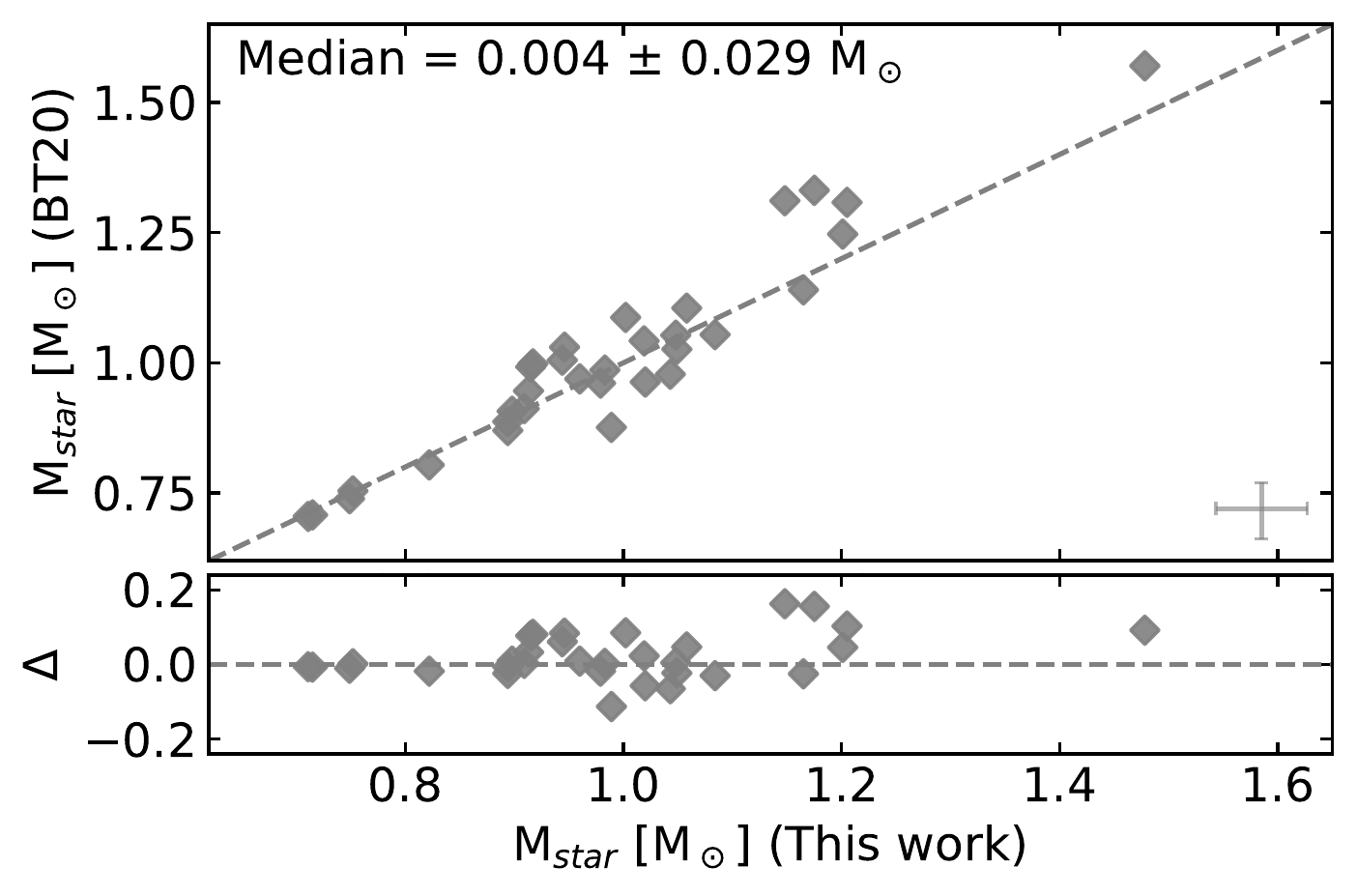}{0.5\textwidth}{}
          }
          \vspace{-10mm}
\gridline{\fig{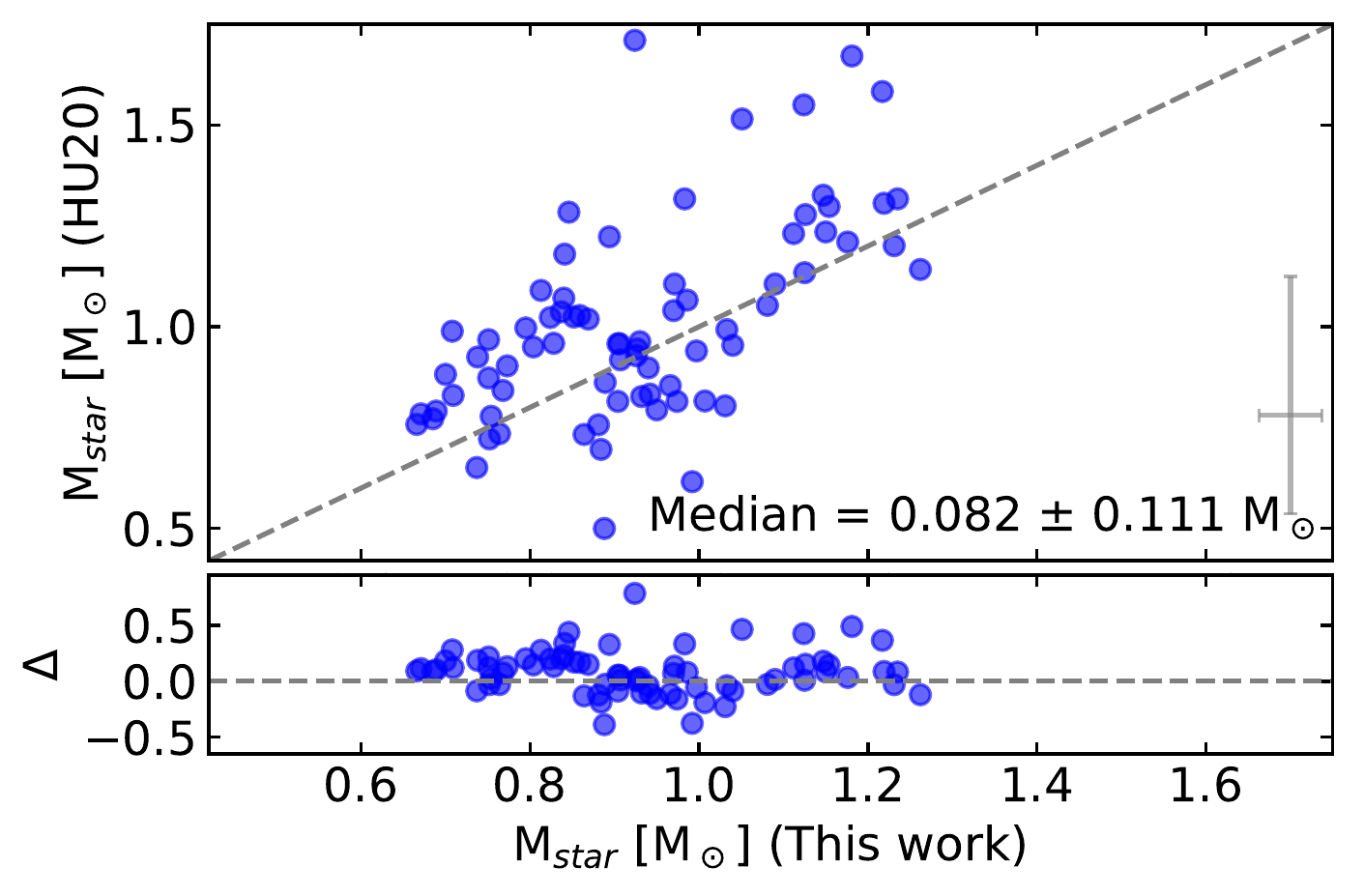}{0.5\textwidth}{}
            %\hspace{-3mm}
          \fig{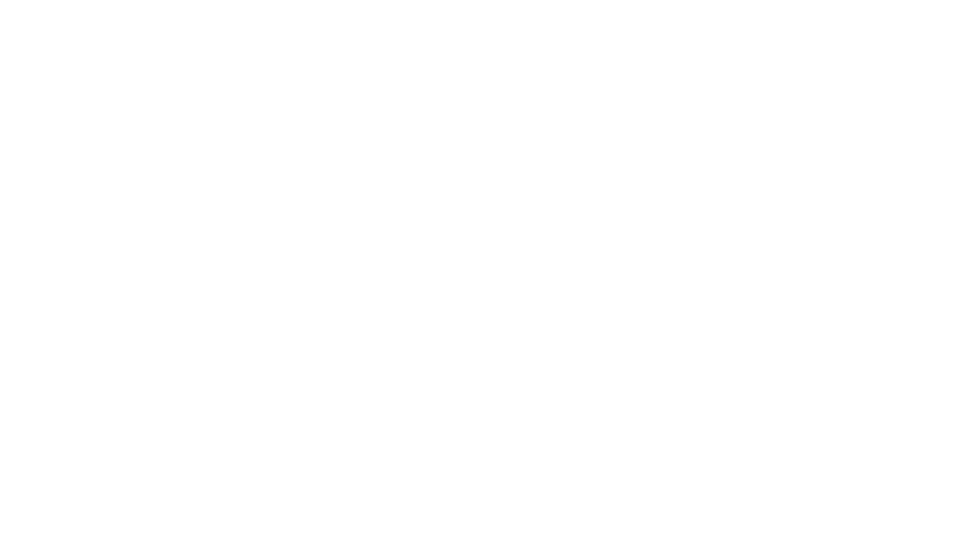}{0.5\textwidth}{}
          }
          \vspace{-8mm}
\caption{Comparisons between the derived stellar masses in this study with those from \citet[H16,][]{huber2016ApJS..224....2H}, \citet[MA18,][]{mayo2018} and \citet[HU20][]{Hardegree2020} for K2 stars (left panels), and \citet[J17,][]{johnson2017} and \citet[BT20,][]{berger2020b} for Kepler 1 stars  (right panels).  The bottom sub-panels show the differences, $\Delta$, between `Other Work - This Work'.}
\label{fig:kep_k2_M}
\end{figure*}

Moving on to discuss the stellar masses derived in this study, comparisons with other results are shown in Figure \ref{fig:kep_k2_M}. 
The mass scale from \cite{mayo2018} has a median difference (and MAD) that are very small when compared to ours, indicating overall good agreement. We note, however, that for stars more massive than M$_{star}\sim$1.1 M$_\odot$, there is significantly more scatter, while for stars with M$<$1.1 M$_\odot$ we find a median mass difference of 0.026 $\pm$ 0.020 M$_\odot$.
The mass comparisons for the Kepler 1 stars (\cite{johnson2017} and \cite{berger2020b}, right panels of Figure \ref{fig:kep_k2_M}), indicate even smaller offsets, where, the median mass differences are less than 3\% and MADs are $\Delta$M$_{star}\sim $0.03 M$_{\odot}$, although there are also outliers in these comparisons. The median difference for stars with less than 1 M$_\odot$ are $-0.009 \pm 0.025$ M$_\odot$ and 0.003 $\pm$ 0.021 M$_\odot$ for \cite{johnson2017} and \cite{berger2020b}, respectively.
The mass-scale of \cite{Hardegree2020}, on the other hand, has a larger systematic offset when compared to ours, being overall more massive than our scale by $\sim$8.2\%; the latter comparison also shows some scatter with an RMS value of 0.22 and a MAD of 0.111 M$_\odot$, which is the largest in Figure \ref{fig:kep_k2_M}.

Asteroseismic masses are, in principle, the most accurate masses presented in Figure \ref{fig:kep_k2_M}, having the smallest expected uncertainties \citep{pinsonneault2018}. When compared to our results, there is a clear systematic difference between the mean mass in \cite{huber2016ApJS..224....2H} and ours, with our masses being smaller than the asteroseismic ones in the median by $\sim$3\%.
We note for example, the significant outlier that appears as much more massive (M$_{star}=$ 1.2 M$_\odot$) when compared to our mass value (M$_{star}\sim$0.7 M$_\odot$) and again refer to the discussion in Section \ref{sec:astero}. As in the comparison with \cite{mayo2018}, there is also more scatter for stars with masses larger than 1.1 M$_\odot$.

%----------------------------------------
%----------------------------------------

%---------------------------------------------------
\begin{deluxetable*}{llccc}
\tablecaption{Planetary Radii\label{tab:Rpl}}
\tablecolumns{11}
\tablenum{6}
\tablewidth{0pt}
\tablehead{
\colhead{Star ID} & \colhead{Planet Name} & $\Delta F$ &   \colhead{$R_{pl}$} & \colhead{$\delta R_{pl}$} \\
\colhead{} & \colhead{}  &  (ppm) & \colhead{($R_\oplus$)} & \colhead{($R_\oplus$)}
}
\startdata
EPIC201577035 & K2-10 b          & 1691.0 & 4.09 & 0.14 \\
EPIC211990866 & K2-100 b         & 122.5  & 1.36 & 0.04 \\
EPIC211913977 & K2-101 b         & 749.0  & 2.17 & 0.08 \\
EPIC211970147 & K2-102 b         & 200.0  & 1.09 & 0.09 \\
EPIC211525389 & K2-105 b         & 1338.0 & 3.86 & 0.14 \\
EPIC220674823 & EPIC220674823 b & 284.7  & 1.78 & 0.06 \\
EPIC220674823 & EPIC220674823 c & 903.0  & 3.17 & 0.10 \\
EPIC211736671 & K2-108 b         & 973.0  & 5.78 & 0.45 \\
EPIC201596316 & K2-11 b          & 916.0  & 2.72 & 0.11 \\
\nodata  & \nodata  & \nodata  & \nodata     \\    
\enddata
\tablecomments{Transit depth ($\Delta F$) collected from \cite{kruse2019} and where available for K2 planets, while for Kepler 1 planets from \cite{thompson2018}. This table is published in its entirety in the machine readable format. A portion is shown here for guidance regarding its form and content.}
\end{deluxetable*}
%---------------------------------------------------

\subsection{Planetary Radii}

\begin{figure*}[!]
\gridline{\fig{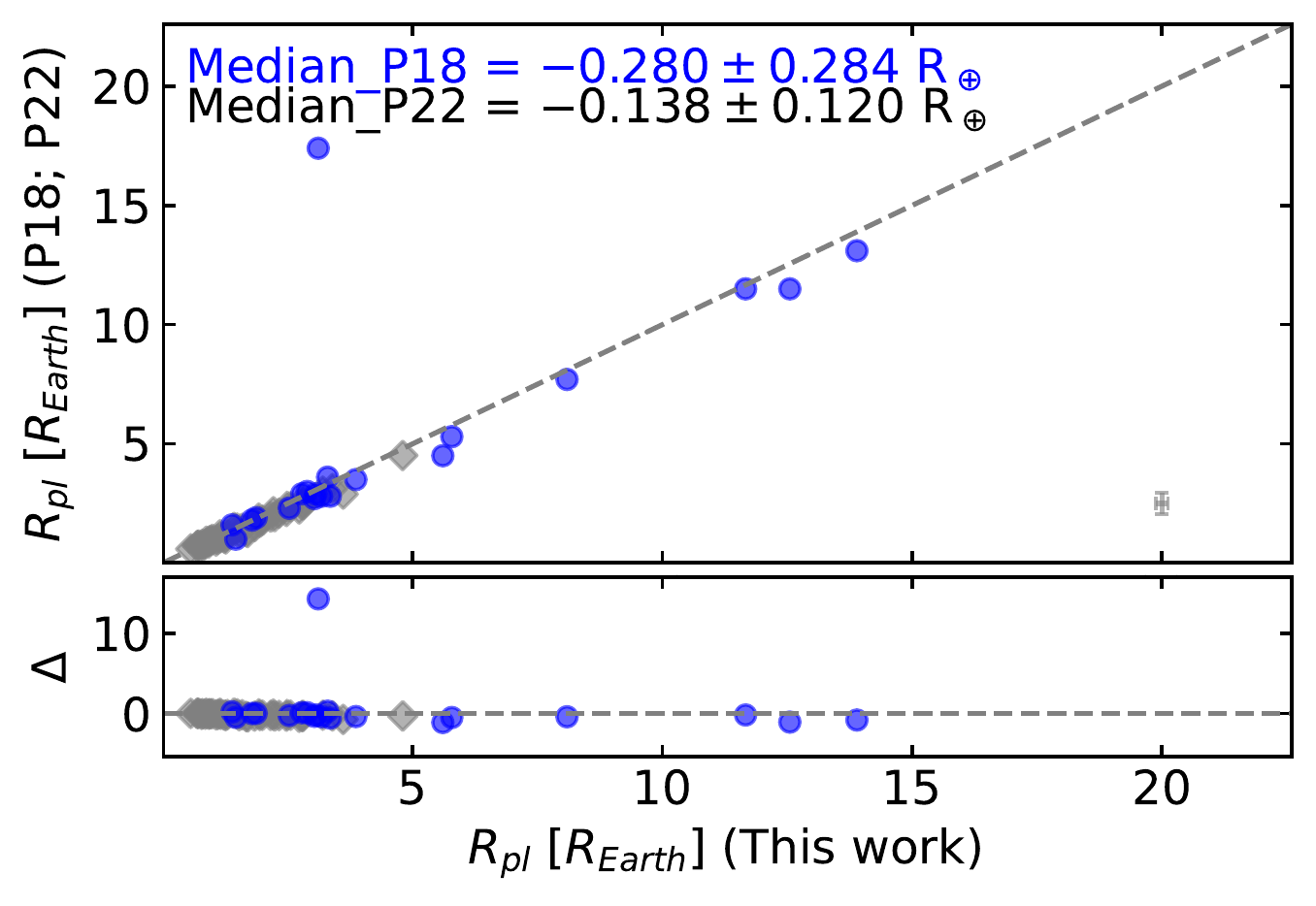}{0.45\textwidth}{}
            \hspace{-3mm}
          \fig{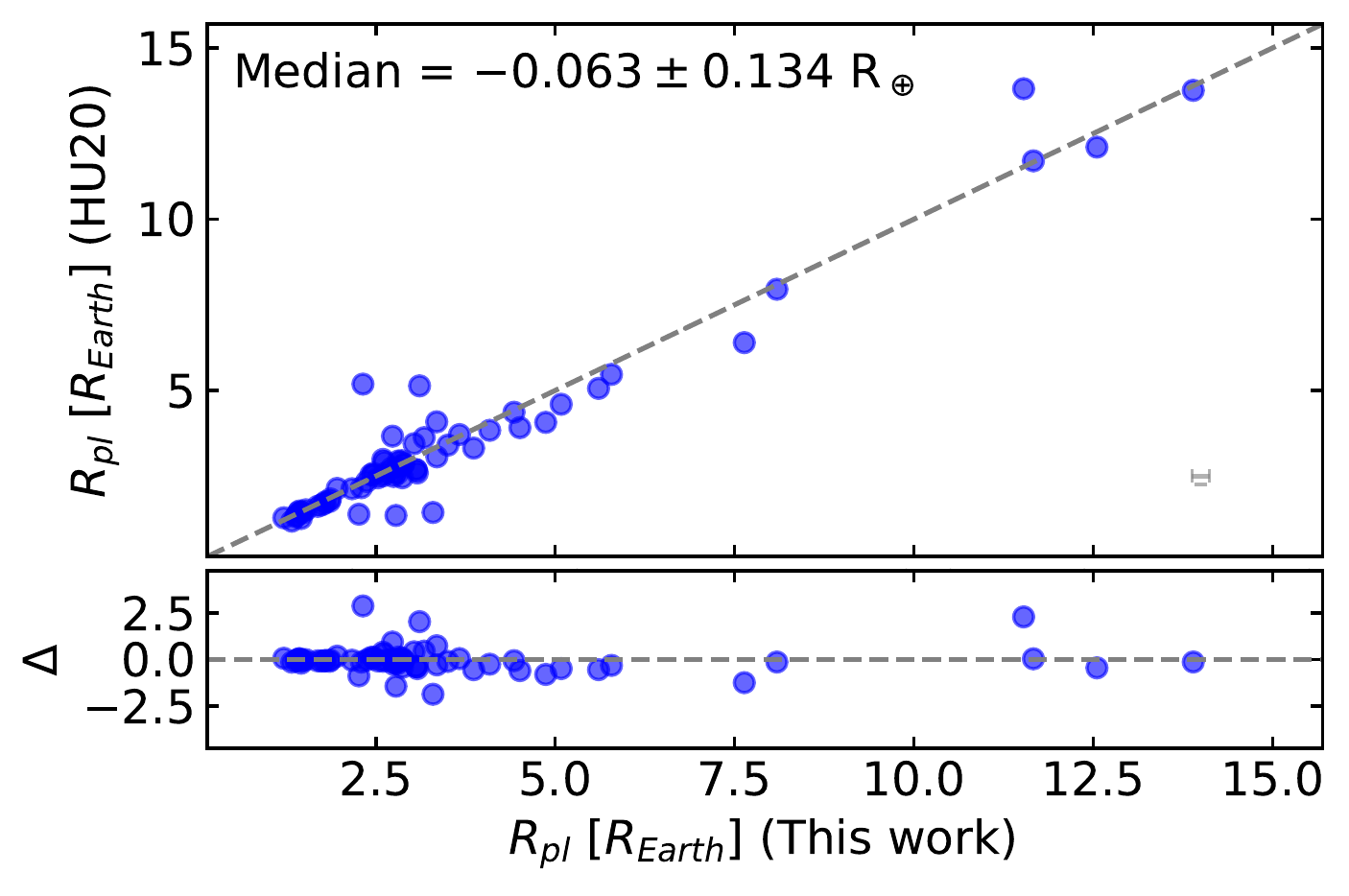}{0.47\textwidth}{}
          }
          \vspace{-11mm}
\gridline{\fig{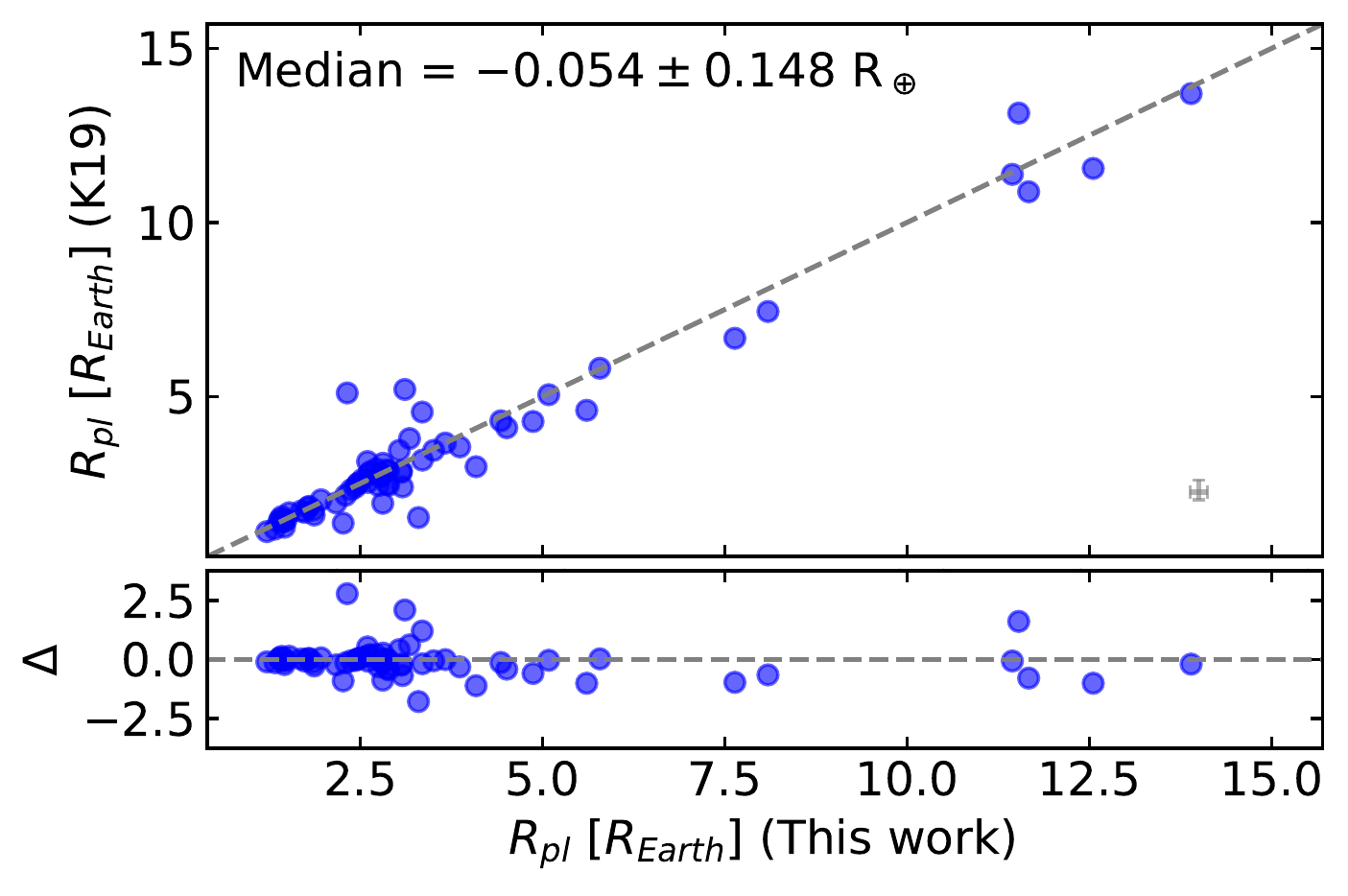}{0.45\textwidth}{}
            \hspace{-3mm}
          \fig{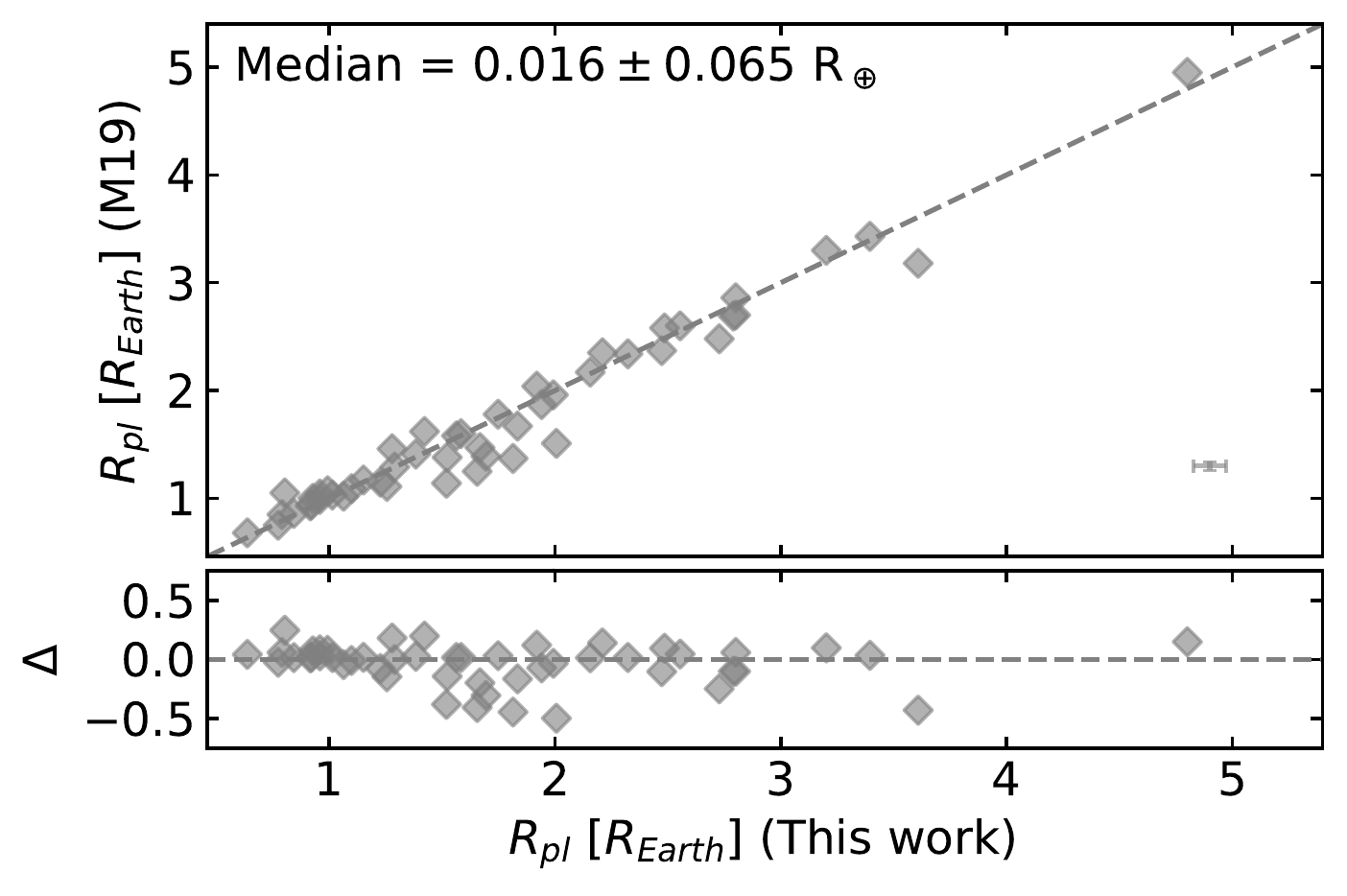}{0.45\textwidth}{}
          }
          \vspace{-12mm}
\gridline{\fig{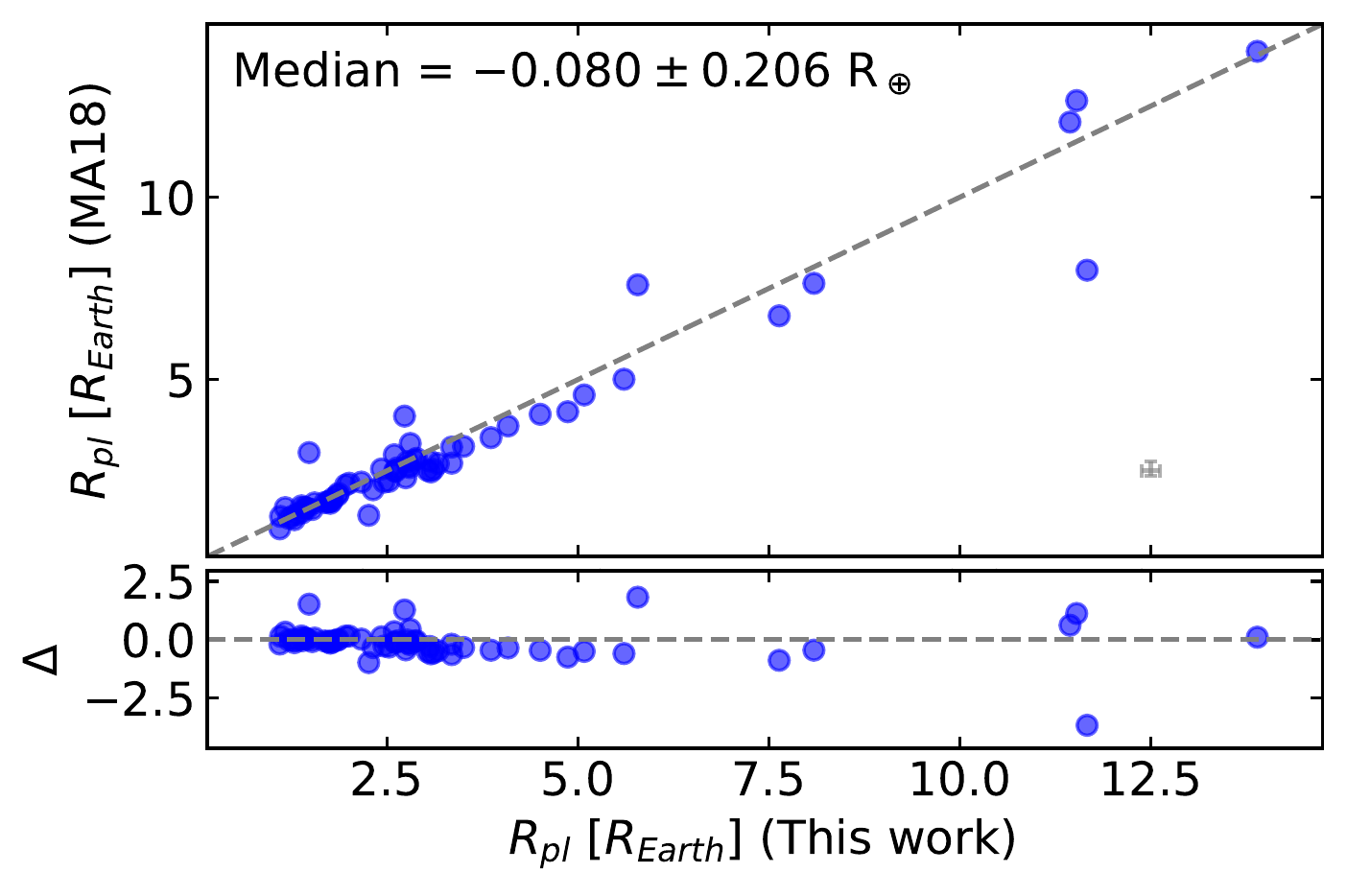}{0.45\textwidth}{}
            \hspace{-3mm}
          \fig{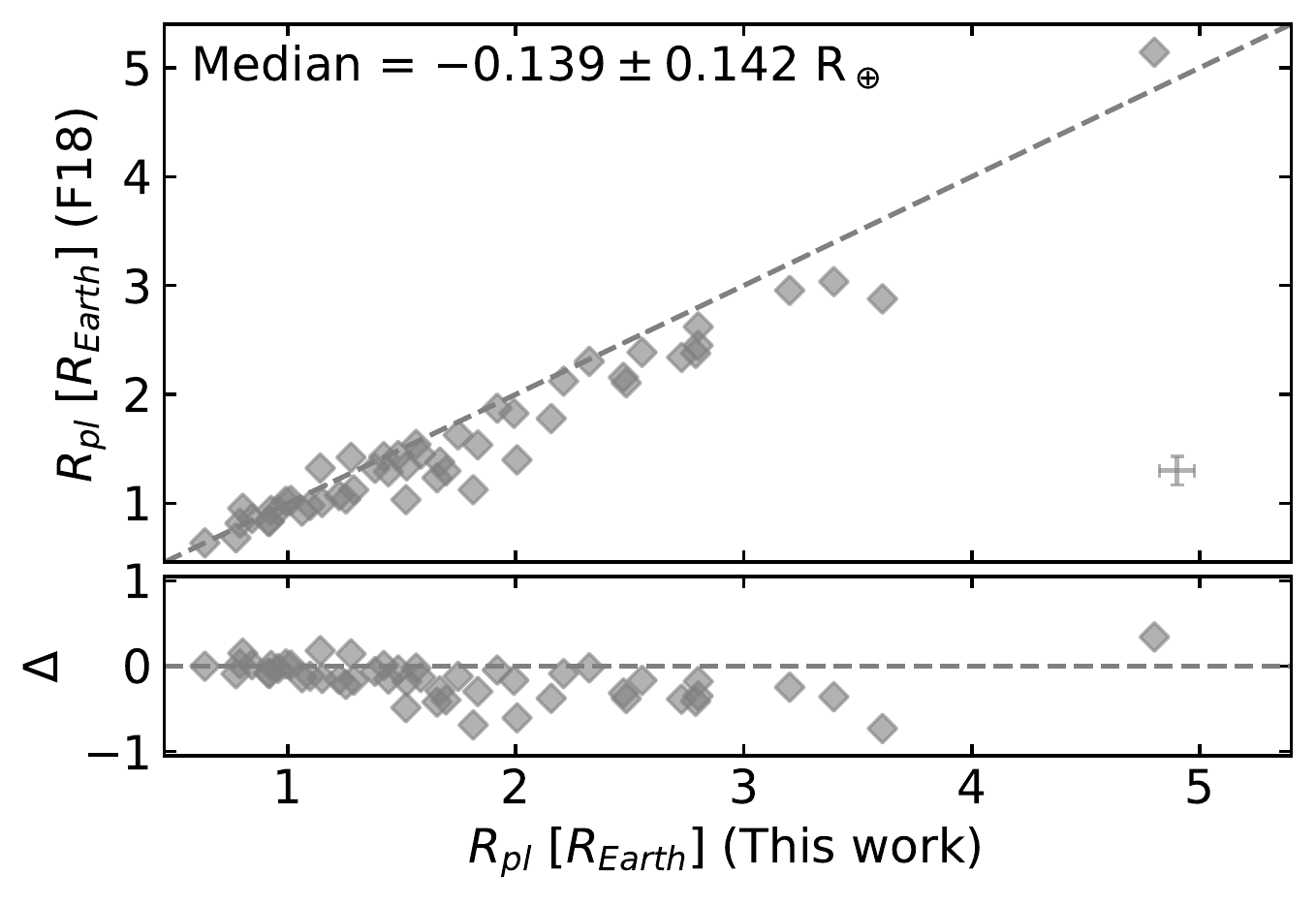}{0.43\textwidth}{}
          }
          \vspace{-12mm}
\gridline{\fig{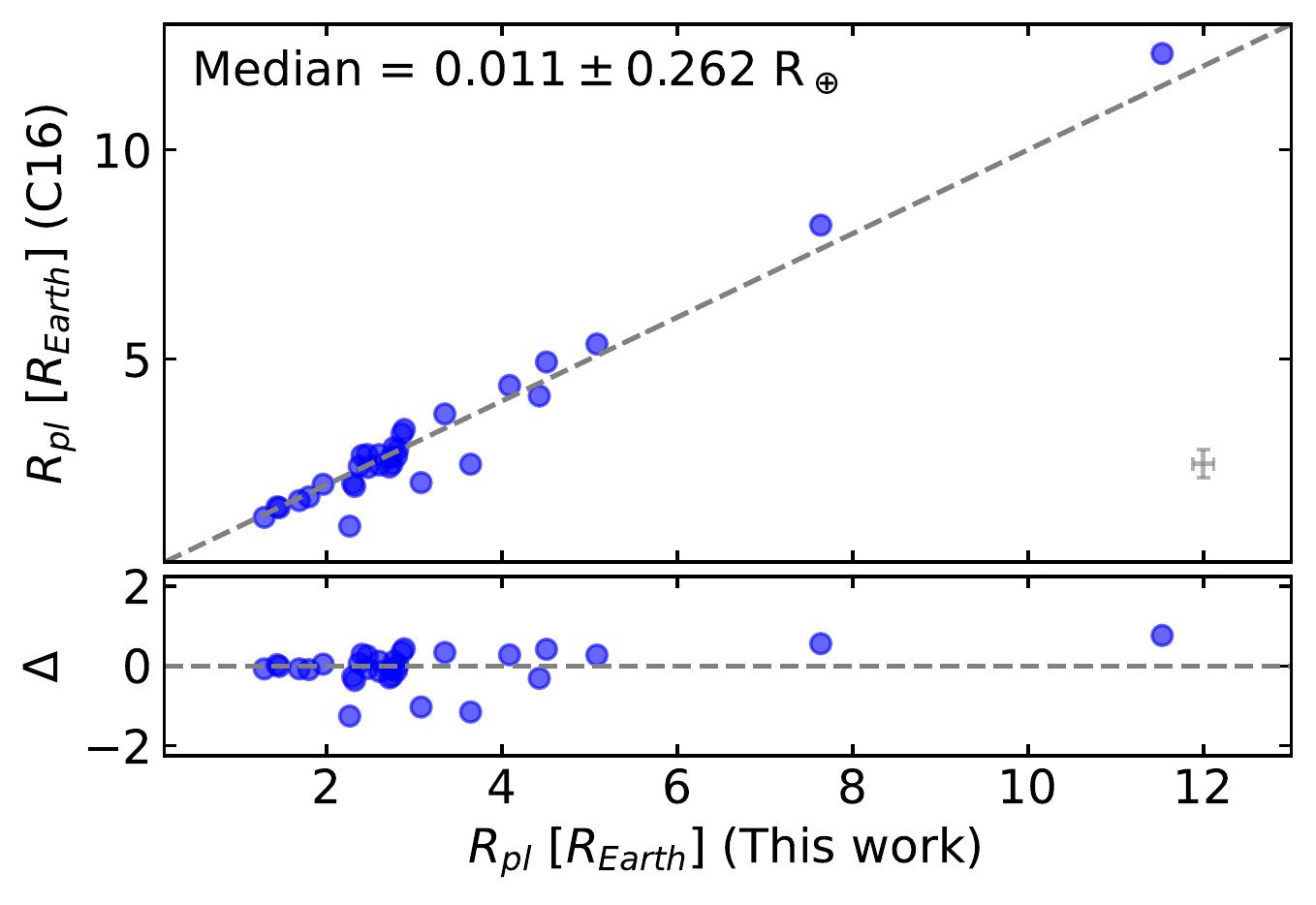}{0.45\textwidth}{}
            \hspace{-3mm}
          \fig{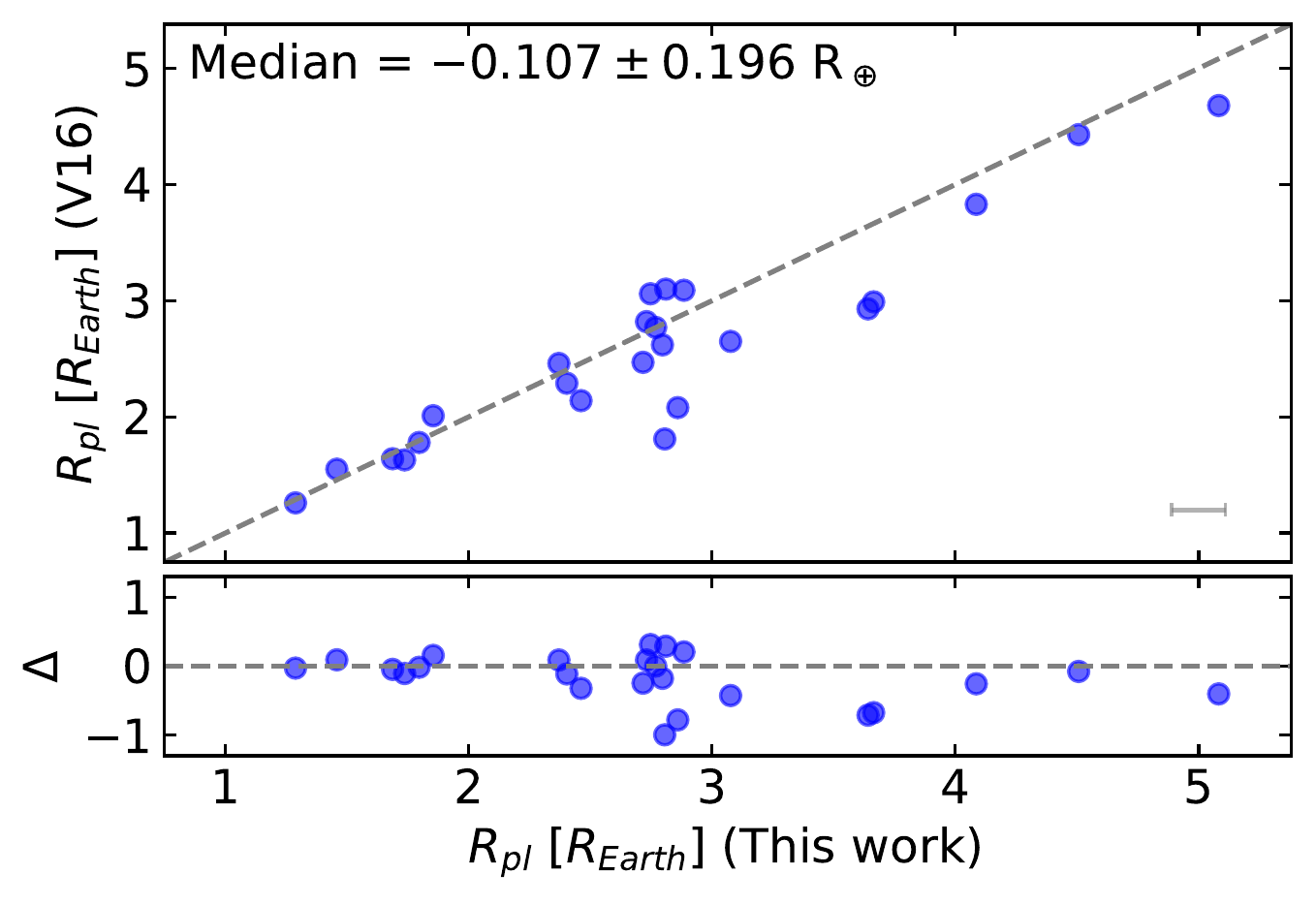}{0.45\textwidth}{}
          }
          \vspace{-7mm}
\caption{Comparison of the derived planetary radii with the literature values. Blue circles represent the planets of K2 host stars and grey diamonds of Kepler 1 host stars; beginning in the top left panel and preceeding clockwise, the comparisons are with \citet[P18,][P22]{petigura2018,petigura2022}, \citet[HU20,][]{Hardegree2020}, \citet[K19,][]{kruse2019}, \citet[M19,][]{martinez2019}, \citet[MA18,][]{mayo2018}, \citet[F18,][]{fulton2018}, \citet[C16,][]{crossfield2016}, and \citet[V16,][]{vanderburg2016}.  The bottom sub-panels present the differences, $\Delta$, of `Other Work - This Work'.}
\label{fig:com_Rpl}
\end{figure*}

The radii of exoplanets orbiting Kepler 1 and K2 host stars have been derived in several studies in the literature and these are compared with our results in the different panels of Figure \ref{fig:com_Rpl}. Here, planetary radii were computed using Equation (\ref{eq:Rpl}) that combines our derived stellar radii with transit depth values available in the literature (Section \ref{sec:radii}), and these are presented in Table \ref{tab:Rpl}, which contains the star identification, the planet name, transit depth, and planetary radii along with the error in R$_{pl}$.

For K2, we compare planetary radii with those in \cite{petigura2018}, \cite{Hardegree2020}, \cite{kruse2019}, \cite{mayo2018}, \cite{vanderburg2016}, and \cite{crossfield2016}. Overall, the results in Figure \ref{fig:com_Rpl} indicate that our planetary radii for K2 hosts are, on the median, larger than in the other K2 works, except for \cite{crossfield2016}; however, the median differences in the planetary radii (``Other Work - This Work") are small, with median systematic differences ((``Other Work - This Work")/``This Work'') of: 
$0.2 \%$ for \cite{crossfield2016}, $\sim$ 2\% for \cite{kruse2019}, $\sim$ 3\% for \cite{Hardegree2020} and \cite{vanderburg2016}, and $\sim$ 4\% for \cite{mayo2018}.
Although the median differences are, in some cases insignificant (at the level of 0--2\%, or small (at the level of 3--4\%), there are some outliers with larger discrepancies in some regimes, as is the case of the comparison with  \cite{vanderburg2016} that reveals larger offsets, in particular for planets with radii larger than $\sim$2.5R$_{\oplus}$. The comparison with \cite{petigura2018} finds larger median differences with our derived planetary radii of $\sim6$\% larger. 

In this comparison (top left panel), there is one significant outlier in the small planet regime, planet K2-183d, for which we find R$_{pl}=3.1$ R$_{\oplus}$, while \cite{petigura2018} find R$_{pl}=17.4$ R$_{\oplus}$. Our result is better agreement  with the planet radius reported in \citet[][R$_{pl}=2.9$R$_\oplus$]{livingston2018} and \citet[][R$_{pl}=2.5$ R$_\oplus$]{mayo2018}, as well as  in \citet[][R$_{pl}=5.1$ R$_\oplus$]{Hardegree2020}, in \citet[][R$_{pl}=5.2$ R$_\oplus$]{kruse2019}.
We also note that the difference in the effective temperature between \cite{petigura2018} and our result is $\Delta T_{\rm eff} = -81$ K and that if we used \cite{petigura2018} R$_{pl}$/R$_{star}$ value we would obtain R$_{pl}$ = 17.56 R$_\oplus$, which is in much closer agreement with \cite{petigura2018}, indicating that the difference in transit depth is responsible for a large part of the difference in R$_{pl}$.

For the Kepler 1 planets studied here, the radii in \cite{petigura2022} and \cite{fulton2018} show the largest offsets relative to our results, with a median systematic difference of 8--9\%, again with our planetary radius-scale being larger. 
For \cite{martinez2019}, there is a much smaller systematic shift in the planetary radii relative to ours of $\sim$1\%, and also having the smallest MAD of all comparisons, indicating that the scales for planetary radii are very consistent in both studies.

One aspect to keep in mind is the fact that the various studies discussed here may have employed different transit depths. Some studies, such as ours, use literature values, while others derive their own. For planets in common, we should note that we used transit depths from \cite{kruse2019} and these were also used by \cite{Hardegree2020}, while \cite{vanderburg2016}, \cite{crossfield2016}, \cite{mayo2018}, and \cite{petigura2018} derived their own transit depths. 
For the Kepler 1 planets, we used transit depths from \cite{thompson2018}, which are also used in \cite{martinez2019} and \cite{petigura2022}, while transit depths in \cite{fulton2018} come from \cite{mullally2015}.

%%%%%%%%%%%%%%%%%%%%%%%%%%%%%%%%%%%%%%%%%%%%%%%%%%%%%%%%%%
%%%%%%%%%%%%%%%%%%%%%%%%%%%%%%%%%%%%%%%%%%%%%%%%%%%%%%%%%%

%===========================================
\section{Discussion} \label{sec:discus}
%===========================================

\begin{figure*}[!]
\epsscale{0.8}
\plotone{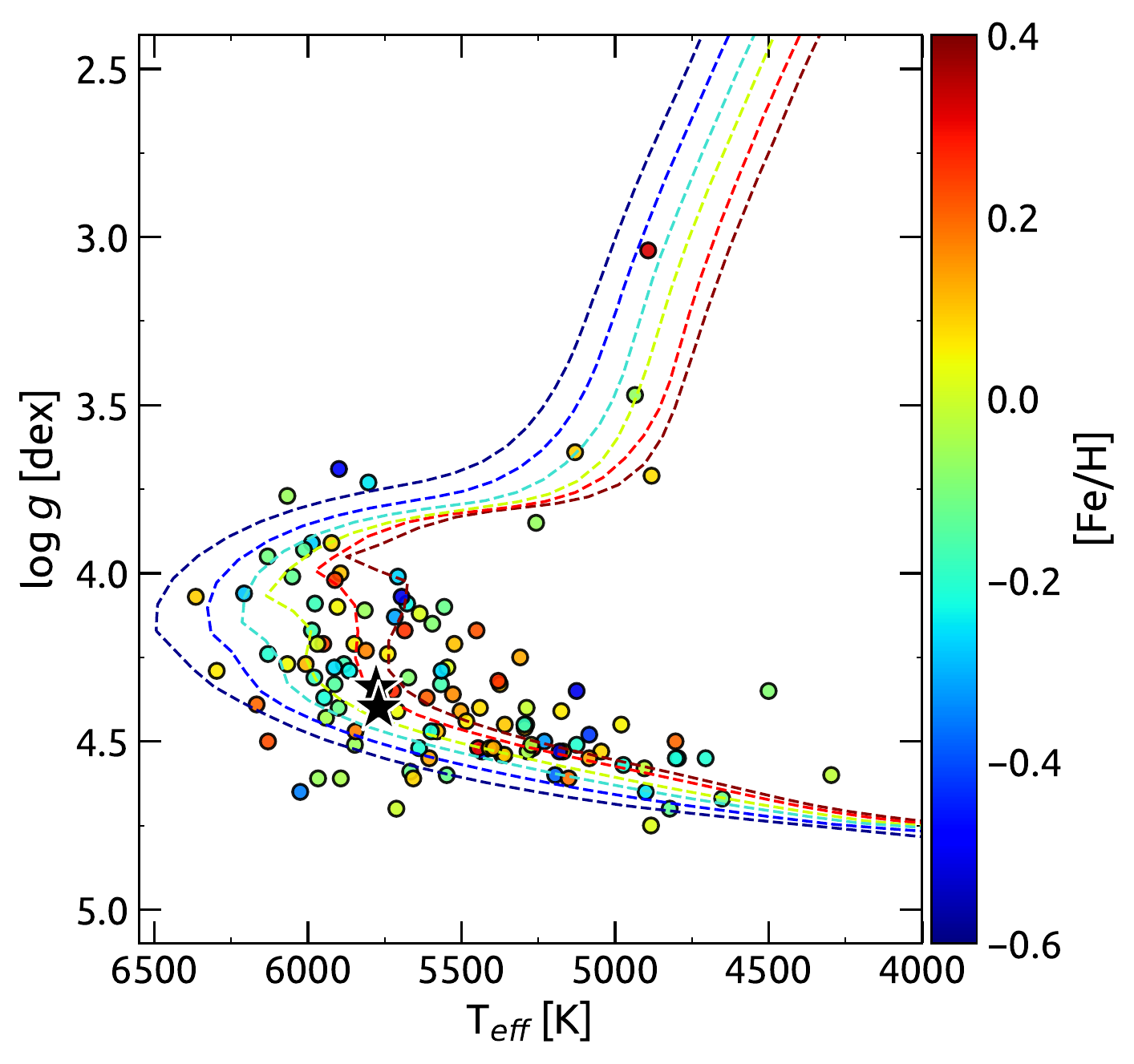}
\caption{Kiel diagram for the sample stars. Dashed lines represent 4.6 Gyr old stellar evolution tracks from  Yonsei-Yale isochrones} for different metallicities: $-0.6$, $-0.4$, $-0.2$, $0.0$, $+0.2$, and $+0.4$ dex.  The colors of circles represent different metallicities as shown by the color bar.
\label{fig:kiel_sample}
\end{figure*}

The stellar parameters obtained for 115 stars analyzed in this study, along with those for the solar proxies Astraea and Parthenope, are summarized in Figure \ref{fig:kiel_sample} as a Kiel diagram, with log $g$ plotted as a function of T$_{\rm eff}$ and the stars shown as filled circles color-coded by their iron abundance; the color bar represents the metallicities. Results for the solar proxies are shown as black stars. 
The dashed lines are the Yonsei-Yale isochrones \citep{yi2001,yi2003ApJS..144..259Y,demarque2004ApJS..155..667D,han2009gcgg.book...33H} corresponding to an age of 4.6 Gyr and metallicities between -0.6 and +0.2 dex, with steps of 0.2 dex. As can be seen from the location of the points in this diagram, most of the stars in our sample are on, or near, the main-sequence, but there are also a small number of stars that are clearly evolved. 

%%%%%%%%%%%%%%%%%%%%%%%%%%%%%%%%%%%%%%%%%%%%%%%%%%%%%%%%%%
%%%%%%%%%%%%%%%%%%%%%%%%%%%%%%%%%%%%%%%%%%%%%%%%%%%%%%%%%%

\begin{figure*}[!]
\gridline{\fig{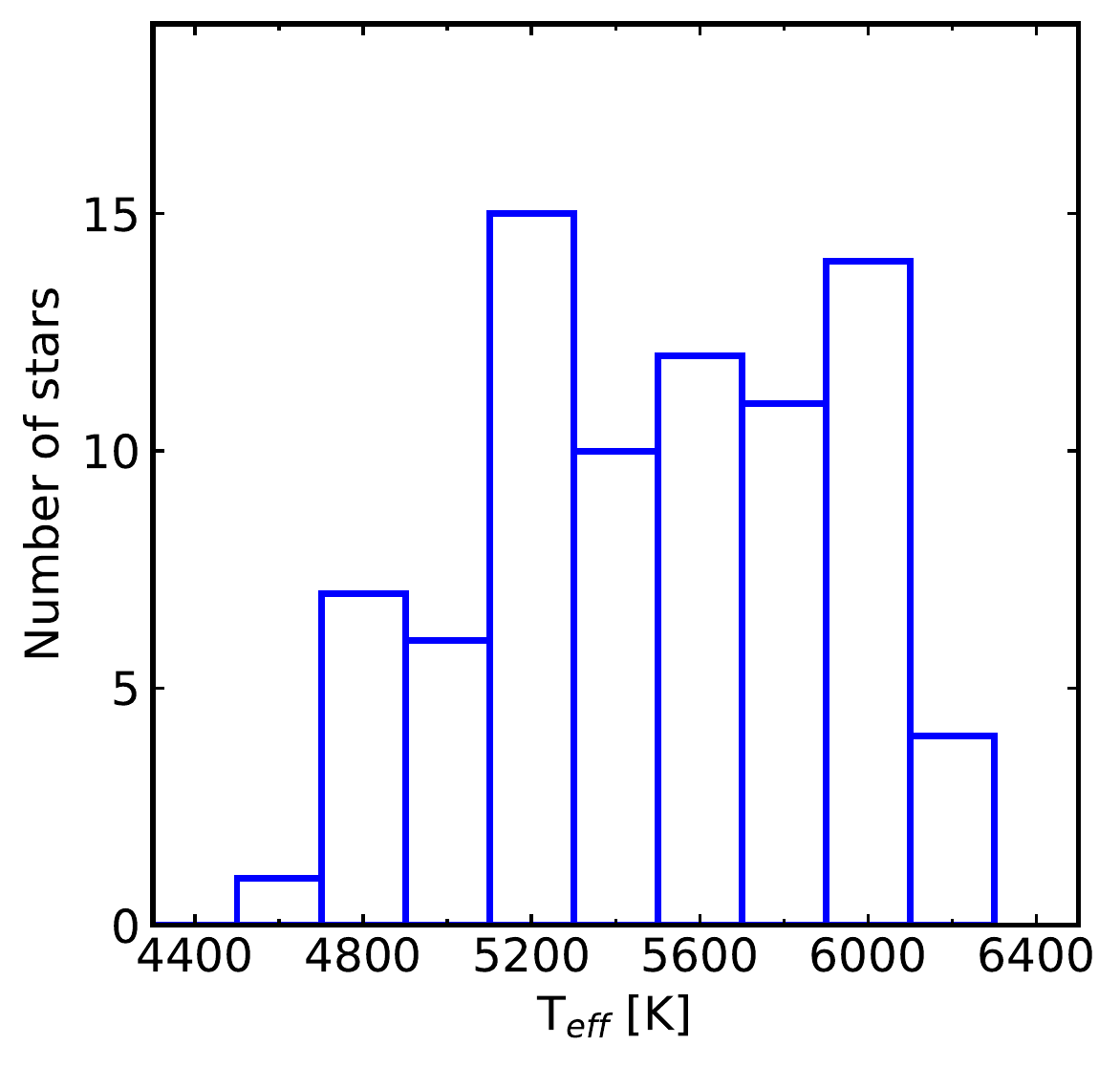}{0.45\textwidth}{}
            \hspace{-3mm}
          \fig{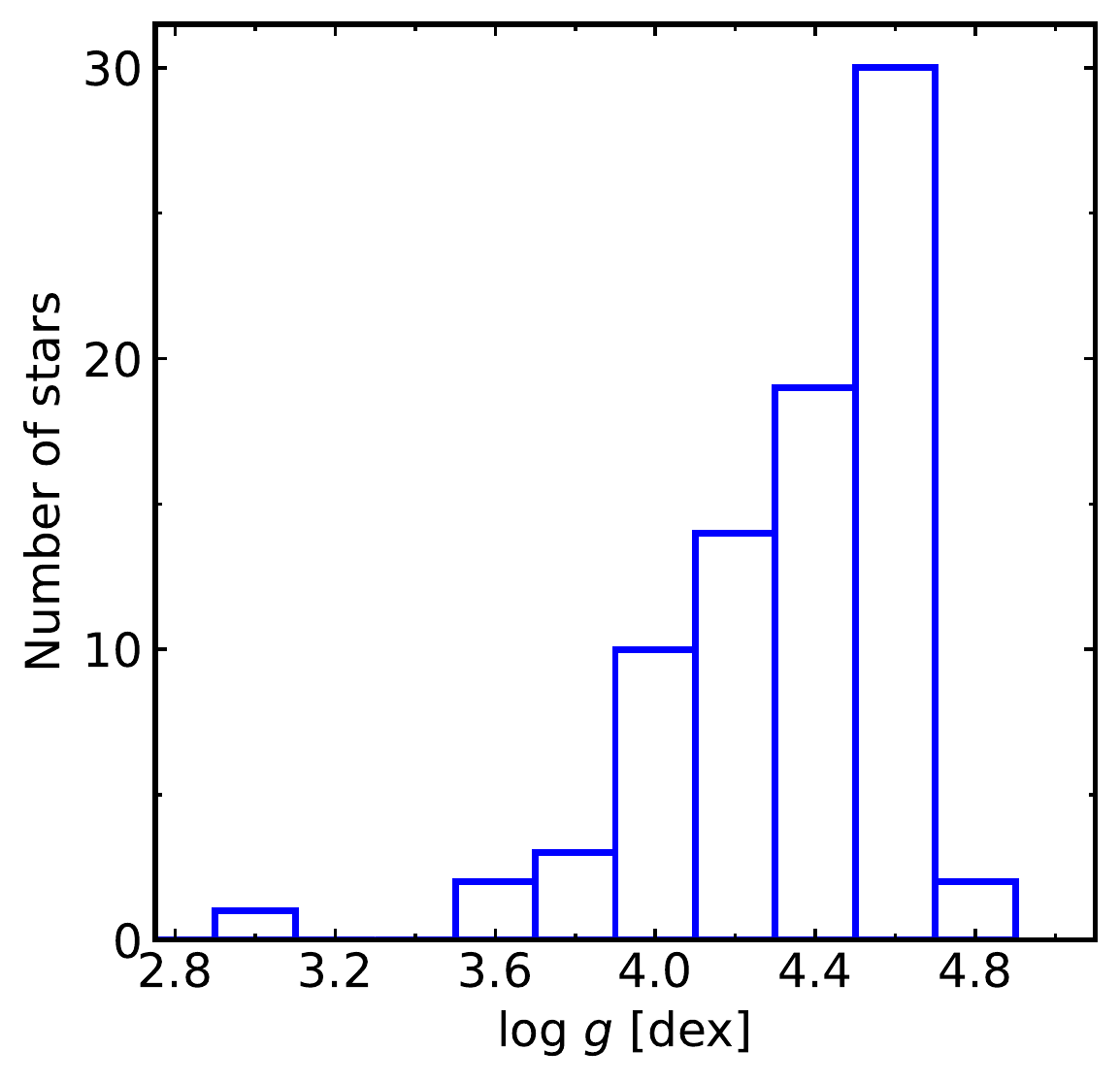}{0.45\textwidth}{}
          }
          \vspace{-7.5mm}
\gridline{\fig{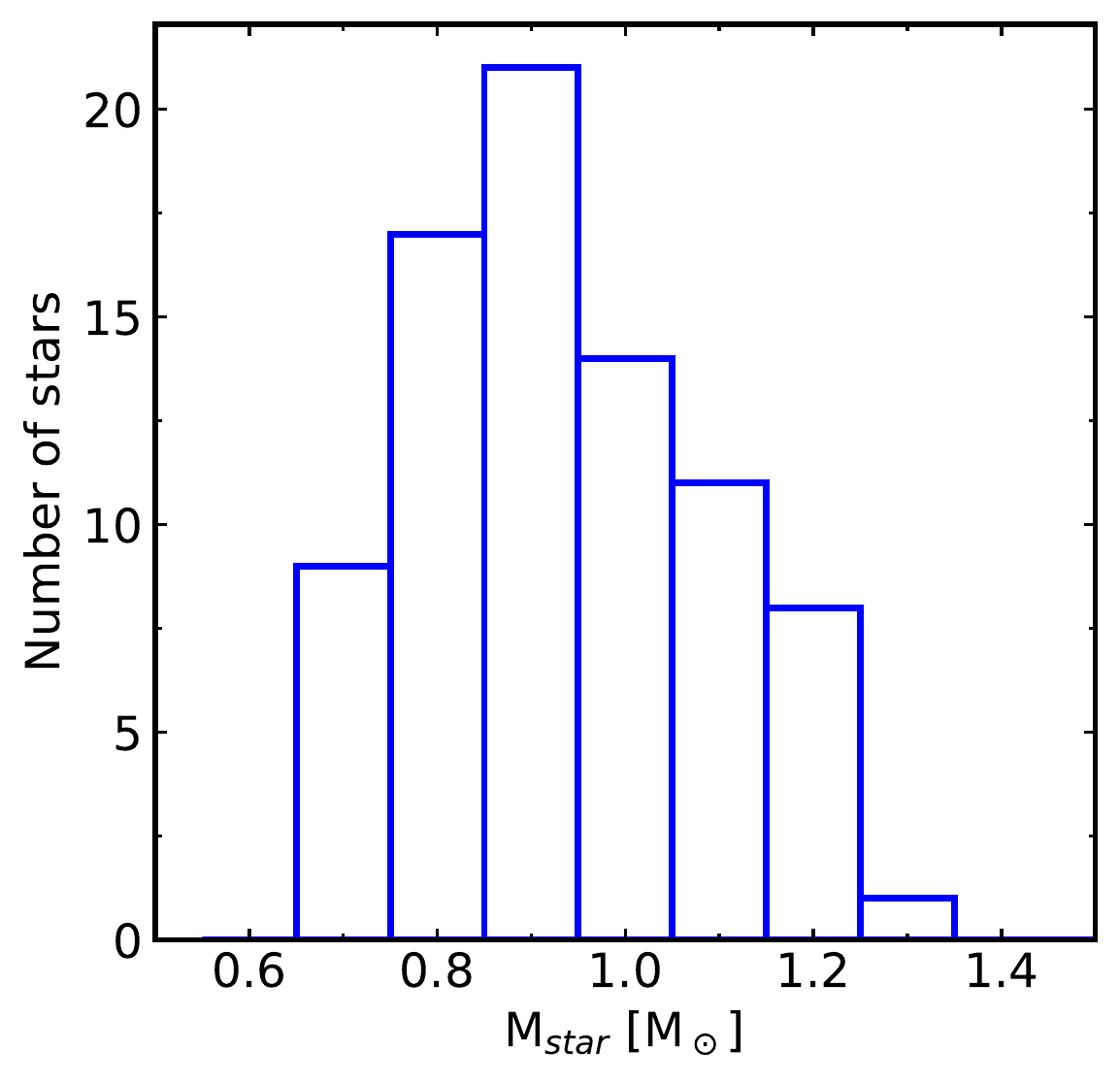}{0.45\textwidth}{}
            \hspace{-3mm}
          \fig{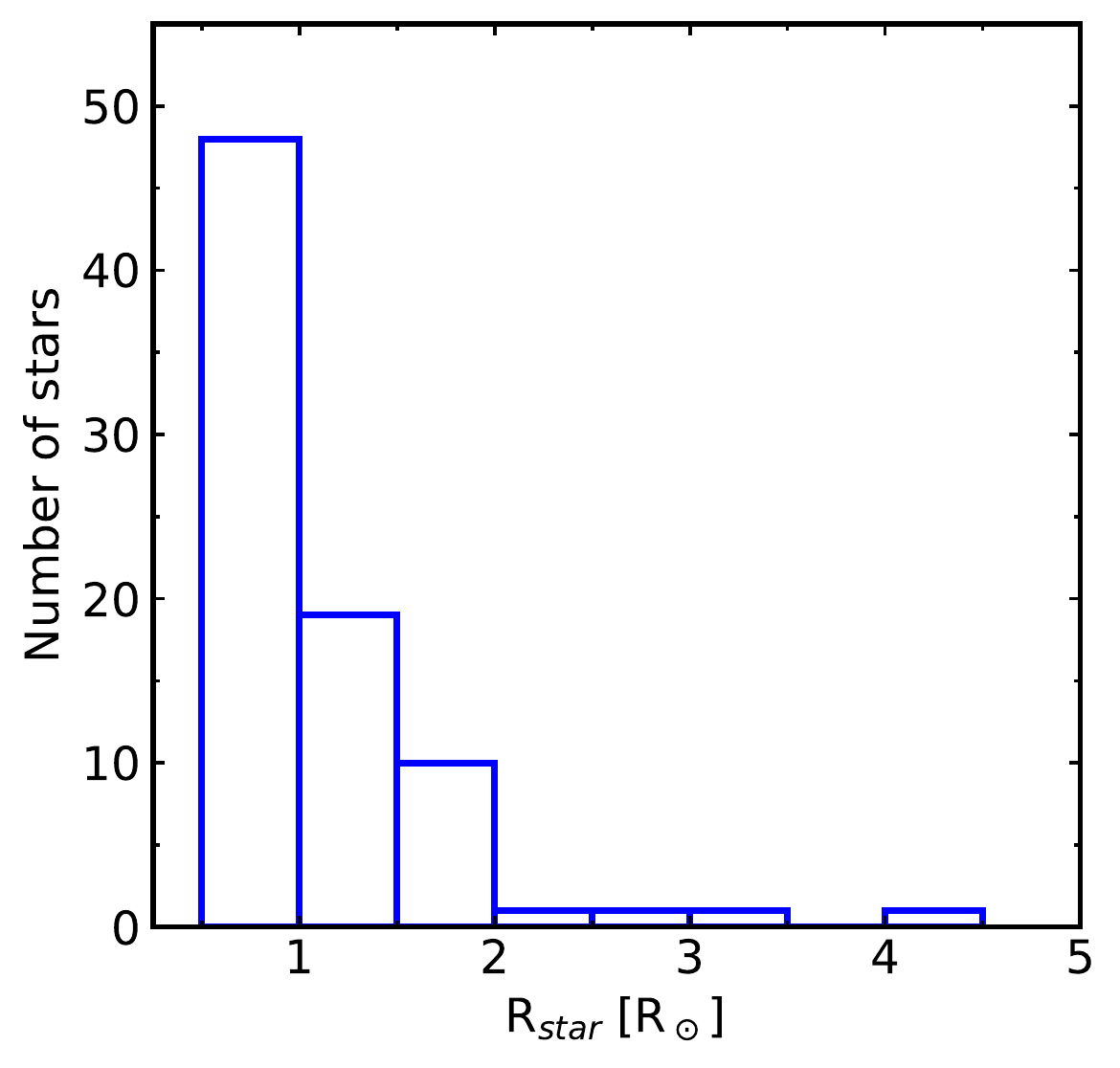}{0.45\textwidth}{}
          }
          \vspace{-9mm}
\caption{Effective temperature, surface gravity, mass, and radius distributions for the K2 stellar sample.}
\label{fig:hist_sp}
\end{figure*}

Focusing initially on our results for the K2 sample, 
Figure \ref{fig:hist_sp} uses histograms to illustrate the distributions of effective temperature and surface gravity in the top left and right panels, respectively.
Our target selection was biased, as it avoided M dwarfs, and most of the K2 stars in this sample have effective temperatures between $\sim$4800 -- 6000 K, with a peak at roughly T$_{\rm eff}\sim$5500 K (median = 5503 K; MAD = 346 K).
The log $g$ distribution of the K2 stars in this study is dominated by stars having values between log $g\sim$ 4.2 -- 4.6, with a median log $g$ = 4.41 (MAD = 0.14; 16th percentile =$-0.36$; 84th percentile =$+0.14$).

The bottom panels of Figure \ref{fig:hist_sp} show the distributions of the derived stellar masses (left bottom panel) and radii (right bottom panel) for the K2 targets. 
The median of the mass distribution of the K2 sample studied here is just below the solar value: M$_{star}=0.93 \pm 0.11$ M$_\odot$, with all K2 stars having masses between M$_{star}=0.6-1.3$ M$_\odot$.  This is a narrower mass range than that found for the 2956 Kepler hosts in \cite{berger2020b}, which covers M$_{star}$=0.5--1.7M$_{\odot}$, noting however that their median stellar mass (M$_{star}=0.99\pm 0.2$ M$_{\odot}$) is just slightly larger than our K2 sample. The K2 C5 sample in \cite{zink2020AJ....160...94Z} has a similar median mass of M$_{star}=0.92^{+0.34}_{-0.18}$ M$_{\odot}$, but which extends over a much larger range in mass, from $\sim$0.1 to 2.5 M$_\odot$.

The stellar radii distribution of the studied K2 sample (shown in the right bottom panel of Figure \ref{fig:hist_sp}), has a median radius of R$_{star}=0.94$ R$_\odot$ (16th percentile $=-0.22$; 84th percentile $=+0.59$), with very few stars in our sample having radii larger than R$_{star} > 2$ R$_\odot$.
An investigation of the mass-radius relation indicates that the 4 stars in our K2 sample having radii R$_{star}>2$ R$_\odot$ and log $g$ values smaller than 3.8 all have masses larger than 1.1 M$_{\odot}$, indicating that they have evolved away from the main sequence over realisitic timescales ($\sim7-8$ Gyr for 1.1 M$_{\odot}$). 

%%%%%%%%%%%%%%%%%%%%%%%%%%%%%%%%%%%%%%%%%%%%%%%%%%%%%%%%%%
%%%%%%%%%%%%%%%%%%%%%%%%%%%%%%%%%%%%%%%%%%%%%%%%%%%%%%%%%%%%%%%%%

\begin{figure*}[!]

\gridline{\fig{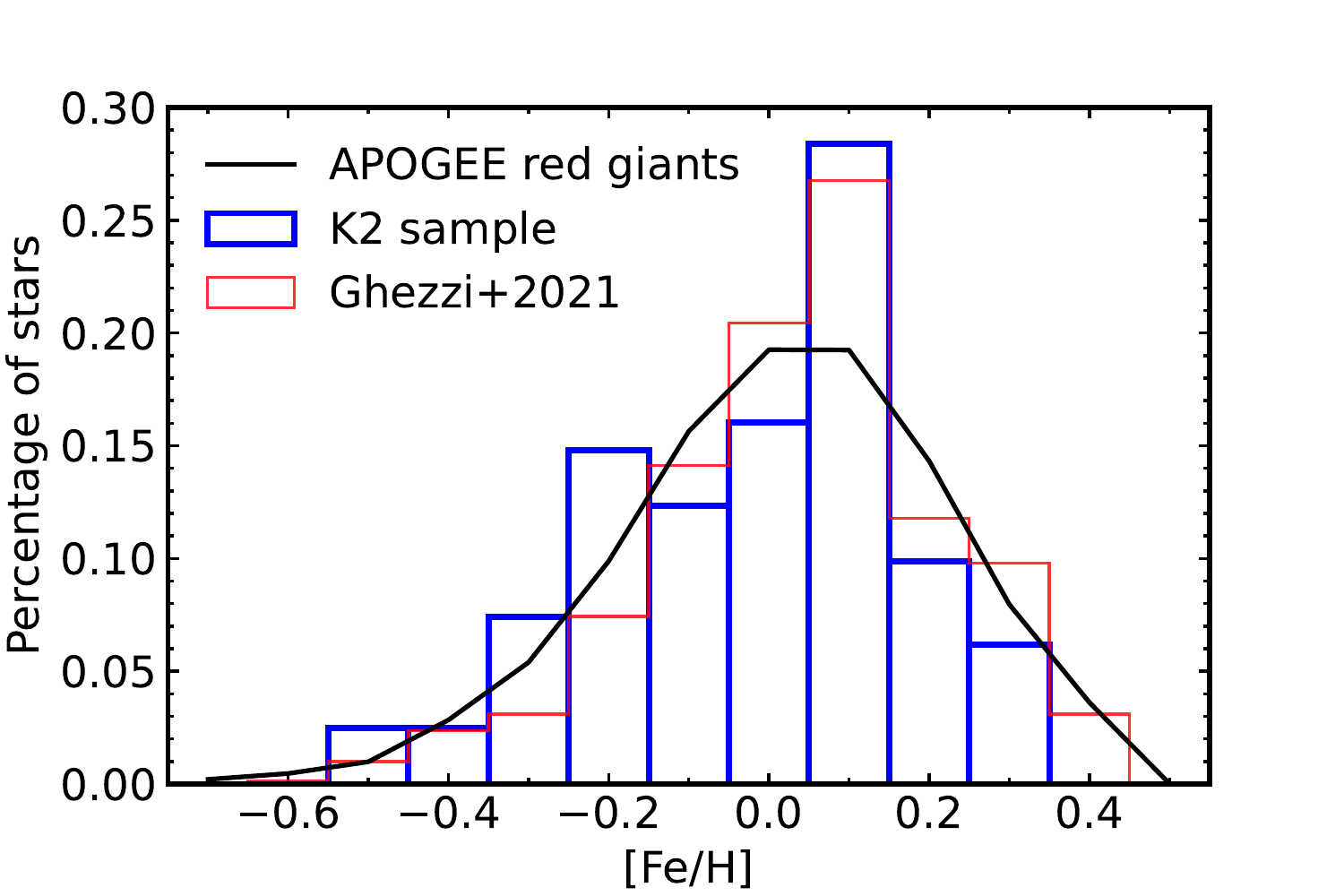}{0.55\textwidth}{}
          \hspace{-9mm}
          \fig{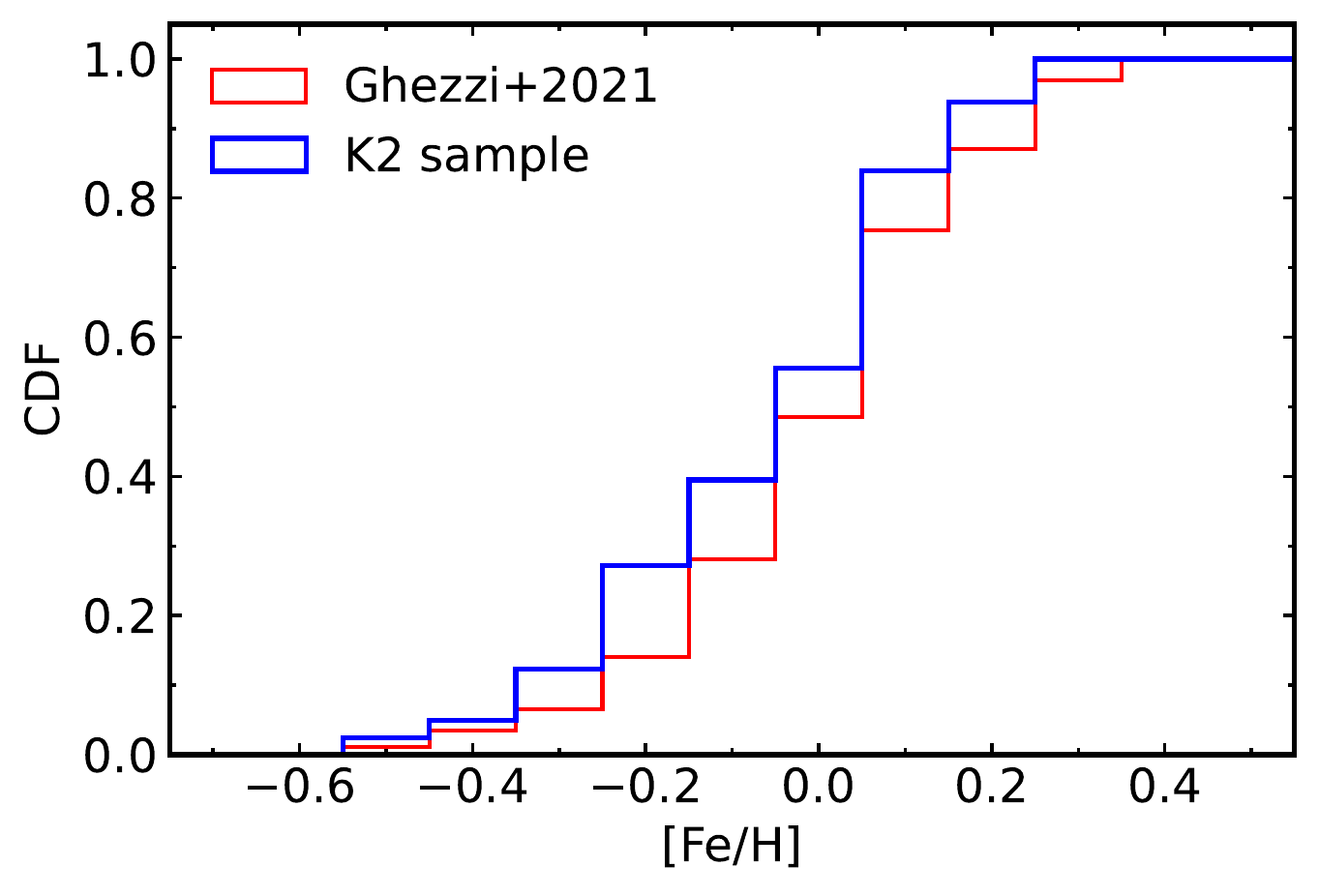}{0.49\textwidth}{}
          }
          \vspace{-7mm}
\caption{Iron abundance distributions (left panel) and cumulative distribution functions (right panel) for different samples of stars. The left panel presents the K2 sample studied here, shown as the blue histogram, and the CKS sample from \cite{Ghezzi2021} is shown in red, while the [Fe/H] distribution for the red giant stars in the local Milky Way disk from the APOGEE survey is shown as the black curve. The cumulative distribution functions for the K2 sample here and the CKS sample from \cite{Ghezzi2021} are compared in the right panel.  Both panels show that the K2 sample studied here is slightly more metal poor than both the CKS sample and the local disk.}
\label{fig:dits_FeH}
\end{figure*}

In Figure \ref{fig:dits_FeH} we show the metallicity distributions on the left panel, along with the cumulative distribution functions on the right panel for the K2 sample (shown in blue) and the CKS sample analyzed in \citet[][shown in red]{Ghezzi2021}. The metallicity distribution of our K2 sample covers the range between  [Fe/H]$=-0.5$ to +0.3 (iron abundances roughly between $7.00 <$ A(Fe) $< 7.80$), with $\sim$12 $\%$ of stars being more metal poor than [Fe/H]$= -0.2$. 
The median ($\pm$MAD) metallicity of the distribution is [Fe/H]=$-0.03 \pm 0.14$ dex (A(Fe)= $7.49 \pm 0.14$ dex) and this value is very close to the metallicity obtained here for the solar-proxy asteroids ($<$A(Fe)$>$=7.52), and also in good agreement with the solar Fe abundance from \cite{magg2022AA...661A.140M} (see Section \ref{sec:proxy}). Overall the range in metallicity of our K2 sample overlaps roughly with that of the Galactic thin disk, although the K2 sample is slightly more metal-poor. 
The comparison with the metallicity distribution of the CKS sample from \cite{Ghezzi2021} also indicates that our K2 sample is more metal poor, which is in line with the finding in \cite{Ghezzi2021} that the CKS metallicity distribution was akin to the metallicity distribution function (MDF) of the solar neighborhood based on stellar samples with Galactocentric distances between 7 kpc $<$ Rg $<$ 9 Kpc from APOGEE and GALAH surveys \citep{hayden2015ApJ...808..132H,hayden2020MNRAS.493.2952H}; 
the metallicity distribution of APOGEE red-giants is also shown as the black curve in Figure \ref{fig:dits_FeH} for comparison. We can see that the distribution of both the K2 sample and the \cite{Ghezzi2021} CKS sample have a peak at [Fe/H]=+0.1 dex that is not well matched by the metallicity distribution of red-giants in APOGEE. Compared to APOGEE, the K2 sample is relatively more metal poor, not extending to [Fe/H]=$+0.4$, having fewer stars in the metal-rich end, and a more significant number of stars in the [Fe/H]$=-0.2$ bin than the APOGEE distribution.
The median metallicity of the CKS sample is [Fe/H]=+0.06 $\pm$ 0.14 \citep{Ghezzi2021}, or, [Fe/H]=$+0.04 \pm 0.11$ \citep{petigura2017}; the K2 sample studied here is more metal-poor by $-0.09$ dex in the median, but we note that it is not as metal-poor as the K2 Campaing 5 sample analyzed in \cite{zink2020AJ....160...94Z}, which has an approximately Gaussian distribution with a median [Fe/H] $=-0.14 \pm 0.18$.

%-----------------------------------------------------------------
%-----------------------------------------------------------------

%------------------------------------------------
\subsection{Planetary Radii and the Radius Gap} \label{sec:pl_radii}
\begin{figure*}
\gridline{\fig{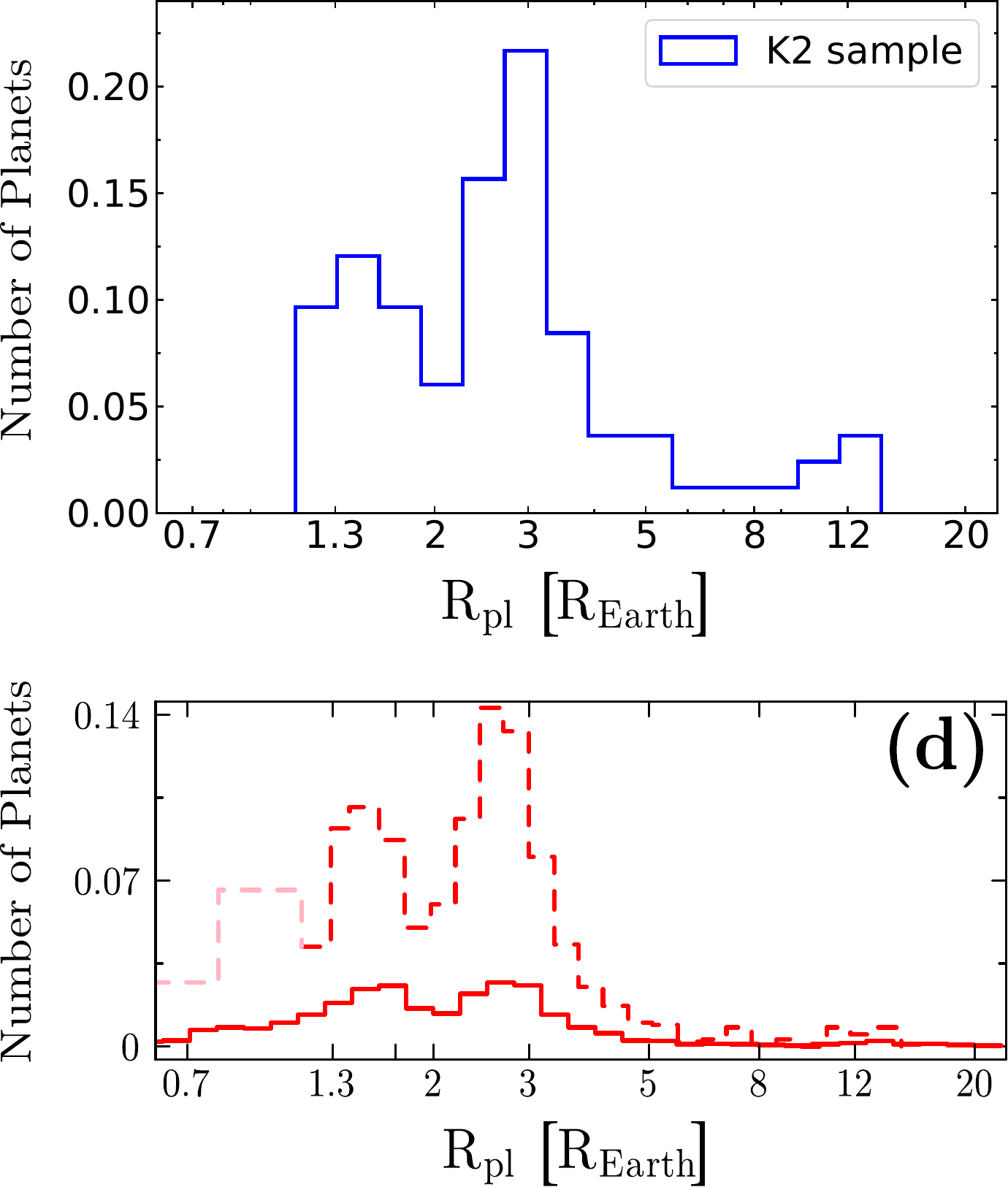}{0.62\textwidth}{}
          }
          \vspace{-7mm}
\caption{The top panel shows the planetary radius distribution for the K2 planet sample studied here, while the bottom panel shows the distribution of the CKS planet sample studied in \cite{martinez2019}, as shown in their Figure 11 panel (d). The red filled line histogram represents their sample and the red dashed line distribution takes into account completeness corrections. The location of the gap in the K2 (top panel) and CKS (bottom panel) radius distributions is approximately the same: R$_{\rm gap}\sim$2 R$_{\oplus}$.}
\label{fig:his_Rpl}
\end{figure*}

%------------------------------------------------
The final distribution of planetary radii derived from the Hydra spectra of K2 host stars contains 85 confirmed planets orbiting 69 stars.  Although this sample is relatively small, the results presented here are derived using an independent methodology that relies on an analysis of a carefully selected set of Fe I and Fe II lines, which
are used to determine fundamental host-star parameters (T$_{\rm eff}$, log $g$, [Fe/H]), and thus represent a useful addition to the growing number of independently-derived K2 planetary radii.
Figure \ref{fig:his_Rpl} (top panel) presents the histogram of K2 planetary radii that result from the derived stellar radii combined with available transit depth catalogs.  The 85 planets plotted in Figure \ref{fig:his_Rpl} reveal a distinct and well-defined radius valley that spans R$_{pl}\sim$1.6$-2.2$ R$_{\oplus}$, with a minimun near 1.9 R$_{\oplus}$.  Figure \ref{fig:his_Rpl} includes all planetary orbital periods which, for this sample, range from P$\sim$0.5 days up to 53 days and it should be noted that, due to the differing observing techniques between Kepler 1 and K2, the sample studied here is biased towards shorter orbital periods when compared to Kepler 1 periods, e.g., the CKS sample.  Nevertheless, the K2 radius valley observed in this sample, with a minimum at R$_{pl}\sim$1.9 R$_{\oplus}$, is very similar to results derived from the CKS sample from several studies \citep[e.g.,][]{fulton2017,berger2018,fulton2018,vaneylen2018,martinez2019}. 

The bottom panel of Figure \ref{fig:his_Rpl} provides a comparison of radii derived for the CKS sample (all Kepler 1 planets) from \citet[][Fig. 11(d)]{martinez2019}. Comparing top and bottom panel of Figure \ref{fig:his_Rpl} illustrates the similarity in the location of the radius valley between the K2 and Kepler 1 samples.
Although the Kepler 1 field focused on a single pointing, encompassing a limited Galactic longitude and latitude (l $\sim$70$^{\rm o}-85^{\rm o}$, b $\sim$10$^{\rm o}-20^{\rm o}$), the K2 fields were constrained by the ecliptic plane and thus ranged over a broader range of Galactic longitudes and latitudes.  The similarity in the position of the radius valley suggests it to be a ubiquitous phenomenon among short-period planets across a broad range of Galactocentric distances in the thin and thick disk populations; \cite{zink2021AJ....162..259Z} arrived at this conclusion after their analysis of planetary radii in the K2 Campaign fields 1$-$8 and 10$-$18.

%%%%%%%%%%%%%%%%%%%%%%%%%%%%%%%%%%%%%%%%%%%%%%%%%%%%%%%%%%
%%%%%%%%%%%%%%%%%%%%%%%%%%%%%%%%%%%%%%%%%%%%%%%%%%%%%%%%%%

\begin{figure*}
\centering
\gridline{\fig{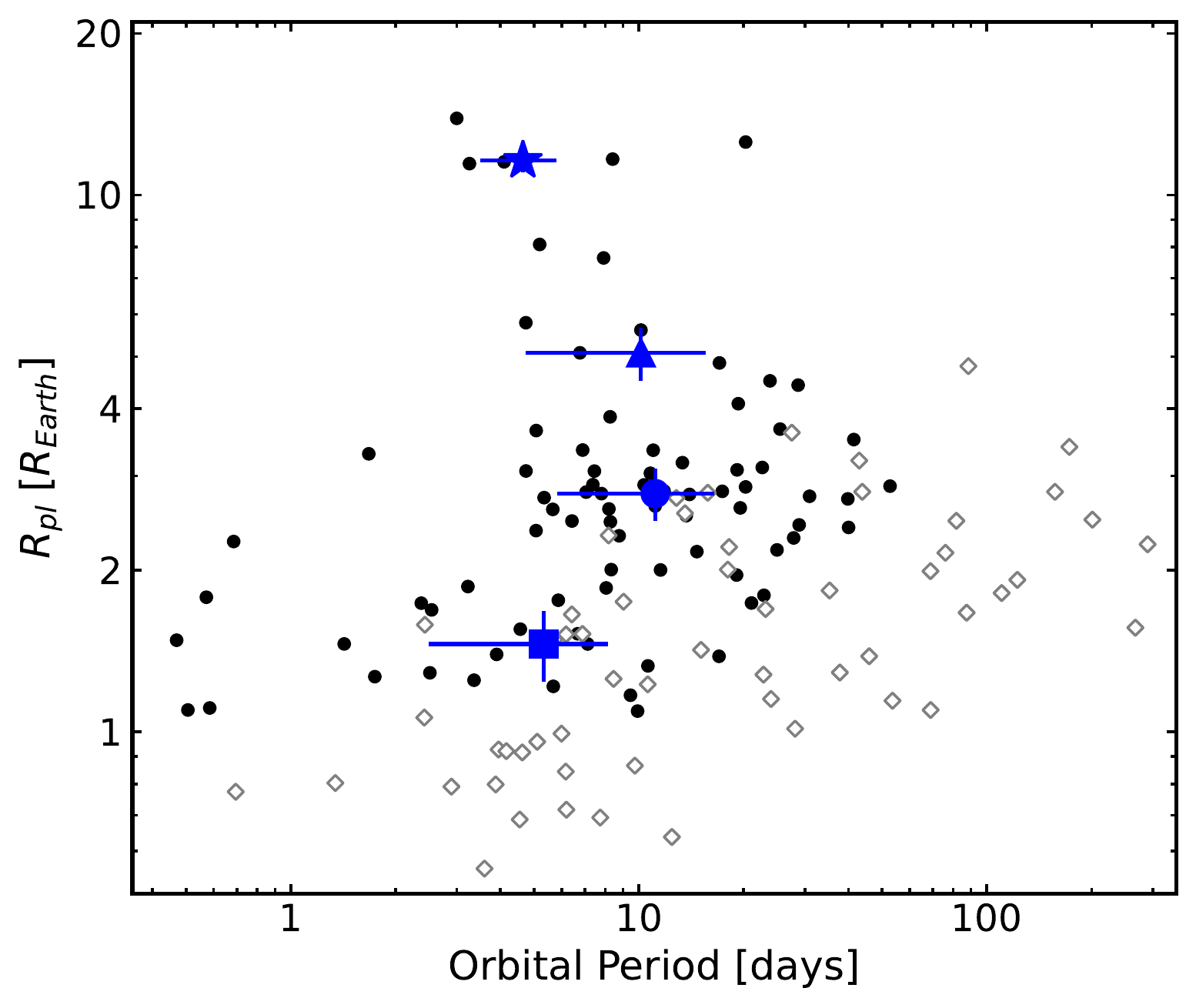}{0.51\textwidth}{}
          }
\gridline{\fig{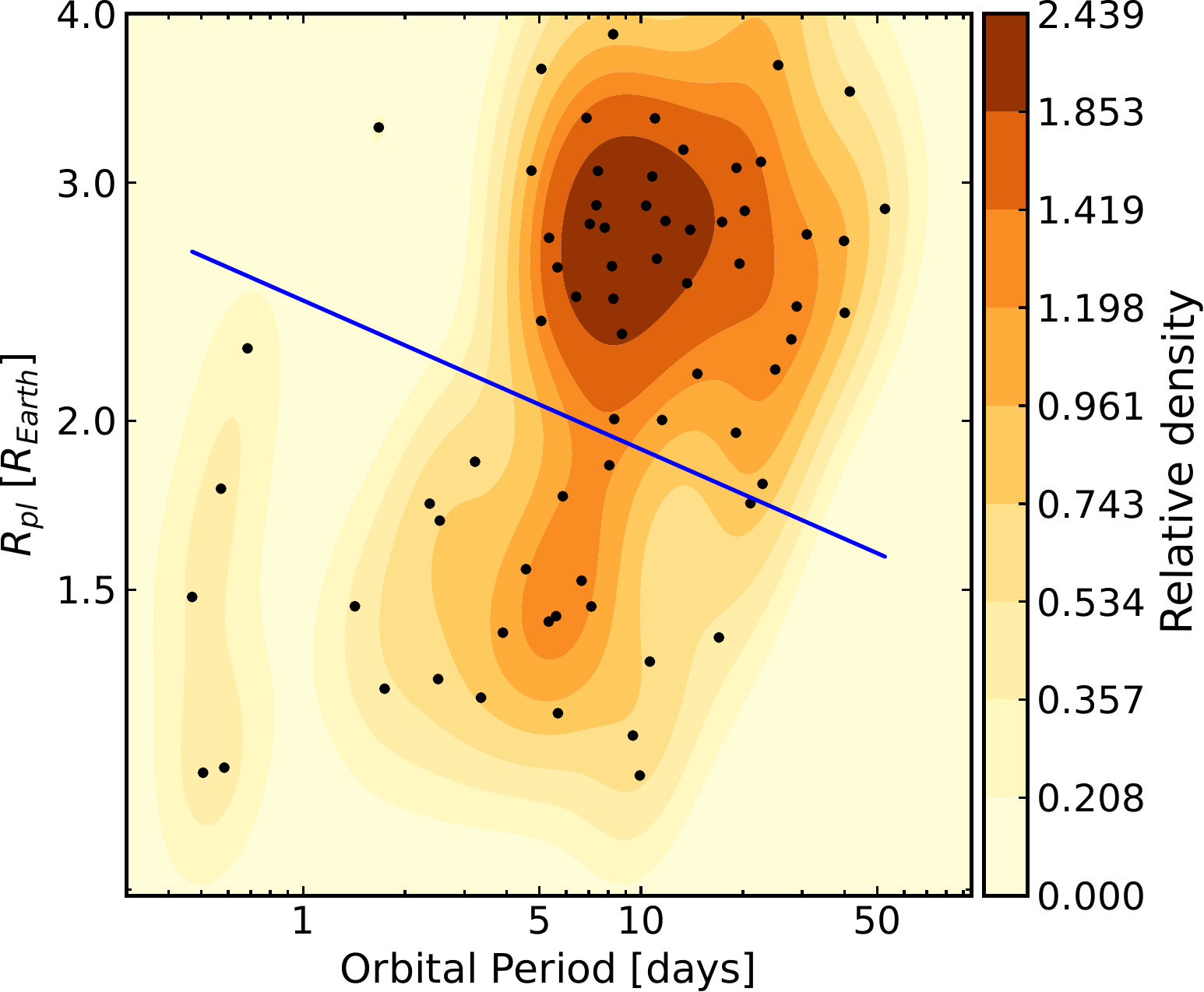}{0.5\textwidth}{}
         \hspace{1mm}
          \fig{Rpl_orbper_4R_k2+kep}{0.5\textwidth}{}
          }
\caption{Planetary radius as a function of planetary orbital period. The top panel shows our sample of K2 planets (filled black circles) and Kepler 1 planets (open diamonds). The blue symbols are the median values of R$_{pl}$ versus orbital period for the Jupiters (8 R$_{\oplus} <$ R$_{pl} \leq 20$ R$_{\oplus}$), Sub-Saturns (4.4 R$_{\oplus} <$ R$_{pl} \leq 8$ R$_{\oplus}$), Sub-Neptunes (2 R$_{\oplus} <$ R$_{pl} \leq 4.4$ R$_{\oplus}$), and Super-Earths (R$_{pl} \leq 2$ R$_{\oplus}$). The bottom left panel shows  K2 planets in our sample with $< 4$ R$_{\oplus}$ and in the bottom right panel we show the same but with the addition of the Kepler 1 planets in our sample. The slope of the radius valley versus orbital period derived in \cite{martinez2019} for the CKS sample is also shown for comparison purposes only. The color bar shows the relative density of detected planets in the $P-R_{pl}$ plane using a Gaussian kernel density estimation (KDE).}
\label{fig:Rpl_orb}
\end{figure*}

With the similarity in the position of the radius valley between K2 and Kepler 1 samples \citep{mayo2018,Hardegree2020,zink2021AJ....162..259Z}, we next investigate planet radius as a function of orbital period, with Figure \ref{fig:Rpl_orb} plotting our planet sample in a period-radius plane. The top panel includes all planets (our full sample) in the different orbital periods, with the K2 planets plotted as black filled circles and the Kepler 1 planets are plotted as open diamonds; blue symbols represent the median and MADs of the planetary radii and orbital period distributions for each planet size domain: super-Earths (R$_{pl} \leq 2$ R$_{\oplus}$, square), sub-Neptunes (2 R$_{\oplus} <$ R$_{pl}$ $\leq 4.4$ R$_{\oplus}$, circle), sub-Saturns (4.4 R$_{\oplus} <$ R$_{pl}$ $\leq 8$ R$_{\oplus}$, triangle), Jupiters (8 R$_{\oplus} <$ R$_{pl}$ $\leq 20$ R$_{\oplus}$, star). These median values are overall similar to those for the CKS sample in \cite{martinez2019},  but here we adopt the limit of 4.4 R$_{\oplus}$ for the transition between sub-Neptunes and sub-Saturns, as discussed in \cite{Ghezzi2021}. 

Planetary radii $< 4$ R$_{\oplus}$ are shown in the bottom panels, where the bottom left panel plots only the sample of K2 planets, while in the bottom right panel includes all K2 planets and those Kepler 1 planets having orbital periods less than 100 days. In both plots the slope of the radius valley is shown as a blue line \citep[we used the slope $-$0.11 from][]{martinez2019}. Although we do not fit a trend of the position of the radius gap, R$_{\rm gap}$, as a function of orbital period (P) to the 139 planets in our sample, the planetary radii that result from the stellar parameters derived here are consistent with the relation of the radius valley following a power law of the form R$_{\rm gap}\propto$ P$^{-0.11}$, as shown in the bottom panels of Figure \ref{fig:Rpl_orb}.

%%%%%%%%%%%%%%%%%%%%%%%%%%%%%%%%%%%%%%%%%%%%%%%%%%%%%%%%%%

%%%%%%%%%%%%%%%%%%%%%%%%%%%%%%%%%%%%%%%%%%%%%%%%%%%%%%%%%%

\subsection{Planetary Radius, Orbital Period and Stellar Metallicity}

\begin{figure*}
\centering
\gridline{\fig{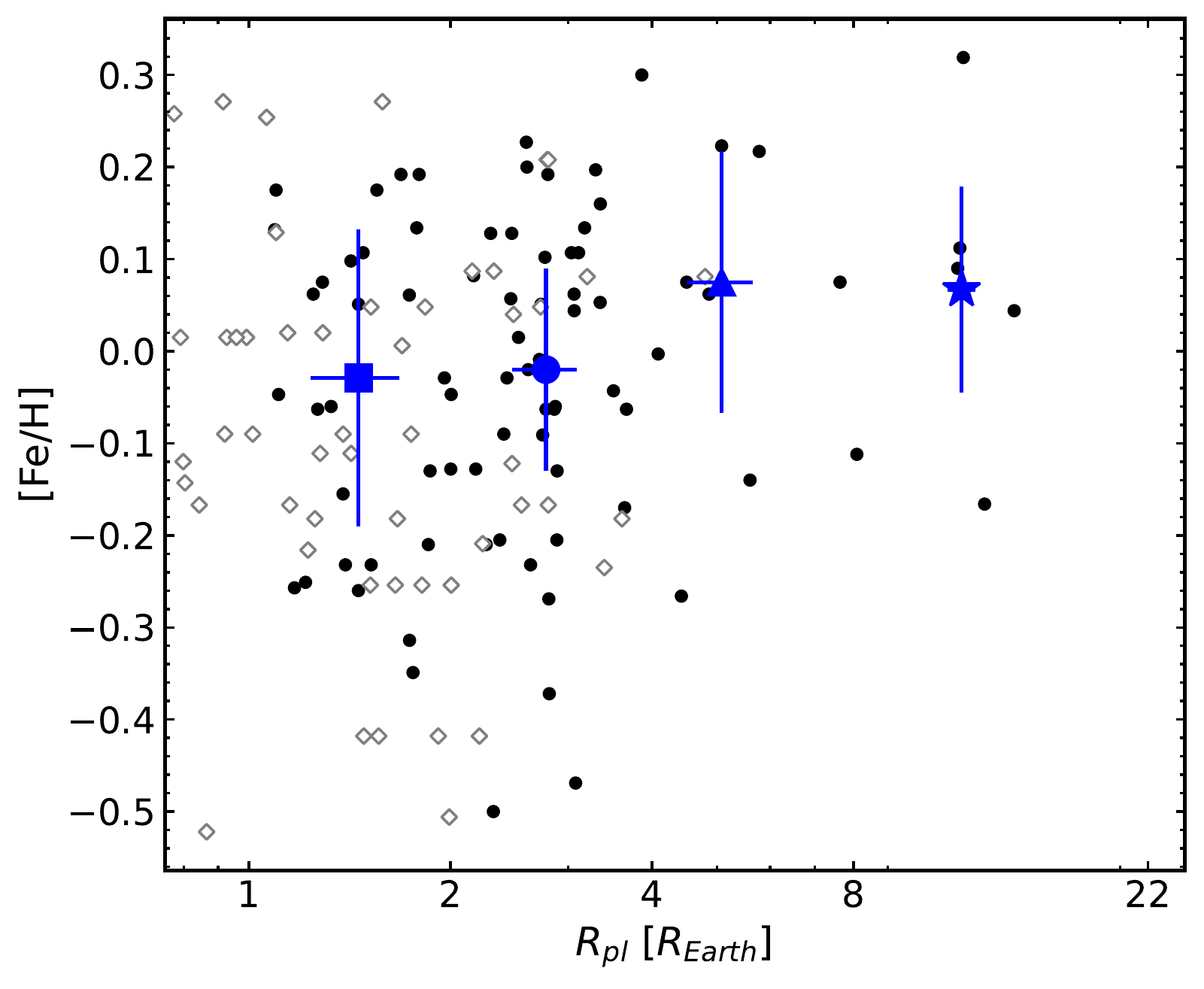}{0.51\textwidth}{}
         \hspace{-3mm}
          \fig{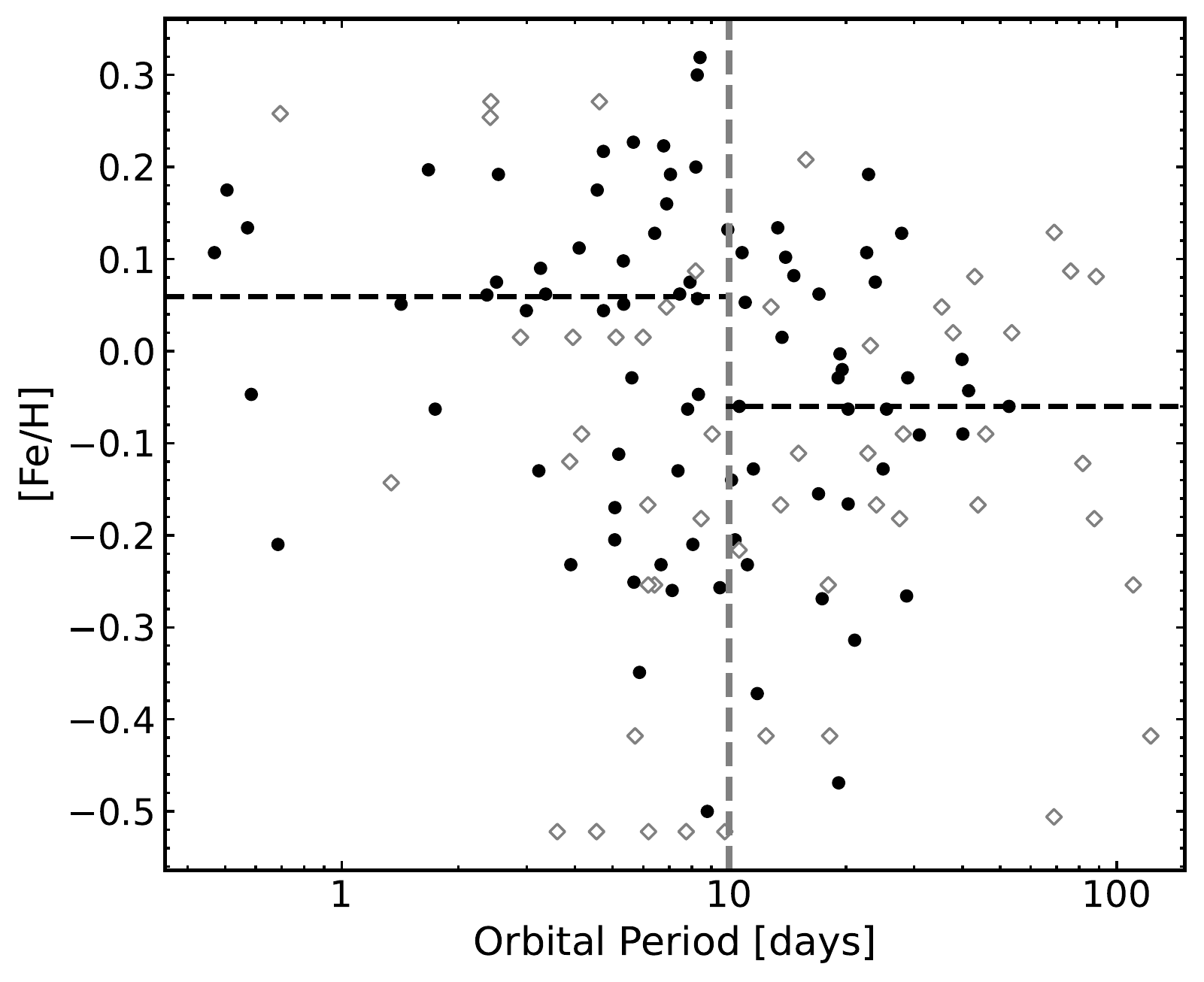}{0.51\textwidth}{}
          }
\caption{Host star metallicities as a function of the planetary radius (left panel) and orbital period (right panel). K2 planets are shown as black filled circles and Kepler 1 planets as open diamonds. Left Panel: The blue symbols are the median host star metallicities for Super-Earths (square), Sub-Neptunes (circle), Sub-Saturns (triangle) and Jupiters (star). Right Panel: the horizontal dashed lines are the median host star metallicities for planets having orbital periods smaller and larger than 10 days.}
\label{fig:FeH_Rpl_orb}
\end{figure*}

The distribution of host star metallicity as a function of planetary radius is shown in the left panel of Figure \ref{fig:FeH_Rpl_orb}. The 85 K2 planets in this study are the solid black circles and the Kepler 1 planets the grey open diamonds. The blue symbols represent the median ($\pm$ MAD) metallicities of K2 planet hosts dividing the sample in the same planet size domains as in Figure \ref{fig:Rpl_orb}: super-Earths, sub-Neptunes, sub-Saturns, and Jupiters. Although the K2 planet sample size is small, we find that, in general, the metallicity of K2 planet hosts increases with planetary radius, but the increase in metallicity is seen in particular for the transition between the small (Super-Earth and Sub-Neptunes) and large (sub-Neptune and sub-Saturn) planet regimes, a result that is similar to what has been found in previous studies of Kepler planets in the literature (\cite{narang2018}, \cite{petigura2018AJ....155...89P}, \cite{Ghezzi2021}, see also \cite{beauge2013}).

As mentioned in the introduction, previous works have investigated correlations between host star metallicity and planet orbital period distributions, concluding that small and hot planets with orbital periods $P\lesssim 8-10$ days appear preferentially around metal-rich stars \citep[e.g.,][]{mulders2016,wilson2018}. The right panel of Figure \ref{fig:FeH_Rpl_orb} shows the host star metallicity as a function of the planet orbital period for our studied sample (symbols are the same as in the left panel).  The horizontal dashed lines represent the mean metallicities for those K2 planets having orbital periods below and above 10 days, and the 10-day boundary is marked as a dashed vertical line. We find that the median host star metallicity for planets with P $<$ 10 days is slightly metal rich, $+0.059 \pm 0.122$ dex, while the median host-star metallicity for planets with P $>$ 10 days is slightly metal poor, $-0.060 \pm 0.106$ dex. Our K2 sample has mostly small planets and only some large planets. If we restrict the sample to that having only planets with $R_{pl}$ $<$ 4.4 R$_\oplus$, we obtain a similar behavior, with a median metallicity of $+0.051 \pm 0.124$ for P $<$ 10 days and $-0.051 \pm 0.090$ for planets with P $>$ 10 days. We note that adopting a boundary at $R_{pl}$=8.3 R$_\oplus$, as found in \cite{wilson2018}, gives similar metallicity differences between the two orbital period regimes.
Summarizing the results obtained here for the K2 host star metallicities and orbital planetary periods are in line with what was found previously for Kepler 1 systems.

%%%%%%%%%%%%%%%%%%%%%%%%%%%%%%%%%%%%%%%%%%%%%%%%%%%%%%%%%%

%%%%%%%%%%%%%%%%%%%%%%%%%%%%%%%%%%%%%%%%%%%%%%%%%%%%%%%%%%

\color{black}
%---------------------------------------------------------
\section{Summary and Conclusions} \label{sec:conclusion}
%---------------------------------------------------------

We present effective temperatures, surface gravities, metallicities, and microturbulent velocities for 81 planet-hosting K2 stars based on a homogeneous spectroscopic analysis using a uniform set of high-resolution spectra from the WIYN/Hydra spectrograph covering a spectral window between $\lambda$6050--6350\AA. Additionally, stellar parameters are presented for 33 planet-hosting Kepler 1 stars.

The calculations were done in LTE and we used Kurucz model atmospheres. The stellar parameters and metallicities were derived from measurements of equivalent widths of Fe I and Fe II lines and from requiring excitation and ionization balance, in addition to requiring the independence of the Fe line abundances with the equivalent widths. Such a methodology has the advantage of directly estimating the microturbulent velocity, which is a needed parameter for an abundance analysis, and being independent of fitting for the projected rotation velocity (v sin i), macroturbulent velocity and the instrumental profile. 

The limited wavelength coverage of the Hydra spectra combined with a resolving power (R= $\lambda/\Delta \lambda$ =18,500) that is lower than that typically found in single object high-resolution spectrographs, demanded a systematic and careful selection of unblended Fe I and Fe II lines combined with a 'boutique' analysis of a benchmark solar twin to assess the reliability of the individual line results. The methodology and line list were then validated via the analysis of two solar proxy spectra (obtained with the Hydra spectrograph) of the asteroids Astraea and Parthenope, which resulted in effective temperatures, surface gravity values and metallicities that are similar to cannonical values in the literature for the Sun \citep[][]{asplund2021AA...653A.141A,magg2022AA...661A.140M}.

Once the line list was vetted, the spectral analysis was done semi-automatically using the code $q^2$ \citep{Ramirez2014} that interpolates atmospheric models, computes iron abundances, and their corresponding errors. Additional stellar properties, such as stellar masses and radii, were also computed using the $q^2$ package, where the input parameters were the derived effective temperatures and metallicities, along with V magnitudes, parallaxes from Gaia EDR3, and their corresponding uncertainties and isochrones from Yonsei-Yale \citep{demarque2004ApJS..155..667D,han2009gcgg.book...33H}

The K2 stellar sample analyzed has distributions of effective temperature, surface gravity, and metallicity that fall mostly between T$_{\rm eff} \sim$ 4800 $-$ 6200 K, log $g$ $\sim$ 3.7 $-$ 4.6, and [Fe/H] $\sim$ --0.5 $-$ +0.3 dex, respectively. These stars all have distances (\cite{bailerjones2021}) within $\sim$900 pc of the Sun, with most of them having stellar radii between $1-2$ R$_\odot$, and only a few stars having radii between 2 - 4 R$_\odot$, their masses vary mostly between $\sim$ M$_{star}=0.7 - 1.2$ M$_\odot$.

Stellar radii and transit depth values are needed to constrain the planetary radii, which is a crucial parameter necessary to unveil planetary composition. Most of the transit depth values for K2 planets here were from \cite{kruse2019}, and the internal precision achieved in the derived planetary radii in this work is 4.44 \%.

The results derived here for T$_{\rm eff}$, log $g$, A(Fe), R$_{star}$, M$_{star}$, and R$_{pl}$ were compared to, and found to be in general agreement with, results obtained in several literature studies that were based on high-resolution spectra and asteroseismology for Kepler 1 stars, as well as results obtained through photometry, spectrum synthesis, and asteroseismology for K2 stars.
More specifically, comparisons between our results and those in spectroscopic studies and surveys for both T$_{\rm eff}$ and log $g$, find agreement within $\sim$100 K and $<0.1$ dex, respectively.  A closer inspection reveals consistent offsets in the sense that the effective temperatures derived here are hotter and the surface gravities slightly lower, with median differences taken over all studies (Other work - This work) of $<\Delta$T$_{\rm eff}> =-37 \pm 38$ K and $<\Delta$log $g>=+0.05 \pm 0.04$ dex.

The possible impact that magnetic stellar activity might have on the derived stellar parameters was investigated by removing Fe I lines that were found to be the most sensitive to Zeeman broadening/enhancement. A subset of 15 stars that were deemed likely to be active, were reanalyzed using these less magnetically sensitive Fe I lines.  No significant differences, beyond expected uncertainties, were found between the results from an analysis that included magnetically sensitive lines when compared to the analysis which excluded such lines.  Based on this exercise, we do not find evidence that the stellar parameters presented here have been biased significantly by underlying stellar activity.

The relations between planet radius and orbital period and metallicity in our small K2 planet sample confirm previous results in the literature for Kepler 1 planets \citep[][]{mulders2016,wilson2018,narang2018,petigura2018AJ....155...89P, Ghezzi2021}. Overall, the metallicity of K2 planet hosts increases with planetary radius, this increase in metallicity is seen in particular for the transition between the small ($<4.4$ R$_\oplus$) and large ($>4.4$ R$_\oplus$) planet regimes. The median K2 host star metallicity for planets with orbital period $<$ 10 days is slightly metal-rich, while the median host-star metallicity for planets with P $>$ 10 days is slightly metal-poor. When we restrict the sample to that having only planets with R$_{pl} <$ 4.4 R$_\oplus$, we obtain a similar behavior.

Previous studies deriving precise values for the planetary radii  \citep{fulton2017,petigura2017,vaneylen2018,martinez2019} have uncovered signatures, such as the radius gap, and the slope in the radius gap with orbital period, which would not be apparent when uncertainties in stellar parameters are higher.
The distribution of K2 planetary radii resulting from the stellar parameters derived here reveals a well-defined radius gap, with a minimum at R$_{pl}\sim$1.9 R$_{\oplus}$.  This gap falls at the same radius value as found for planets orbiting stars found by Kepler 1; the similarity between differing samples of exoplanet-hosting stars inhabiting larger swathes of volume across the Galaxy points to the radius gap as a common feature of short-period (P$<$100 days) exoplanetary systems. Such a conclusion concerning K2 planetary radii was also reached in \cite{zink2020AJ....160...94Z}.

Although the number of K2 planetary radii derived here is relatively small, the radius gap as a function of orbital period defined by this sample is in agreement with the decreasing value of R$_{pl}$ with increasing orbital period (P) found in previous studies of Kepler 1 planets \citep[e.g.,][]{vaneylen2018,martinez2019}, the minimum of the radius gap follows a trend of R$_{\rm gap}$ $\propto$ P$^{-0.11}$. 

This study adds to the list of K2 hosts with stellar parameters from high-resolution spectra, which is crucial for the field of exoplanet studies.

\clearpage

%\begin{acknowledgments}

We thank the referee for detailed comments that helped improve the paper. We thank Ivan Ramirez for his support regarding the use and handling of the qoyllur-quipu code. V. L-T. acknowledges the financial support from Coordena\c{c}\~{a}o de Aperfei\c{c}oamento de Pessoal de Nível Superior (CAPES). S.C.S was supported by David Delo Research Professor and Dana Foundation grants from the University of Tampa.
Based on observations at Kitt Peak National Observatory at NSF’s NOIRLab (NOIRLab Prop. ID 2019A-0334; PI: V. Smith), which is managed by the Association of Universities for Research in Astronomy (AURA) under a cooperative agreement with the National Science Foundation. The authors are honored to be permitted to conduct astronomical research on Iolkam Du’ag (Kitt Peak), a mountain with particular significance to the Tohono O’odham.  Data presented were obtained at the WIYN Observatory, operated by NOIRLab, under the NN-EXPLORE partnership of the National Aeronautics and Space Administration and the National Science Foundation. This work was supported by a NASA WIYN PI Data Award, administered by the NASA Exoplanet Science Institute. This research has made use of the NASA Exoplanet Archive, which is operated by the California Institute of Technology, under contract with NASA under the Exoplanet Exploration Program.

%\end{acknowledgments}

%\bibliographystyle{astron}
%\bibliographystyle{yahapj}
%\bibliography{references}

\bibliography{manuscript}{}
\bibliographystyle{aasjournal}

\end{document}